\documentclass[12pt]{article}

\pdfoutput=1

\usepackage{amssymb,amsmath}
\usepackage{fullpage}
\usepackage{graphicx}
\usepackage{amsmath}		
\usepackage[margin=0.8in]{geometry}
\usepackage{setspace}
\usepackage{color}
\usepackage{fancyhdr}
\usepackage{collcell}
\usepackage{datatool}
\usepackage{environ}
\usepackage{latexsym}
\usepackage{amssymb}
\usepackage{epsfig,amsmath,graphics}
\usepackage{epstopdf}
\usepackage{verbatim}
\usepackage{wasysym}
\usepackage{feynmp-auto}
\usepackage{authblk}
\usepackage{xcolor}
\usepackage{enumitem}
\usepackage[utf8]{inputenc}
\usepackage{slashed}
\usepackage{cite}
\usepackage[toc,page]{appendix}
\usepackage[normalem]{ulem}
\usepackage{hyperref}

\usepackage{empheq}

    \newlength\fsep
    \newsavebox\widebox

\definecolor{deepgreen}{rgb}{0.2,0.8,0.3}

\definecolor{deepblue}{rgb}{0.2,0.2,0.8}

\definecolor{deepred}{rgb}{0.8,0.2,0.2}

\usepackage[skip=0pt]{caption}

\def\r{\right)}
\def\l{\left(}

\newcommand{\Wf}{\Psi[\phi(\vec{x})]}

\title{Cosmohedra
    \includegraphics[scale=0.03]{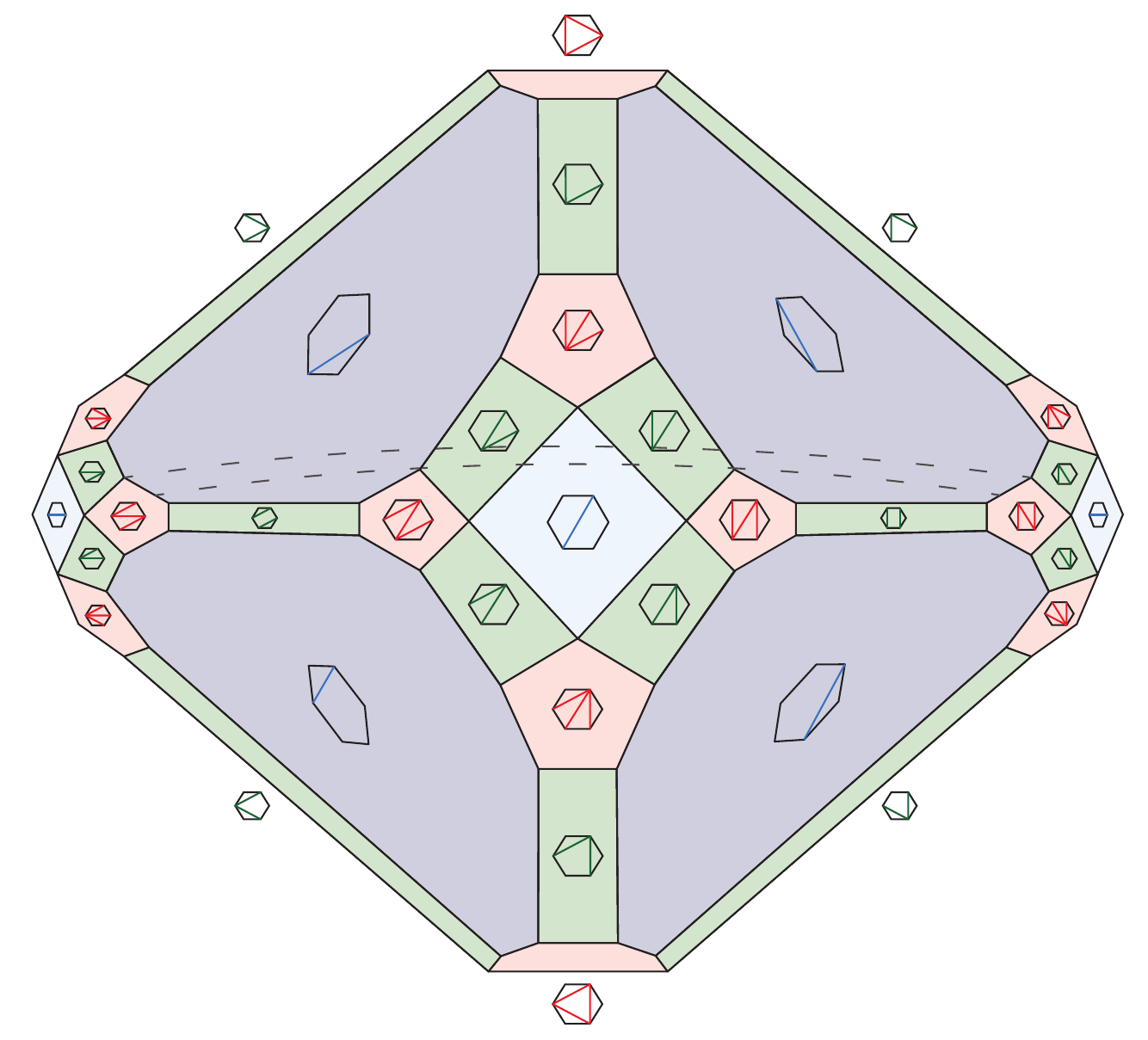}}
\author[1]{Nima Arkani-Hamed}
\author[2]{Carolina Figueiredo}
\author[3]{Francisco Vazão}
\affil[1]{\small School of Natural Sciences, Institute for Advanced Study, Princeton, NJ 08540}
\affil[2]{\small Department of Physics, Princeton University, Princeton, NJ 08540}
\affil[3]{\small Max-Planck-Institut f\"ur Physik, Werner-Heisenberg-Institut, Boltzmannstr, 8, 85748 Garching, Germany}

\begin{document}
\hspace{380pt}{MPP-2024-265}

%This make title command is this way so I can have the preprint number, otherwise it creates a new page.
{\let\newpage\relax\maketitle}

\vspace{5mm}

\begin{abstract}

It has been a long-standing challenge to find a geometric object underlying the cosmological wavefunction for Tr($\phi^3$) theory, generalizing associahedra and surfacehedra for scattering amplitudes. In this note, we describe a new class of polytopes -- ``cosmohedra'' -- that provide a natural solution to this problem. The faces of associahedra capture the combinatorics of non-overlapping chords of the momentum polygon, reflecting all partial factorizations of amplitudes.  Cosmohedra are far richer -- instead of non-overlapping chords, their faces capture the ``russian doll'' structure of non-overlapping subpolygons that determine the wavefunction. We show that cosmohedra are intimately related to associahedra, obtained by ``blowing up" faces of the associahedron in a simple way. We give a full combinatorial description of cosmohedron faces and their factorization properties, and provide an explicit realization in terms of facet inequalities that further ``shave" the facet inequalities of the associahedron. We also discuss a novel way for computing the wavefunction from cosmohedron geometry that extends the usual connection with polytope canonical forms. We illustrate cosmohedra with examples at tree-level and one loop; the close connection to surfacehedra suggests the generalization to all loop orders. Moving beyond the  wavefunction, we  briefly describe ``cosmological correlahedra" for full correlators, which are one higher-dimensional polytopes, interpolating between associahedra and cosmohedra on opposite facets in an extra direction associated with the total energy. We speculate on how the existence of cosmohedra might suggest a ``stringy" formulation for the cosmological wavefunction/correlators, generalizing the way in which the Minkowski sum decomposition of associahedra naturally extend particle to string amplitudes. 

\end{abstract}

%It has been a long-standing challenge to find a geometric object underlying the cosmological wavefunction for Tr($\phi^3$) theory, generalizing associahedra and surfacehedra for scattering amplitudes. In this note we describe a new class of polytopes -- ``cosmohedra'' -- that provide a natural solution to this problem. cosmohedra are intimately related to associahedra, obtained by ``blowing up" faces of the associahedron in a simple way, and we provide an explicit realization in terms of facet inequalities that further ``shave" the facet inequalities of the associahedron. We also discuss a novel way for computing the wavefunction from cosmohedron geometry that extends the usual connection with polytope canonical forms. We illustrate cosmohedra with examples at tree-level and one loop; the close connection to surfacehedra suggests the generalization to all loop orders. We also briefly describe ``cosmological correlahedra" for full correlators. We speculate on how the existence of cosmohedra might suggest a ``stringy" formulation for the cosmological wavefunction/correlators, generalizing the way in which the Minkowski sum decomposition of associahedra naturally extend particle to string amplitudes.

\newpage

\setcounter{tocdepth}{1}
\tableofcontents

%%%%%%%%%%%%%%%%%%%%%%%%%%%%%%%%%%%%%%%%%%%%%%

\section{Amplitudes/Wavefunction, Associahedra/Cosmohedra}

Over the last two decades, many remarkable combinatorial structures underlying flat-space scattering amplitudes have been discovered. At first such structures were identified for simpler toy model theories such as $\mathcal{N}=4$ super Yang-Mills \cite{Amplitudhedron,GrassGeoScatAmp,SMatDuality,WittenTwistorString}, as well as for a simple theory of colored scalars \cite{ABHY,CurveInt,CurveInt2,TropLag} -- Tr($\phi^3$) theory -- but, over the past few years, these have been extended to real world theories \cite{Zeros,Gluons,NLSM,CHY}. 

There are many reasons to expect similar structures to underlie the cosmological wavefunction of the universe, not least because it has long been appreciated that flat-space scattering amplitudes are contained on the ``total energy singularity" of the cosmological wavefunction \cite{CosmoCollider,Maldacena:2011nz, Raju:2012zr,Benincasa:2018ssx, Arkani-Hamed:2018bjr}. Much of the effort in looking for such structures has focused on a simple toy model of conformally coupled scalars with general polynomial interactions in a general FRW cosmology \cite{CosmoPolytopes,Anninos:2014lwa,CosmoBoot,DiffEq_CosmoCorr,KinFlow,Hillman:2019wgh,GeometryCosmoCorr,Benincasa:2024lxe,Goodhew:2020hob,CosmoReview,DualCosmoPolyTriangs,KinFlowLoop,IRSub,MelvillePajer,PajerUnitarityLocality,GoodhewCutCosmoCorr,LoopInts,PokrakaDEs,CosmoSMatrix,SimpCosmoCorr,Sachs2,CosmoTreeTheo,Meltzer,Bittermann,CespedesScott}. 
The cosmological wavefunction in these examples is simply related to the flat-space wavefunction. In turn, for any single graph $G$, the flat-space wavefunction was recognized as being determined by the canonical form of a ``cosmological polytope" determined by $G$, in parallel with the way in which amplitudes are determined by positive geometries \cite{CosmoPolytopes}. 

However, all the magic in amplitudes is seen not one graph at a time, but in the sum over all graphs. This is true not only in gauge theories and gravity, but even in the simplest toy model of colored scalars, the Tr($\phi^3$) theory. At tree-level, the amplitudes for this theory were associated with the canonical form of a famous polytope -- the associahedron. More recently, this theory has been given a new formulation at all loop orders, using ideas related to counting problems attached to curves on surfaces~\cite{CurveInt,CurveInt2}; one aspect of this story is an extension of associahedra at tree-level to ``surfacehedra" at all loop orders. And very surprisingly, this seeming ``toy model" secretly contains realistic theories -- the amplitudes for pions and gluons can be obtained from the ``stringy" Tr$(\phi^3)$ amplitudes by a simple kinematic shift \cite{Zeros,Gluons,NLSM,circles,UnivSplit}. 

It is thus obviously important to look for a combinatorial/geometric structure for cosmology, not one graph at a time, but naturally combining all diagrams together into a single object. Indeed, the discovery of the associahedron for Tr($\phi^3$) theory, already at tree-level, suggested the search for the ``cosmohedron" that would compute the flat-space wavefunction for Tr($\phi^3$) theory. The most obvious thought is to try and ``glue together" cosmological polytopes for the different diagrams into the cosmohedron, but efforts in this direction were not successful. And while there have been many other significant developments exposing the mathematical structures underlying cosmological observables -- from generalizations of the cosmological polytope for computing correlators to the discovery of ``kinematic flow" differential equations directly describing FRW correlators \cite{DiffEq_CosmoCorr,KinFlow} -- the basic question of ``what object combines all diagrams together in cosmology?" has remained open for many years. 

\begin{figure}[t]
    \centering   \includegraphics[width=\linewidth]{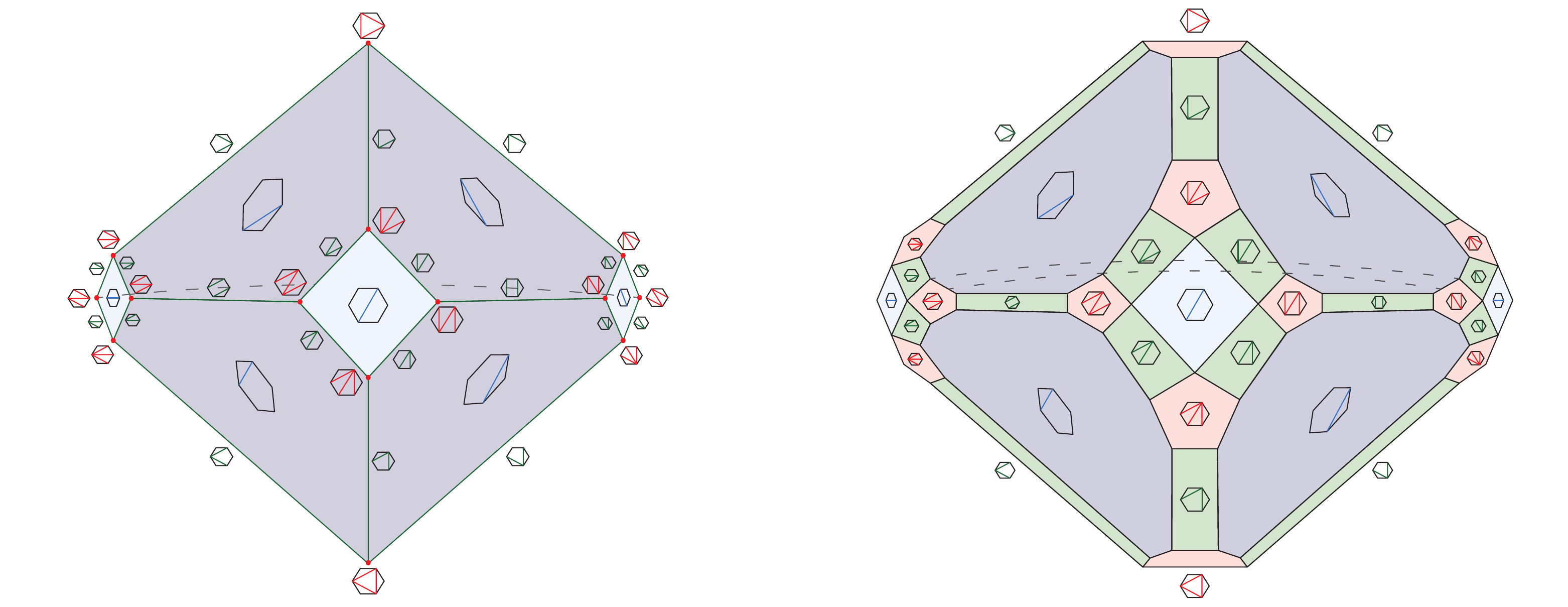}
    \caption{Associahedron (left) and cosmohedron (right) at 6-points}
    \label{fig:sixcosmo}
\end{figure}
In this note, we present a solution to this problem. We will describe ``cosmohedra'', which are polytopes that entirely capture the combinatorial structure of the wavefunction of Tr$(\phi^3)$ theory, in exact parallel with the associahedron for amplitudes. Indeed, as we will see, cosmohedra are very closely connected to associahedra, and can be obtained from associahedra by ``blowing up" the faces on associahedra in a natural way. We will work for simplicity mostly at tree-level, but will give some explicit examples at one-loop, but the way in which cosmohedra are connected to the amplitude geometries obviously extends to all loop orders, and we will explicitly describe some examples at one-loop. 

We also briefly describe ``cosmological correlahedra", which capture all contributions not just for the wavefunction, but for the full correlator. This object is a one-higher dimensional polytope sandwiched between an associahedron on a ``top" facet and a cosmohedron on the ``bottom" facet. 

While cosmohedra are very rich objects, their description, both combinatorially as well as how they are cut out by inequalities, are strikingly simple. At tree-level, for the $n$-pt wavefunction we consider a $n$-gon. The faces of the cosmohedron are associated with collections, $P$,
of non-overlapping sub-polygons, which satisfy the rule that if any subpolygon $p$ in $P$ contains another subpolygon of $P$ inside it, then all of $p$ must be covered by subpolygons in $P$. The cosmohedron captures all such collections in its face structure, where $P^\prime$ is a face of $P$ if $P \subset P^\prime$, that is if $P^\prime$ is a refinement of $P$. This mirrors the combinatorial definition of the associahedron, where faces are labeled by collections of non-overlapping chords, $C$, instead of non-overlapping sub-polygons. 

The inequality definition of the cosmohedron can also be given in one line. We begin with the inequalities cutting out the associahedron: for every chord $I \equiv (i,j)$ of the $n$-gon, we associate a variable $X_I$. The $X$'s are constrained by the ABHY conditions \cite{ABHY}; there is a realization of the associahedron for choice of triangulation, but one especially simple set is determined by $X_{i,j} + X_{i+1,j+1} - X_{i,j+1} - X_{i+1,j} = c_{i,j}$, where $c_{i,j}$'s are the set $\{c_{1,3},c_{1,4},\cdots,c_{1,n-1},c_{2,4},c_{2,5}, \cdots, c_{2,n-1}, \cdots c_{3,5},$ $ \cdots, c_{n-3,n-1}\}$. Under this constraint, the associahedron is cut out by the equations $X_I \geq 0$ for all $I$. The cosmohedra are then obtained by further ``shaving" the associahedron, by imposing an additional set of inequalities. There is a facet of the cosmohedron for every partial triangulation, or what is the same, every collection of non-overlapping chords, $C$. The inequalities defining the cosmohedron are then simply $\sum_{I \subset C} X_I \geq \sum \delta_{P}$, where the sum is over all the subpolygons defined by $C$, and $\delta_P$ are positive constants that are taken to be much smaller than the $c_{i,j}$'s. The $\delta_P$ must satisfy the crucial inequalities $\delta_P + \delta_{P^\prime} \leq \delta_{P \cup P^\prime} + \delta_{P \cap P^\prime}$, (with $\delta_{{\rm full}} = 0$ for the full polygon is defined to be zero).  This is all that is needed to define and study cosmohedra at tree-level. Everything  else that follows in this note consists only of  motivation, exposition and examples.    

Our aim in this note is to define and explore the most basic properties of cosmohedra. We have tried to make the presentation largely self-contained, without assuming previous knowledge of the physics of the cosmological wavefunction or the combinatorial geometry of amplitudes. We give a lightning introduction to these two topics in appendices \ref{app:pertwf} and \ref{appendixAssoc}. Cosmohedra also feature a new geometry associated to a single diagram --``graph associahedra" -- different from the cosmological polytopes of \cite{CosmoPolytopes}, whose properties and relation with cosmological polytopes we describe in greater detail in appendix \ref{app:CosmoPolytopes_GraphAssoc}. As we will see, there are many physical and mathematical novelties associated with cosmohedra and cosmological correlahedra, and a great deal remains to be understood about these objects both physically and mathematically. We believe the existence of these remarkable objects is a clear indication of a new world of ideas extending combinatorial geometries to cosmology, and we hope this note will help stimulate further developments.

\section{Russian dolls, subpolygons and flat space $\Psi$}

\label{sec:Def_Wf}

In this note, we will be studying the wavefunction for a theory of colored scalars interacting via a cubic interaction with the following action:
\begin{equation}
    \mathcal{S}[\phi]=\int d^d x d\eta \, \frac{1}{2}\text{Tr}\left( \partial \phi\right)^2 - \frac{\lambda_3(\eta)}{3}\text{Tr}\, \phi^3 ,
    \label{eq:action}
\end{equation}
where the background spacetime is flat, but we allow for general time-dependent couplings $\lambda_3(\eta)$. In particular, for $\lambda_3(\eta) = \lambda_3\, a(\eta)^{-(d-1)/2+2}$ we can do a Weyl rescaling to obtain the action of conformally coupled scalars in an FRW cosmology with scale factor $a(\eta)$ \cite{CosmoPolytopes,CosmoLightStates,CosmoReview}. In the rest of the note, we will focus on the flatspace case $\lambda_3(\eta)=const.$ since, as explained in \ref{sec:CosmoFlat} and in greater detail in appendix \ref{app:pertwf}, starting with the flatspace answer we can obtain the cosmological one via a simple integral transform. 

After performing the path integral that defines $\Psi$, we can write it as follows:
\begin{figure}[t]
    \centering \includegraphics[width=\linewidth]{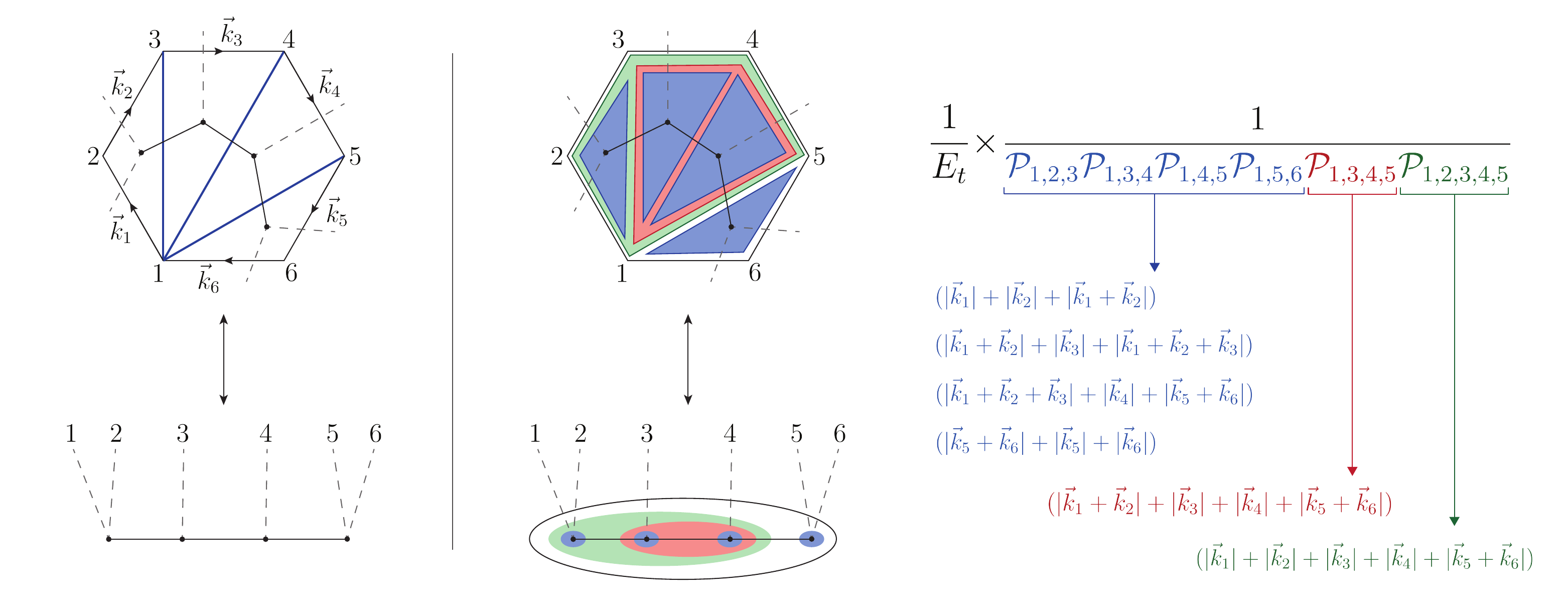}
    \caption{(Left) Triangulation of the $\vec{k}$ $6$-gon and the respective dual cubic diagram. (Right) Russian doll on the momentum $6$-gon and respective tubing. Associated with each Russian doll, $\mathcal{R}$, a factor of 1 over the product of the perimeters of the subpolygons entering in $\mathcal{R}$. Any russian doll always contains the full polygon whose perimeter is the sum of the $|\vec{k}_i|$ which we call the total energy $E_t$. }  
    \label{fig:RD_tubings}
\end{figure}
\begin{equation}
    \Psi = \exp\left\{ \sum_{n\geq 2} \int \prod_{i=1}^n\, d^d k_i \, \Psi_n[\vec{k}_{i} ]\, \delta^d\left(\textstyle{\sum_i} \vec{k}_i\right)  \right\},
\end{equation}
where $\Psi_n[\vec{k}_{i} ]$ are called the \textit{wavefunction coefficients} that capture the contributions to the path integral with $n$ field insertions at the boundary $\eta=0$. The $\Psi_n[\vec{k}_{i} ]$'s admit a perturbative expansion in $\lambda_3$ that we review in appendix \ref{app:pertwf}. As it was first shown in \cite{CosmoPolytopes}, for a particular ordering of the external states, stripping out the color-factor Tr$(\phi_1 \phi_2 \cdots \phi_n)$, we can write $\Psi_n[\vec{k}_{i} ]$ as a sum over all the diagrams compatible with the ordering. From each diagram, we consider all the possible collections of compatible subgraphs -- \textit{tubings} (see figure \ref{fig:RD_tubings} middle bottom, for an example of a tubing of 6-point tree diagram)-- and from each tubing get a factor of the product of one over the energies entering each subgraph. So we can write
\begin{equation}
    \Psi_n = \sum_{\text{diagrams, } \mathcal{D}} \left( \sum_{\text{tubings } t \text{ of } \mathcal{D} } \left(\prod_{\text{subgraph, } s\in t} \frac{1}{E_b}\right)\right).
\end{equation}
To compare the wavefunction with the amplitude, it is useful to recast the tubing picture in terms of subpolygons living inside the spatial-momentum polygon. Since we have spatial momentum conservation, $\sum_{i=1}^n \vec{k}_i=0$, we can draw the momentum $\vec{k}_{i}$ tip to toe, according to the ordering in study, and from it, we obtain a closed polygon -- the spatial momentum $n$-gon. Now we can specify a tree-level diagram by looking at a given \textit{triangulation} of the $n$-gon -- $i.e.$ a way of dividing the $n$-gon into triangles using internal chords of the $n$-gon (see figure \ref{fig:RD_tubings}, left, for an example for $n=6$). 

So in this language, a graph is associated with a triangulation of the momentum $n$-gon, which determines a set of triangles, and quite nicely, a tubing is then a maximal collection of non-overlapping subpolygons\footnote{where non-overlapping means that their edges don't cross, but they can be, however, fully contained inside each other.} -- which we call a \textit{Russian doll} -- containing the set of triangles determined by the triangulation (see figure \ref{fig:RD_tubings} right). Note that this means that for a given graph, any tubing of the graph will always contain the tubes enclosing the vertices -- the triangles -- as well as a big tube which encloses the full graph -- the full momentum polygon. For each subgraph in the tubing, we get a factor of the sum of the energies, $E_i=|\vec{k}_i|$, entering the subgraph. At the level of the momentum $n$-gon/Russian doll this corresponds precisely to the perimeter of the subpolygon associated with the corresponding subgraph --  $\mathcal{P}_{i,\dots,j}$ is the perimeter of subpolygon with vertices $\{i,\dots,j\}$, (figure \ref{fig:RD_tubings},right).

Thus, we have our most basic expression for the wavefunction, given as a sum over maximal sets ${\bf P}$ of non-overlapping sub-polygons: 
\begin{equation}
\Psi = \sum_{{\bf P}} \prod_{P \subset {\bf P} } \frac{1}{{\cal P}_P}.
\end{equation}
Since every ${\bf P}$ includes the triangles of a triangulation, we can also write this as a sum over all diagrams/triangulations $T$, together with a sum over all the ``russian dolls" associated with the diagram $R_T$, as 
\begin{equation}
\Psi = \sum_{T} \sum_{R_T} \prod_{P \subset R_T} \frac{1}{{\cal P}_P}.
\label{eq:wf_TRussianDolls}
\end{equation}
This formulation makes manifest the extra complexity of the wavefunction as compared with the amplitude (which we get back to in section \ref{sec:CompWf_Amp}) -- while for the amplitude we get a simple sum over cubic diagrams/triangulations, and we get a single factor for each, in the wavefunction we have an extra sum over all the russian dolls compatible with the triangulation. So we have that, manifestly, while Tr$(\phi^3)$ amplitudes, $\mathcal{A}_n$, are about maximal collections of \textit{non-overlapping chords}, that define full triangulations of the $n$-gon, the wavefunction, $\Psi_n$ is about maximal collections of \textit{non-overlapping subpolygons}, that define full russian dolls on the $n$-gon. 

Nonetheless, when we go on the residue of the total energy, $E_t=0$, the wavefunction highly simplifies and gives us precisely the flatspace amplitude:
\begin{equation}
   \mathop{\mathrm{Res}}_{E_t=0} \Psi_n^{\text{tree}} = \mathcal{A}_{n}. 
\end{equation}
So far we have given more emphasis to the tree-level case, but all of the above discussion extends to all loops. Already at tree-level, we can replace the momentum $n$-gon by a disk with $n$ marked points on the boundary (following the appropriate color-ordering), and where each boundary component is assigned a momentum $\vec{k}_i$. The subpolygons were then defined by collections of boundary edges and internal chords, whose perimeter was just the sum of the length of each of these. In the disk case, the subpolygons correspond to subsurfaces bounded by boundary components as well as internal curves going from marked points to marked points. We can determine the perimeter of the subsurfaces as the sum of the absolute values of the curves/boundary components bounding the subsurface, where the momentum associated to a given curve is read by \textit{homology}. But as with the story of ``surface kinematics"~\cite{SurfaceKin} for amplitudes on surfaces, it is fruitful to think of more general kinematic variables  associated with the curve on the surface (in general, up to {\it homotopy}), instead of relating it to a set of momenta. In the context of the wavefunction at tree-level, this means that we can think of the perimeters of each subpolygon as independent variables. 

At $n$-points one-loop, the surface we get is instead a punctured disk with $n$-marked points. In this case, to provide a basis of homology on top of assigning momentum to the boundary components of the disk, we also have to give momentum to one of the curves starting in a boundary marked point and ending on the puncture -- this corresponds to the spatial \textit{loop momentum}. Once we have done this, we can again read off the momentum of any curve, $\vec{k}_C$, on the surface by homology. Finally, just like at tree-level, we can list all possible cubic graphs by considering all the possible triangulations of the punctured disk, and the wavefunction is then given as a sum over all russian dolls -- which are now maximal collections of non-overlapping subsurfaces -- where to each subsurface we get a factor of its perimeters -- the sum of $|\vec{k}_C|$ for each curve $C$ bounding the subsurface. The same picture holds at all orders in the topological expansion, where for each order we have a different surface. And again, we will consider more general kinematic variables for the wavefunction as being labelled by subsurfaces of the surface bounded by curves up to homotopy, which can be specialized to the perimeters when written in terms of momenta determined by homology.  Most of this note will focus on the tree-level wavefunction, but in section \ref{sec:Loop} we explain how our results extend to loop-level. 

Finally, there is an obvious recursive expression for the wavefunction~\cite{DiffEq_CosmoCorr}, trivially generalizing the recursive expression as a sum over cuts for single graphs given in \cite{CosmoPolytopes}. We can phrase this at any loop order in terms of the perimeters ${\cal P}_S$ for any surface $S$, as 
\begin{equation}
\Psi_{S} = \frac{1}{{\cal P}_S} \sum_{\text{curves } C} \Psi_{S/C}
\label{eq:wf_CutRec}
\end{equation}
where we sum over all curves $C$, and $S/C$ is the simpler surface obtained by cutting $S$ along $C$. 

Before proceeding, let's review how we can connect this formulation of the flat-space wavefunction to the wavefunction of more general FRW cosmologies described in the beginning of this section. 

\subsection{Flat Space $\to$ Cosmological Wavefunction}
\label{sec:CosmoFlat}

As explained in appendix \ref{app:pertwf}, for the case where the cubic coupling has some general time-dependence $\lambda_3(\eta)$, it is useful to analyze each Fourier mode, $\lambda_3(\varepsilon)$,  separately. In which case, for each cubic vertex, $\lambda_3(\varepsilon)$ produces a shift in the energies entering the vertex by $\varepsilon$. So, for a general graph, we have that the energies associated to each cubic vertex are shifted by the energies $\varepsilon_i$ associated to the couplings. This can be rephrased in terms of the perimeters of subpolygons as follows: the perimeters associated to the triangles, $t_i$ entering the triangulation are shifted by the respective energy, $\mathcal{P}_{t_i} \to \mathcal{P}_{t_i} + \varepsilon_i$, and for a generic subpolygon, $P$, we have $\mathcal{P}_P \to \mathcal{P}_P + \sum_{t_i \subset P} \varepsilon_i$.

Therefore, having obtained the wavefunction for a single graph, $G$, it is easy to obtain the corresponding cosmological wavefunction, in the following way 
\begin{equation}
\Psi^{{\rm Cosm}}_G = \int_{-\infty}^\infty \left(\prod_{{\rm triangles} \, t_i} d \varepsilon_i \, \lambda_3(\varepsilon_i) \right) \Psi^{{\rm Flat}}_G \left({\cal P}_P \to {\cal P}_P + \sum_{t_i \subset P}\varepsilon_i \right),
\end{equation}
where we associate a shift $\varepsilon_i$ with all the $(n-2)$ triangles $t_i$ in the triangulation, and shift every perimeter ${\cal P}_P$ of a sub-polygon $P$ by the sum of the $\varepsilon_i$ for all the triangles $t_i$ contained in $P$, as described earlier. The precise form of $\lambda_3(\varepsilon)$ depends on the time-dependence that we are interested in studying, but already here we see that the combinatorics associated to the flat-space wavefunction coefficients port literally to those of the cosmological wavefunction, with the difference that for the latter we need to further perform this shift and integral.  

We can now describe the same procedure for the full wavefunction, given by the sum over all graphs. The obvious challenge is that the shifts $\varepsilon_i$ seem to be different from graph-to-graph, and this doesn't give us a universal $(n-2)$ dimensional ``$\varepsilon$ integrand'' we simply integrate to get the cosmological wavefunction. 

Fortunately, there is a beautiful solution to this problem, which also arose in labeling general interactions for colored Lagrangians in \cite{TropLag}. 
Let us choose a base triangulation that defines our surface,  and label the triangles in this base triangulation as $(t_1, \cdots, t_N)$. Then, as explained in \cite{TropLag}, {\it every other triangle} on the surface is canonically associated with {\it one} of these $t_i$. In a similar way, any subpolygon $P$ is associated with a collection of triangles, $T_i$, that triangulate it, that can ultimately map it to a collection of triangles in the base triangulation. Therefore, having made a choice of base triangulation, we can unambiguously associate a $\varepsilon_i$ shift to every subpolygon, and we find 
\begin{equation}
\Psi^{{\rm Cosm}} = \int_{-\infty}^\infty \left(\prod_{{\rm triangles} \, t_i} d \varepsilon_i \lambda_3(\varepsilon_i) \right) \Psi^{{\rm Flat}} \left({\cal P}_P \to {\cal P}_P + \sum_{T_i \subset P} \lambda_i \right)
\end{equation}
where here we can choose any triangulation of the subpolygon $P$ we like, as the sum $\sum_{T_i \subset P} \lambda_i$ will be the same for all of them. This map will be explained in more detail in \cite{CosmoFull}.

\subsection{Complexity of the wavefunction vs. amplitude}
\label{sec:CompWf_Amp}

Finally, it is interesting to compare and contrast the ``size" of the amplitude compared with the wavefunction, as this is a qualitative feature of the objects that is very clearly reflected in the geometries we will be discussing. The number of diagrams for amplitudes $A_n$ are famously given by the Catalan numbers, and satisfy an obvious recursion. The number of terms for the wavefunction $\Psi_n$ satisfy a very similar recursion relation, straightforwardly derived from the sum over cuts recursive formula. The recursion relations for $A_n$ and $\Psi_n$ are very similar, 
\begin{equation}
A_n = \sum_{k=3}^{n-1} A_k A_{n+2 - k}, \quad \Psi_n = \sum_{k=3}^{n-1} (k-1) \Psi_k \Psi_{n+2-k}.
\end{equation}
The extra factor of $(k-1)$ for the $\Psi_n$ recursion makes a dramatic difference in the asymptotics. While $A_n$ grows exponentially at large $n$ as $A_n \to 4^n$, $\Psi_n \to n!$ grows factorially. Meanwhile, the number of poles in the amplitude is given by the total number of internal chords for the amplitude, which scales as $n^2$, while the number of poles of the wavefunction is given by the total number of subpolygons, which scales like $2^n$. Thus, the associahedron is an object with polynomially many facets, and exponentially many vertices, while the object encoding the combinatorics for the wavefunction -- the cosmohedron -- will be an object with exponentially many facets and factorially many vertices. 

More precisely, we have the asymptotics for the number of vertices 
\begin{equation}
\# {\rm Vertices}({\rm Assoc})_n \to \frac{4^n}{n^{3/2} \sqrt{\pi}}, \quad \# {\rm Vertices}({\rm Cosmo})_n \to c n^4 n! 
\label{eq:Vert_count}
\end{equation}
 where $c = 0.05\cdots$ is a constant. Meanwhile, the number of facets of the associahedron is $n(n-3)/2$, while as we will see the total number of facets of the cosmohedron is equal to the total number of all (not just complete) triangulations of the $n$-gon, and we have 
\begin{equation}
\# {\rm Facets}({
\rm Assoc})_n \to \frac{n^2}{2},\quad  \# {\rm Facets}({\rm Cosmo})_n \to \frac{c}{n^{3/2}} (3 + \sqrt{8})^n
\label{eq:Facet_count}
\end{equation}
where $c = 0.04 \cdots$ is a constant.

\section{Graph associahedra}
\label{sec:Graph_Assoc}

Before proceeding to the full wavefunction, we are going to start by looking at the contributions from each graph separately, and understand how to encode the combinatorics of the different russian dolls we obtain. A geometric description of the contribution of each graph for the wavefunction was first proposed through the so-called ``cosmological polytopes" in \cite{CosmoPolytopes}. In this section, we will discuss a different geometry that captures the combinatorics of a given graph, that naturally arises when we think of the kinematic variables as naturally associated with perimeters of subpolygons (and assume these are independent). In appendix \ref{app:CosmoPolytopes_GraphAssoc}, we explain how graph associahedra are related to cosmological polytopes.

For any triangulation $T$, we start by producing an associated dual graph $G_T$, placing a node in the middle of the face of every triangle, and connecting nodes if the corresponding triangles share an edge. The russian dolls associated with this graph are ``tubings" of the graph. As highlighted earlier, each russian doll term will include the total energy/the perimeter of the full $n$-gon -- the biggest ``tube" containing the entire graph -- as well as the perimeters of each triangle in the triangulation -- the small tubes encircling each node in the graph. It is thus not necessary to keep track of these tubes, as they are in common to all tubings. Each term in the russian doll sum is then associated with a maximal collection of non-overlapping tubes, where we do not include the total tube or the small tubes encircling each node. 

This leads us to define the ``graph associahedron", ${\cal A}_G$, associated with any graph to be a polytope whose face structure reflects these tubings: the facets of the ${\cal A}_G$ are individual tubes, the vertices are complete tubings, and faces of general dimension are partial tubings $\tau$ with the obvious notion of inclusion: if $\tau^\prime \subset\tau$, the faces for $\tau^\prime$ belongs to that of $\tau$. When the graph comes from the triangulation of a $n$-gon, the graph associahedron ${\cal A}_G$ is $(n-4)-$dimensional. It is non-trivial that such a polytope exists, and we will shortly give an explicit description of it in the course of defining the cosmohedron, but we can illustrate with some simple examples. 

We note that our definition of the graph associahedron is different from the one standard in the mathematical literature \cite{PostnikovGA1,PostnikovGA2}, which also captures the combinatorics of tubings but with a different set of rules than ours. We nonetheless give them the same name because ``our" graph associahedron for a graph $G$ turns out to be exactly the same as the usual graph associahedron for a different graph $\tilde{G}$. Beginning with our triangulation, instead of putting nodes in the interior of each triangle, nodes are placed in the middle of each internal edge of a triangle, and two nodes are connected if the corresponding edges meet at a vertex. The ``standard" graph associahedron for $\tilde{G}$ turns out to be the same as ``our" graph associahedron for $G$.

Before proceeding to examples, we would like to point out that graph associahedra have a simple, beautiful factorization property on their facets. Consider any single tube T of a graph $G$, then we have 
\begin{equation}
{\rm Facet}_{T}({\cal A}_G) = {\cal A}_T \times {\cal A}_{G/T},
\label{eq:fact_graphAssoc}
\end{equation}
where $G/T$ is the graph obtained by shrinking all of $T$ to a single point. 

\subsection{5-points example}

At 5-points, let's start by fixing the triangulation to be that containing chords $\{(1,3),(1,4)\}$ (all the remaining triangulations are simply given by cyclic rotations of this one). 
\begin{figure}
    \centering   \includegraphics[width=\linewidth]{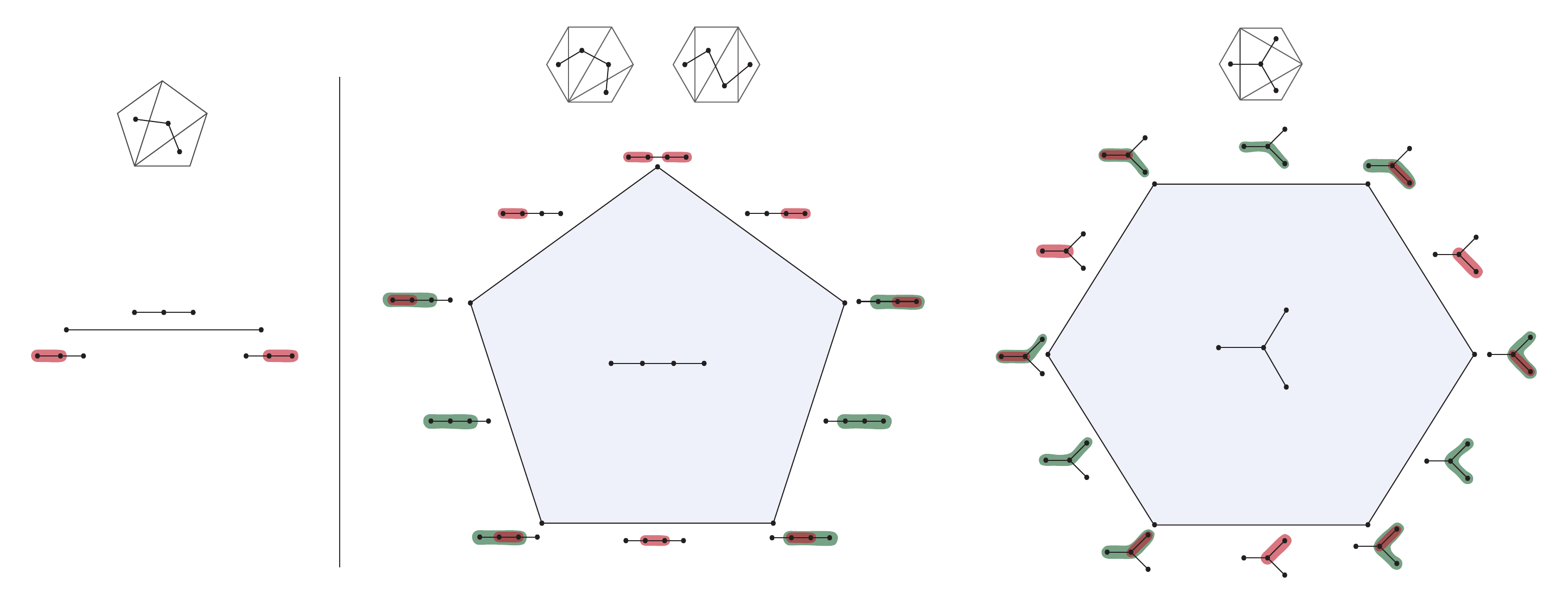}
    \caption{5(left) and 6(right) point graph associahedron. When drawing the graph we omit the external legs to make manifest that for the purpose of the combinatorics of tubings what matters is the topology of the graph with just the internal edges.}
    \label{fig:5_6GraphAssoc}
\end{figure}
In this case, we can factor out from all the russian dolls a factor of $1/(E_t \mathcal{P}_{1,2,3}\mathcal{P}_{1,3,4}\mathcal{P}_{1,4,5})$, and after doing this we get that the contribution to $\Psi^{(5)}$ coming from this graph is simply:
\begin{equation*}
    \Psi^{(5)}_{\{(1,3),(1,4)\}} = \frac{1}{E_t \, \mathcal{P}_{1,2,3} \, \mathcal{P}_{1,3,4} \, \mathcal{P}_{1,4,5}} \left( \frac{1}{\mathcal{P}_{1,2,3,4}}+\frac{1}{\mathcal{P}_{1,3,4,5}} \right),
\end{equation*}
which means we only have two terms. Therefore, we can associate each term with the boundary of a one-dimensional line-segment (see figure \ref{fig:5_6GraphAssoc}, left). Each vertex of the line segment is then associated with a tube which encloses either the left and middle sites, or the right and middle sites. 

\subsection{6-points examples}

At 6-points, there are now different types of triangulation to consider. Let's start with the simplest analog of what we had at $5$-points, $i.e.$ the triangulation containing chords $\{(1,3),(1,4),(1,5)\}$. In this case, we have $5$ different terms which we can write as:
\begin{equation}
\begin{aligned}
    \Psi^{(6)}_{\{(1,3),(1,4),(1,5)\}}  =& \frac{1}{E_t \, \mathcal{P}_{1,2,3}\, \mathcal{P}_{1,3,4}\, \mathcal{P}_{1,4,5}\, \mathcal{P}_{1,5,6}} \left( \frac{1}{\mathcal{P}_{1,2,3,4}\, \mathcal{P}_{1,2,3,4,5}}  +\frac{1}{\mathcal{P}_{1,2,3,4}\, \mathcal{P}_{1,4,5,6}}  \right. \\
    & \left. +\frac{1}{\mathcal{P}_{1,4,5,6}\, \mathcal{P}_{1,3,4,5,6}}+\frac{1}{\mathcal{P}_{1,3,4,5}\, \mathcal{P}_{1,3,4,5,6}}+\frac{1}{\mathcal{P}_{1,3,4,5}\, \mathcal{P}_{1,2,3,4,5}}\right)
\end{aligned}   
\label{eq:graphAssoc6pt_1}
\end{equation}
so we see there are five different subpolygons entering inside the brackets -- $\mathcal{P}_{1,2,3,4},\mathcal{P}_{1,3,4,5},\mathcal{P}_{1,4,5,6}$ squares and $\mathcal{P}_{1,2,3,4,5},\mathcal{P}_{1,3,4,5,6}$ pentagons -- each of which can be associated to an edge of a pentagon, such that each of the five vertices where two edges meet gives one of the terms inside brackets in \eqref{eq:graphAssoc6pt_1} (see figure \ref{fig:5_6GraphAssoc}, center).

So we have that the graph associahedron for $ \Psi^{(6)}_{\{(1,3),(1,4),(1,5)\}}$ is a pentagon. Obviously, the same is true for all the six triangulations that are cyclic rotations of $\{(1,3),(1,4),(1,5)\}$. In addition, it is easy to check that the same is true for the 6 triangulations obtained by cyclic rotations of triangulations $\{(1,3),(3,5),(1,5)\}$ and $\{(1,3),(1,4),(4,6)\}$, $i.e.$ that for all these triangulations once we factor out $E_t$ and the perimeters of the triangles in the triangulation, the rest of the wavefunction has precisely 5 terms that can be associated to vertices of a pentagon in exactly the same way we did for the case of $ \Psi^{(6)}_{\{(1,3),(1,4),(1,5)\}}$. This is ultimately because the associated dual graph for these triangulations where we omit the external legs is also a chain (see figure \ref{fig:5_6GraphAssoc}, middle).

However, if instead we consider triangulation $\{(1,3),(3,5),(1,5)\}$ (or $\{(2,4),(4,6),(2,6)\}$) we have that after we factor out the common part, we are still left with six terms:

\begin{equation}
\begin{aligned}
    \Psi^{(6)}_{\{(1,3),(3,5),(1,5)\}}  =& \frac{1}{E_t \, \mathcal{P}_{1,2,3}\, \mathcal{P}_{3,4,5}\, \mathcal{P}_{1,5,6}\, \mathcal{P}_{1,3,5}} \left( \frac{1}{\mathcal{P}_{1,2,3,5}\, \mathcal{P}_{1,2,3,4,5}}  +\frac{1}{\mathcal{P}_{1,2,3,4,5}\, \mathcal{P}_{1,3,4,5}}  +\frac{1}{\mathcal{P}_{1,3,4,5,6}\, \mathcal{P}_{1,3,4,5}}\right. \\
    & \left. +\frac{1}{\mathcal{P}_{1,3,4,5,6}\, \mathcal{P}_{1,3,5,6}}+\frac{1}{\mathcal{P}_{1,3,5,6}\, \mathcal{P}_{1,2,3,5,6}}+\frac{1}{\mathcal{P}_{1,2,3,5,6}\, \mathcal{P}_{1,2,3,5}}\right)\, ,
\end{aligned}   
\label{eq:graphAssoc6pt_2}
\end{equation}
therefore, for this type of triangulation the graph associahedron is given by a hexagon, where each edge is associated with one of the six subpolygons appearing inside brackets and each vertex associated to one of the terms in \eqref{eq:graphAssoc6pt_2} (see figure \ref{fig:5_6GraphAssoc}, right). Indeed, in this case, the dual graph (after removing the external legs) has a star topology which is different from that of the chain that we found for the remaining triangulations of the hexagon. 

\subsection{7-points examples}

\begin{figure}[t]
    \centering
    \includegraphics[width=\linewidth]{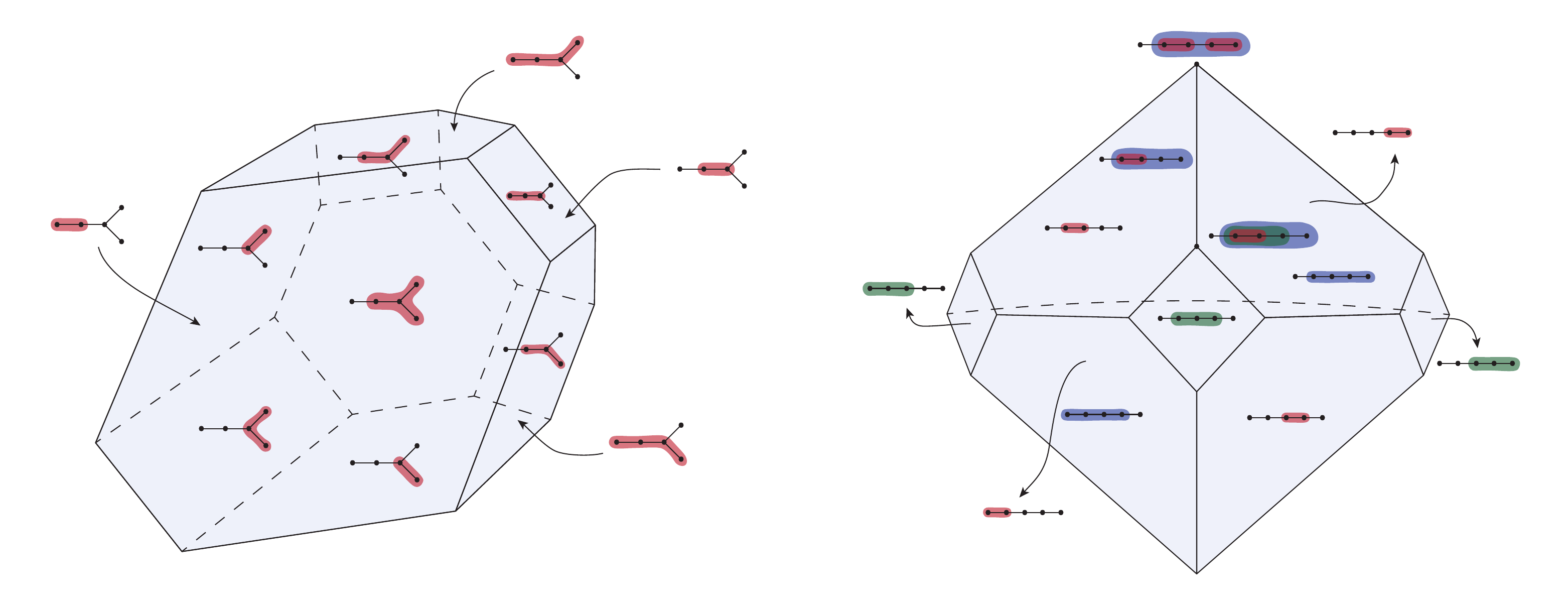}
    \caption{(Left) 7-point graph associahedra for triangulations from the cyclic classes $\{(1,3),(1,4),(4,6),(1,6)\},\, \{(1,3),(3,5),(1,5),(1,6)\}$. (Right) 7-point graph associahedra for triangulations from the cyclic classes $\{(1,3),(1,4),(1,5),(1,6)\},$ $\{(1,3),(1,4),(1,5),(5,7)\},$ $\{(1,3),(1,4),(4,7),(5,7)\},\, \{(1,3),(3,7),(3,6),(4,6)\}$. }
    \label{fig:7GraphAssoc}
\end{figure}
At $7$-points, there are a total of $42$ triangulations. These amount to $6$ different cyclic classes of triangulations. From these 6 cyclic classes (represented by one of its triangulations), there are four, 
\begin{align*}
    &\{(1,3),(1,4),(1,5),(1,6)\},\, \{(1,3),(1,4),(1,5),(5,7)\}, \\
    &\{(1,3),(1,4),(4,7),(5,7)\},\, \{(1,3),(3,7),(3,6),(4,6)\},
\end{align*} 
which produce the graph topology corresponding to a chain with four nodes, for which the graph associahedron we can see in the right of figure \ref{fig:7GraphAssoc}. If we pick $\Psi_{\{(1,3),(1,4),(1,5),(1,6)\}}^{(7)}$, the wavefunction coefficient for this graph will have 14 terms. Two of these terms are represented in figure \ref{fig:7GraphAssoc} (two highlighted vertices), after factorizing $E_t$ and the perimeters of the triangles, they are:
$$
\frac{1}{\mathcal{P}_{134567} \, \mathcal{P}_{1345} \, \mathcal{P}_{1567}}, \, \frac{1}{\mathcal{P}_{134567} \, \mathcal{P}_{13456} \, \mathcal{P}_{1345}},
$$
where the first term corresponds to the tubing at the top of the figure, and the second term corresponds to the tubing in the middle. The triangulations coming from cyclic rotations produce the same graph associahedron, as well as all the other triangulations in the remaining 3 cyclic classes. Since now the facets of the associahedron are two-dimensional, it is easier to illustrate the factorization properties of the facets. For example, the facets associated to the pentagon subpolygons (the green tubes in figure \ref{fig:7GraphAssoc}) are squares, since they are the product of a segment--the graph associahedron of three-site chain--with another segment, seeing that when we shrink the green tubes to a node, we obtain three site chains. All the other facets in figure \ref{fig:7GraphAssoc} (right) are pentagons. Considering they always result from the factorization of the five-site chain into a four-site chain, whose graph associahedron is a pentagon, and a two-site chain whose graph associahedron is a point.

The other two cyclic classes, 
$$
\{(1,3),(1,4),(4,6),(1,6)\},\, \{(1,3),(3,5),(1,5),(1,6)\},
$$
produce the graph topology we see at the left in figure \ref{fig:7GraphAssoc}. The graph associahedron in this case has $18$ vertices, which match the number of terms in the wavefunction coefficient associated with these type of triangulations.

\section{Combinatorial cosmohedra}

So far, we have understood how to encode the combinatorics of russian dolls graph-by-graph -- via the graph associahedron. Now we want to understand how to put the information of all graphs together to find an object that captures the combinatorics of the full wavefunction. 

From the amplitudes side, we already know of an object that precisely captures the combinatorics of triangulations of $n$-gons -- the associahedron, Assoc$_n$. The Assoc$_n$ is a $(n-3)$-dimensional polytope whose faces are labelled by collections of non-overlapping chords of the $n$-gon, such that the vertices label all possible triangulations of the $n$-gon. Combinatorially, the associahedron is defined as follows: Let us consider a set of non-overlapping chords $C$ of the polygon, which defines a partial triangulation. We say that $C^\prime$ is a refinement of $C$ if as sets, we have $C \subset C^\prime$. Then the associahedron is a polytope that has faces for each $C$ such that 
\begin{equation}
C^\prime \text{ is a face of }C\text{ if }C \subset C^\prime.
\label{eq:AssocDef}
\end{equation} 
Note, by convention, the interior of the associahedron is the empty set, and the co-dimension one facets, $(n-4)$-dimensional, are associated with single chords. We review the associahedron in more detail in appendix \ref{appendixAssoc}, as well as the relevant embedding of the polytope in kinematic space, usually called the ABHY associahedron \cite{ABHY}.

Now, the goal is to understand how we can use the structure of the associahedron -- that captures all the diagrams in a single object -- to get an object that puts all the graph associahedra together therefore describing the combinatorics of russian dolls for all the cubic graphs -- the cosmohedron. 

We noted the interesting feature that the graph associahedron for the triangulation of a $n-$gon is $(n-4)$ dimensional. This is precisely the dimension of a {\it facet} of a $(n-3)$ dimensional polytope. This suggests a natural strategy for discovering the cosmohedron. We begin with the $n$-pt associahedron, which is $(n-3)$ dimensional. Each of its vertices corresponds to a complete triangulation, and for the cosmohedron, we would like to associate a $(n-4)$ dimensional graph associahedron with each of these. Thus, we should take the associahedron and ``blow up" each of its vertices into a full $(n-4)$ dimensional facet, with the shape of its corresponding graph associahedron. 

\begin{figure}[t]
    \centering   \includegraphics[width=\linewidth]{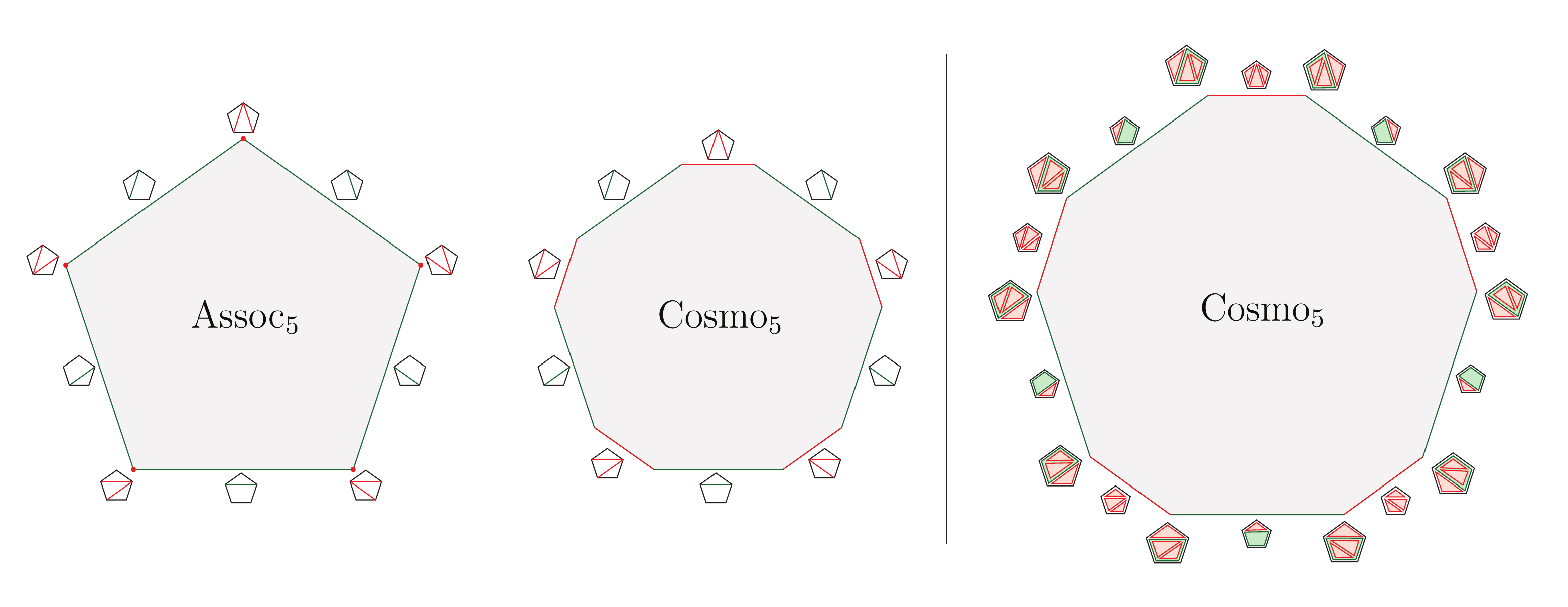}
    \caption{(Left) Associahedron (pentagon) and cosmohedron (decagon) at 5-points. (Right) $5$-point cosmohedron with respective labelling of facets in terms of relevant sub-polygons. }
    \label{fig:pentadeca}
\end{figure}

It is easy to implement this idea for $n=5$: in this case the associahedron is a pentagon and we simply ``blow up" each vertex of the $n=5$ pentagon, into the graph associahedron which is a one-dimensional interval. By doing this, we get a decagon, whose vertices are now labeled by russian dolls for the $n=5$ wavefunction, as shown in figure \ref{fig:pentadeca}. 

Note that while previously for the case of the $n=5$ associahedron we had that the edges were partial triangulations, now for the cosmohedron we have that the edges are partial and full triangulations. Each such partial/full triangulation contains a set of sub-polygons, and the Russian dolls obtained at a given vertex contain the sub-polygons entering the union of those appearing on the edges that meet on the respective vertex. 

The picture is much more interesting for $n=6$. Here, as we saw earlier, 12 of the 14 triangulations have graphs that are four-site chains, whose graph associahedron are pentagons. The two triangulations $\{(1,3),(3,5),(1,5)\}, \{(2,4),(4,6),(2,6)\}$ instead have hexagons as their graph associahedra. Obviously, in order to ``blow up" these vertices into pentagons and hexagons, we will have to introduce many new faces as well, and it is not a priori obvious that the resulting object will reproduce the combinatorics the cosmohedron is meant to capture. But it does! This ``blow up" of the six-point associahedron to the cosmohedron is shown in figure \ref{fig:sixcosmo}.

Let's now define the cosmohedron combinatorial for general $n$. We saw that faces of the associahedron were associated with collections of non-overlapping chords following face structure defined by \eqref{eq:ABHY}. The story for the cosmohedron turns out to be very similar. Instead of collections of non-overlapping chords $C$, we consider collections of non-overlapping sub-polygons $P$. Here, as usual, two sub-polygons are non-overlapping if none of their chords cross; one can be contained in another, or they can be disjoint. There is one more ``russian doll" condition we impose on the collection $P$. If $X,Y$ are sub-polygons in $P$ with $Y$ contained in $X$, then there must be other sub-polygons inside $X$ so that $X$ is fully covered by sub-polygons.

Having defined our subsets $P$, the defining property of the cosmohedron is exactly as it was for the associahedron. The cosmohedron has faces for all $P$, such that
\begin{equation}
P^\prime \text{ is a face of } P \text{ if } P \subset P^\prime.
\label{eq:cosmodef}
\end{equation}

\begin{figure}[t]
    \centering   \includegraphics[width=\linewidth]{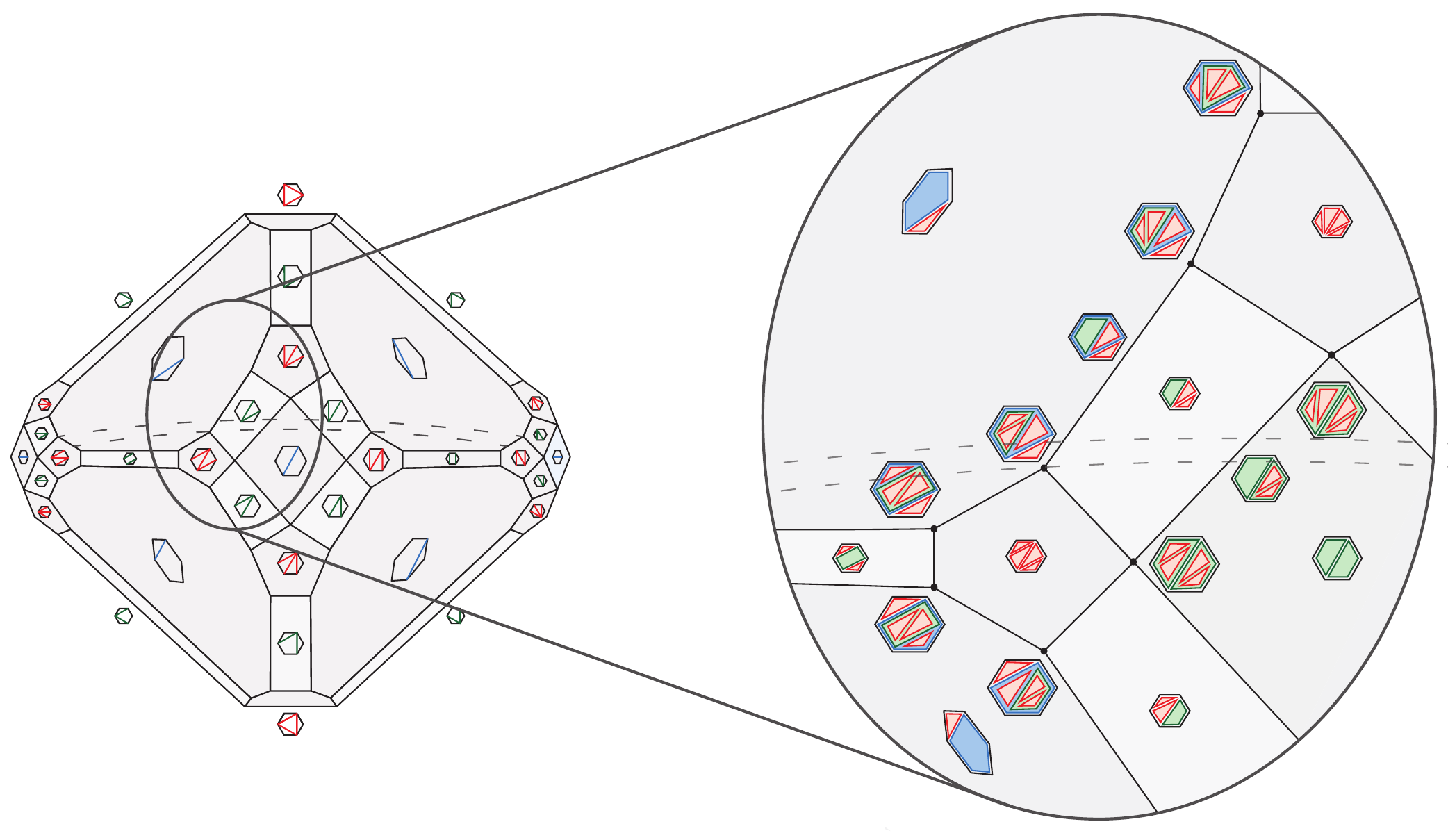}
    \caption{Cosmo$_6$ with labelling of different codimension facets in terms of relevant subpolygons}
    \label{fig:6pt_detail}
\end{figure}

The interior of the cosmohedron can be thought as associated with $P=(1,2,\cdots, n)$ the full polygon. The co-dimension-1 facets of the cosmohedron are associated with $P^\prime$ that correspond to the sub-polygons in any partial triangulation of the $n$-gon. This combinatorial rule for the labelling of the faces is illustrated in figure \ref{fig:6pt_detail} for $n=6$ example. 

\subsection{Faces and Factorization}
\label{sec:FacetFact}

One of the most fundamental properties of associahedra is that faces of associahedra are given by products of lower associahedron -- reflecting the feature of tree-level amplitudes that factorize into lower point amplitudes when we go near a pole. We now describe the analog of this phenomenon for the cosmohedron. 

Let us first discuss facets of the cosmohedron. These are associated with a collection of non-overlapping chords $C$ that give a partial triangulation of the $n$-gon. Given $C$, we have a collection of sub-polygons $\{P_C\}$. We also get a dual graph $G_C$, obtained by putting a vertex in the middle of every sub-polygon and connecting vertices when the corresponding sub-polygons share an edge. 
Then, we have 
\begin{equation}
{\rm Facet}_C[\text{Cosmo}_n] = \prod_{P_i \subset P_C} \text{Cosmo}_{P_i} \times {\cal A}_{G_C}.
\label{eq:factstat}
\end{equation}
Since our $6$-point cosmohedron is three-dimensional, it provides a good illustration of facet factorization. The red facets in figure \ref{fig:sixcosmo} correspond to full triangulations. Thus, the facets will be exactly the graph associahedron of the corresponding graph, since all subpolygons are triangles. The green facets will correspond to partial triangulations with two cords, for this example the subpolygons involved will always be a square and two triangles. This means we can insert three nodes in each of the subpolygons, and the only graph we can form is the three-site chain. Then, the facets will always be squares since the cosmohedron associated with the $4$-point wavefunction is a line interval, and so is the graph associahedron of the three-site chain. Finally, we have the blue facets, which correspond to a partial triangulation with one cord. There are two types of blue facets, the one where the cord splits the hexagon into a pentagon and a triangle, and the one where the cord splits the hexagon into two squares. For both types, the dual graph is the two-site chain, whose graph associahedron is a point. For the first type (darker blue), we get the factorization of the $5$-point cosmohedron and the $3$-point (which is a point), thus these facets will be decagons, which corresponds to the $5$-point cosmohedron. The second type (lighter blue) will be the result of the factorization into two $4$-point cosmohedra, which are segments, resulting in square facets.

\subsection{The geometry of recursive factorization}

\begin{figure}[t]
    \centering   \includegraphics[width=\linewidth]{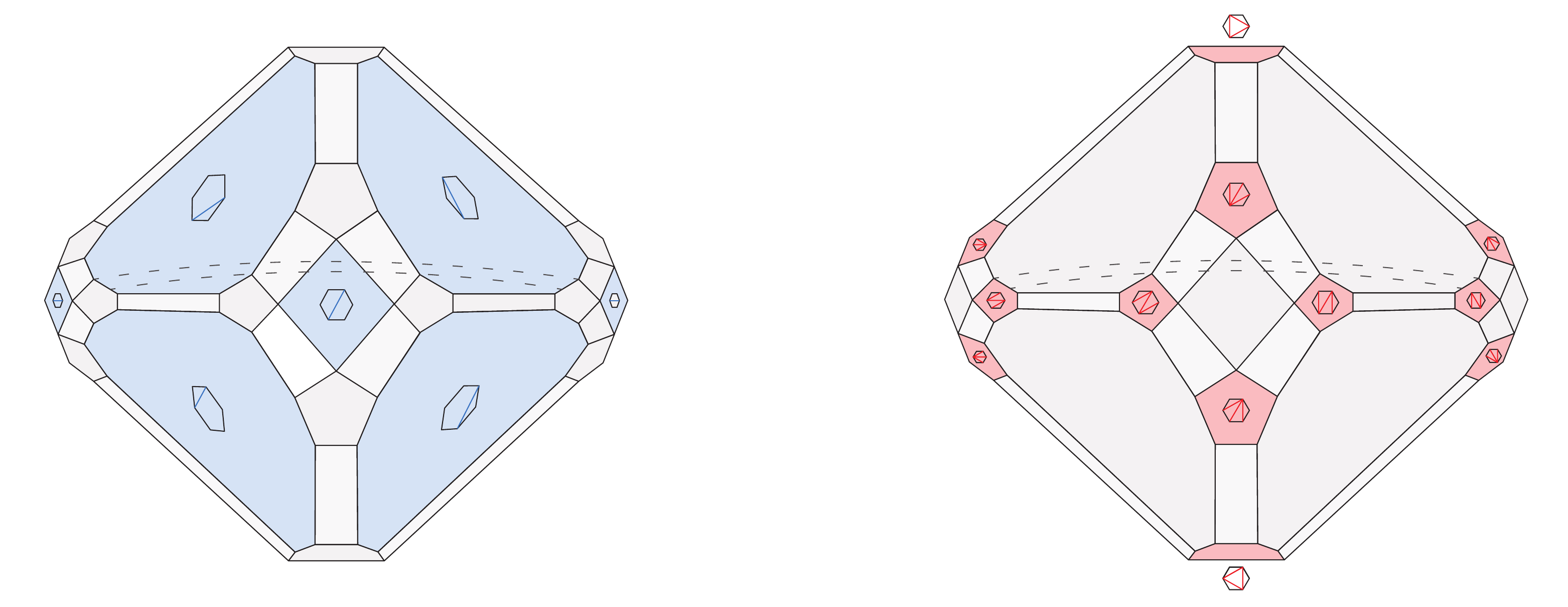}
    \caption{(Left) Set of facets corresponding to partial triangulations with a single chord that by themselves contain \textit{all} vertices of the cosmohedron. (Right) Set of facets corresponding to full triangulations that also touch \textit{all} vertices.}
    \label{fig:russianrec}
\end{figure}

In section \ref{sec:Def_Wf}, we explained how there are two equivalent representations of the wavefunction -- one as a sum over diagrams and their respective russian dolls \eqref{eq:wf_TRussianDolls}, the other via the recursive representation in terms of cuts  given in \eqref{eq:wf_CutRec}.

We would now like to point out how the geometry of the cosmohedron makes both representations of the wavefunction completely obvious. Let's do this by looking at the three-dimensional cosmohedron (see figure \ref{fig:russianrec}). Recall that every term in the russian doll expansion of the wavefunction is associated with a vertex of the cosmohedron. 

Now, the point is that there are a number of natural ways of attaching any vertex of the cosmohedron uniquely to some facet of the cosmohedron. We can consider the ``maximal" facets of the cosmohedron that correspond to complete triangulations T, and associate a vertex corresponding to a given russian doll with its corresponding triangulation. In this way, the collection of all vertices can be organized into first collecting all the facets associated with triangulations $T$, and then looking at the vertices of each facet, as given in \eqref{eq:wf_TRussianDolls}. This is obviously the first representation or what we called the russian doll picture. But there is another interesting way of associating vertices with facets: every vertex can also be naturally attached to one of the ``minimal" facets of the cosmohedron corresponding to a single chord. The corresponding facet is just the product of cosmohedra for the left and right factors on the cut. Hence, we can run through all the vertices by summing over all these facets, and then take the vertices on them. This way of collecting the vertices gives us the recursive computation of the wavefunction in terms of the sum over cuts, as in \eqref{eq:wf_CutRec}. The russian doll and cut-recursive picture of the polytope are shown in figure \ref{fig:russianrec}. Of course, we can uniquely associate vertices to facets in other ways interpolating between the two extremes we have discussed, corresponding to different ways of running the recursive sum over cuts, but deciding to represent some of the lower wavefunction factors directly as a sum over russian dolls. 

\section{``Cosmologizing" the Feynman fan}
\begin{figure}[t]
    \centering
    \includegraphics[width=\linewidth]{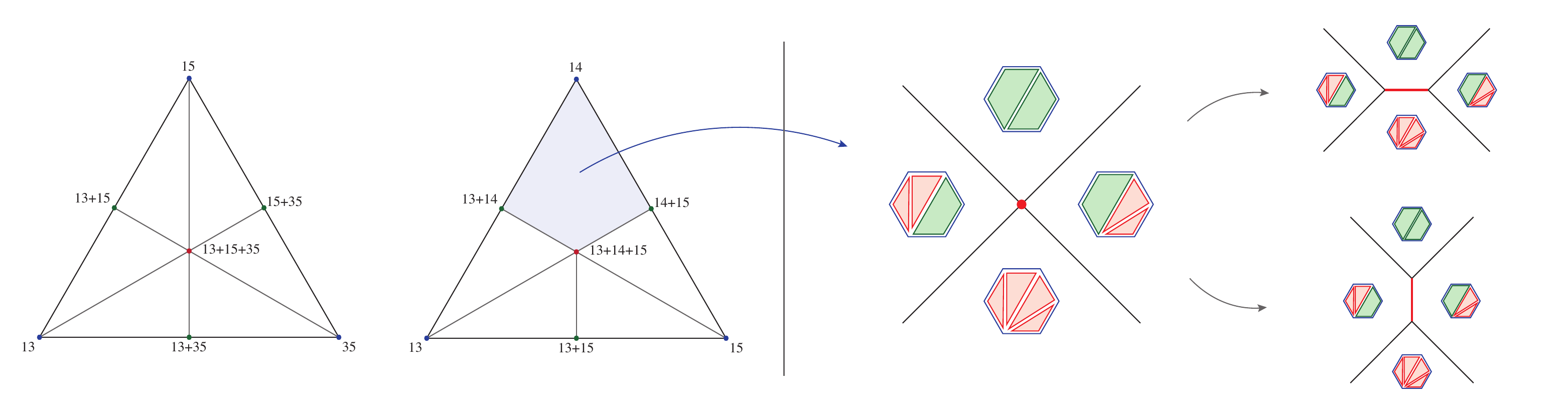}
    \caption{(Left) "Cosmologizing" the $n=6$ associahedron fan to obtain the Cosmo$_6$ fan. In light blue, we highlight the cone corresponding to the non-simple vertex. (Right) Labelling of the four-facets meeting at the non-simple vertex, as well as the two possible ``blow up"s into simple vertices. In both cases, we create a new edge (marked in red) that is already labelled by a full russian doll.}
    \label{fig:non-simple}
\end{figure}

We now want to explain how we can systematically obtain the $\text{Cosmo}_n$ from the $n$-point associahedron, and to do this it is useful to start by looking at the respective dual fans.

There is a very simple picture for the fan of the cosmohedron, beginning with the fan of the associahedron. The $g$-vectors for all the curves $(i,j)$ of the associahedron divide the $(n-3)$-dimensional $g$-vector space into cones, each of which corresponds to a triangulation/diagram. The cosmohedron ``cosmologizes" these cones by further subdividing them into smaller cones in a natural way. 

Let us consider the example of the cone of the 6-point associahedron bounded by the curves $\{(1,3),(1,5),(3,5)\}$. 
Since we only care about the direction of the rays, we can represent this cone projectively by a two-dimensional triangle with $(g_{1,3},g_{3,5},g_{1,5})$ as its vertices (see figure \ref{fig:non-simple}, left only blue vertices). Now, to ``cosmologize'' this cone, we begin by adding rays corresponding to all possible subset sums of the rays $\{(1,3),(3,5),(1,5)\}$ bounding the parent cone. Thus, we have the original rays $g_{1,3},g_{3,5},g_{1,5}$ together with $g_{1,3} + g_{3,5}, g_{3,5} + g_{1,5}, g_{1,3} + g_{3,5}$ (in green) and $g_{1,3} + g_{3,5} + g_{1,5}$ (in red). These add midpoints and the barycenter of the two-dimensional triangle, corresponding to the original cone. Now, we build new cones in the obvious way, by joining the vertices and midpoints of the triangle with the barycenter, as shown in figure \ref{fig:non-simple}. In this way, we produce many new cones, that correspond to the russian doll vertices of the cosmohedron. Note that the central ray $(g_{1,3},g_{3,5},g_{1,5})$ is bounded by six cones. The corresponding facet of the cosmohedron is a hexagon, which is the correct graph associahedron for the graph associated with this triangulation. 

\begin{figure}[t]
    \centering   \includegraphics[width=\linewidth]{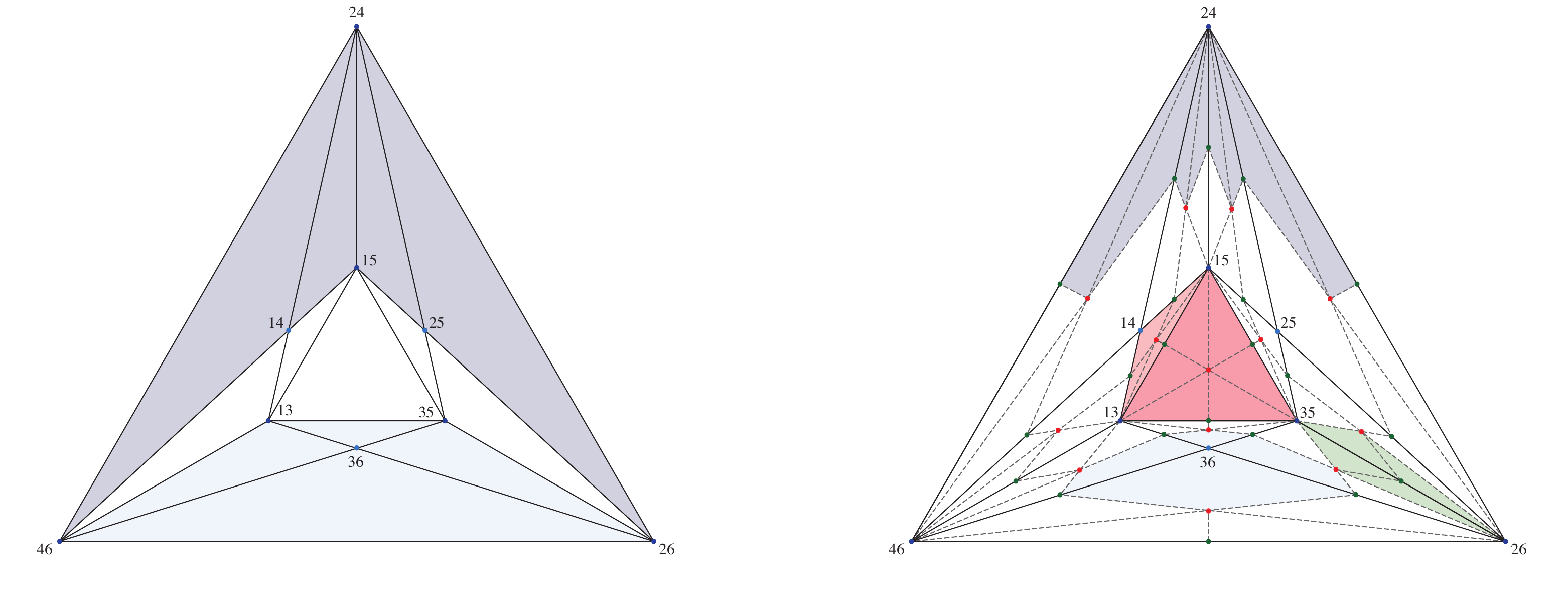}
    \caption{(Left) Fan of the 6-point associahedron. (Right) Fan of the 6-point cosmohedron that can be obtained by ``cosmologizing" the associahedron one.}
    \label{fig:fan6}
\end{figure}
The situation is more interesting if we start with a different triangulation of the associahedron, say bounded by $\{(1,3),(1,4),(1,5)\}$, where the graph associahedron is a pentagon rather than a hexagon. We again begin with the parent rays and add all the subset sums associated with them, giving us again the vertices $(g_{1,3},g_{1,4},g_{1,5})$, midpoints $(g_{1,3} + g_{1,4}, g_{1,3} + g_{1,5}, g_{1,4} + g_{1,5})$ and barycenter $g_{1,3} + g_{1,4} + g_{1,5}$ (see figure \ref{fig:non-simple}, middle). This time to produce the cones, we connect all these points to the barycenter, {\it except} we don't include an edge connecting $g_{1,4}$ to $g_{1,3} + g_{1,4} + g_{1,5}$, as highlighted in figure \ref{fig:non-simple}. This means that we have a cone bounded by four rays $(g_{1,3} + g_{1,4}, g_{1,4}, g_{1,4} + g_{1,5}, g_{1,3} + g_{1,4} + g_{1,5})$, so the corresponding vertex of the cosmohedron belongs to four facets (as illustrated on the right of figure  \ref{fig:non-simple}), reflecting the fact we have already mentioned, that the cosmohedron is \textit{not} a simple polytope. Note also that there are five cones touching the ray at the center, so that the facet of the cosmohedron associated with triangulation $\{(1,3),(1,4),(1,5)\}$ is a pentagon, which is correctly the graph associahedron of the corresponding diagram. 

The combinatorics for the full fans of the six-point associahedron and cosmohedron are shown in figure \ref{fig:fan6}. The fan is three-dimensional, and the figure shows all the cones of the fan, except for the cone in the back. 
For the associahedron (left of figure \ref{fig:fan6}), we see the familiar nine rays corresponding to the facets, and 14 cones corresponding to the vertices of the associahedron. We have shaded the five cones meeting at a ray corresponding to the pentagonal faces (dark blue) and the four cones meeting at a ray corresponding to the square faces of the associahedron (light blue); for the pentagon, the fifth cone is located on the back triangle and is not shaded to avoid clutter.  For the cosmohedron (right of figure \ref{fig:fan6}), we add the midpoints on all edges and barycenters, and connect them with edges as shown in the picture. We have highlighted the collection of cones that give the decagon (dark blue), hexagon (dark pink), pentagon (light pink) and square (light blue and green) facets of the cosmohedron. Only eight of the ten cones of the decagon are visible in the picture, the remaining two are on the back triangle of the fan and are again not shaded in the picture to avoid clutter. 

In summary, we can obtain the cosmohedron fan by starting with the associahedron fan as follows: look at a cone of the associahedron fan, which is defined by a collection of g-vectors $g_C$, each associated to a chord $C$ entering the triangulation $T$ dual to the cone, take all possible subsets $\mathcal{S}=(C_1,C_2, \cdots, C_k)$ with all $C_i\in T$, of any length $k=1,2,\cdots, n-3$, and to each such subset add a ray:
\begin{equation}
    g_\mathcal{S} = \sum_{C \, \in \, \mathcal{S}} g_C.
\end{equation}
This defines all the rays of the cosmohedron fan, and therefore the facets of the cosmohedron. Collections of these rays give us cones that specify the vertices of the cosmohedron. But these cones are not always simplices -- cosmohedra are not simple polytopes.

\subsection{Cosmohedra are \textit{not} simple polytopes}
\label{sec:CosmoNotSimple}
As we have highlighted earlier, and seen in the $n=6$ example, cosmohedra are \textit{not} simple polytopes. This is to be contrasted with the associahedron which indeed is a simple polytope (as we can see from its fan construction as well as in figure \ref{fig:sixcosmo} for the $n=6$ case).

As we will explain now, this feature turns out to be extremely crucial to have an object that reproduces the combinatorial feature of russian dolls (as described in \eqref{eq:cosmodef}), and therefore that encodes the information of the wavefunction. 

Let's say instead we ``blow-up" all the non-simple vertices to obtain a simple polytope. For simplicity let's look at the case of $n=6$, which is the first case this happens, and look at the non-simple vertex associated with cone $\{g_{1,4},g_{1,3}+g_{1,4},g_{1,4}+g_{1,5},g_{1,3}+g_{1,4}+g_{1,5}\}$ highlighted in figure \ref{fig:non-simple}. In these vertices, the four faces meet -- $\{(1,4)\}$, $\{(1,3),(1,4)\}$, $\{(1,4),(1,5)\}$ and $\{(1,3),(1,4),(1,5)\}$ -- and the union of their respective subpolygons forms the russian doll containing triangles $\{(1,2,3),(1,3,4),(1,4,5),(1,5,6)\}$ and the two squares $\{(1,2,3,4),(1,4,5,6)\}$. Now there are two ways in which we can blow up this vertex, one way is by adding an edge connecting rays $g_{1,4}$ and $g_{1,3}+g_{1,4}+g_{1,5}$ -- the object we obtain in this case corresponds to the full barycentric subdivision of the associahedron, which we will later on denote by \textit{Permuto-cosmohedron}; another way is by adding an edge connecting rays $g_{1,3}+g_{1,4}$ and $g_{1,4}+g_{1,5}$. At the level of the polytope, the first type of blow up would lead to the object on the top right of figure \ref{fig:non-simple} while the second one leads to the one on the bottom right of figure \ref{fig:non-simple}.

However, note that in both cases, the object we obtain after the ``blow-up'' does \textit{not} encode the combinatorics of russian dolls correctly. This is because if we look at the new edge (represented in red in figure \ref{fig:non-simple}), it is labelled by the union of the subpolygons of the facets that meet along it, which in both cases means that it is already labelled by the full russian doll associated with the original non-simple vertex. 

This is an important difference between the cosmohedron and the associahedron. We will now proceed to discuss the realization of the geometry that precisely reproduces the combinatorics of the cosmohedron. As we will see, this embedding starts from the kinematic embedding of the associahedron as in \cite{ABHY} and adds some extra inequalities that precisely shave off this polytope exactly in the way that produces the correct polytope with non-simple vertices. 

\section{Cosmic realizations}

Let's now discuss the embedding of the cosmohedron in kinematic space. In appendix \ref{appendixAssoc}, we described how to carve out the associahedron in the space of planar propagators $X_{i,j}$ via a set of inequalities. For the case of the cosmohedron, in addition to the inequalities cutting out the associahedron, we have an additional set of inequalities that further ``shave off" the different codimension faces of the associahedron. 

Recall that for the Cosmo$_n$, we have a facet associated with every partial triangulation, given by a set of non-overlapping chords $C$. Therefore, for each $C$ we have an inequality of the form 
\begin{equation}
\sum_{c \in C} X_c \geq \epsilon_C,
\label{eq:ineqs}
\end{equation}
where we take 
\begin{equation}
\epsilon_C \ll c_{i,j},
\label{eq:eps_cij}
\end{equation}
for any $(i,j)$, where $c_{i,j}$ are the non-planar Mandelstam that enter the embedding of the associahedron, defining the position of the different facets (see appendix \ref{appendixAssoc}). With this constraint, we have that these new inequalities are only ``shaving off" faces of the associahedron. 

The $\epsilon_C$ must satisfy certain relations and hierarchies for these new inequalities to correctly cut out the cosmohedron from the underlying associahedron, all of which relate $\epsilon$'s with the sets $C, C^\prime$ to those of the union $(C \cup C^\prime)$ and the intersection $(C \cap C^\prime)$. 
We must have {\it inequalities} 
\begin{equation}
\epsilon_C + \epsilon_{C^\prime} < \epsilon_{C \cup C^\prime} + \epsilon_{C \cap C^\prime} ,
\label{eq:IneqEps}
\end{equation}
when $C \cap C^\prime$ is empty or is  entirely  to the  left  or right  of $C, C^\prime$, and {\it equalities}
\begin{equation}
\epsilon_C + \epsilon_{C^\prime} = \epsilon_{C \cup C^\prime} + \epsilon_{C \cap C^\prime},
\label{eq:epsEq}
\end{equation}
otherwise. The equalities \eqref{eq:epsEq} force the existence of non-simple vertices. Since the facets containing a given vertex have at most $(n-3)$ $X_{i,j}$ variables in their respective facet inequalities \eqref{eq:ineqs}, then any non-simple vertex is obtained by requiring that more than $(n-3)$ facet inequalities to be saturated.  This imposes an equality of the type of equation \eqref{eq:epsEq}.

It is possible to further simplify conditions \eqref{eq:epsEq} and \eqref{eq:IneqEps}. The equalities are guaranteed if we express $\epsilon_C$ as a sum over variables $\delta_P$ attached to every subpolygon of the partial triangulation given by $C$. In other words, we take 
\begin{equation}
\epsilon_C = \sum_{P \, \text{of } \, C} \delta_P.
\label{eq:eps_to_polys}
\end{equation}
In turn, the inequalities for the $\epsilon_C$ are guaranteed by very similar inequalities for the $\delta_P$:
\begin{equation}
\delta_P + \delta_{P^\prime} < \delta_{P \cap P^\prime} + \delta_{P \cup P^\prime}
\label{eq:ineqs_polys}
\end{equation}
In this expression, we must further ensure that the $\delta$ for the full polygon, $\delta_{(12\cdots n)}$, is set to zero. It is simple to parametrize $\delta_P$'s that satisfy these constraints. For instance, any convex function of the number of edges $(\#P)$ of $P$, that vanishes when $\#P = n$, will satisfy these inequalities. A simple choice is 
\begin{equation}
\delta_P = \delta (n - \#P)^2.
\label{eq:param_polys}
\end{equation}
Here $\delta$ is a uniform small factor that we can make as small as we like to ensure that $\delta_P$ and hence $\epsilon_C$'s are all much smaller than the $c_{i,j}$ cutting out the underlying associahedron \eqref{eq:eps_cij}.

This then defines the embedding of the cosmohedron, which automatically also defines an embedding for the graph associahedra -- which we introduced in section \ref{sec:Graph_Assoc}, to encode the combinatorics of russian dolls graph by graph. To explicitly read off the embedding, all we need to do is to go on a facet corresponding to the full triangulation that is dual to the graph we want to consider. In appendix \ref{sec:EmbedGraphAssoc}, we give the resulting set of the inequalities that directly carve out the graph associahedron.

We will now give an example of what the set of equalities/inequalities are for the case of the Cosmo$_6$. 

\subsection{6-point example}
\begin{figure}[t]
    \centering
\includegraphics[width=\linewidth]{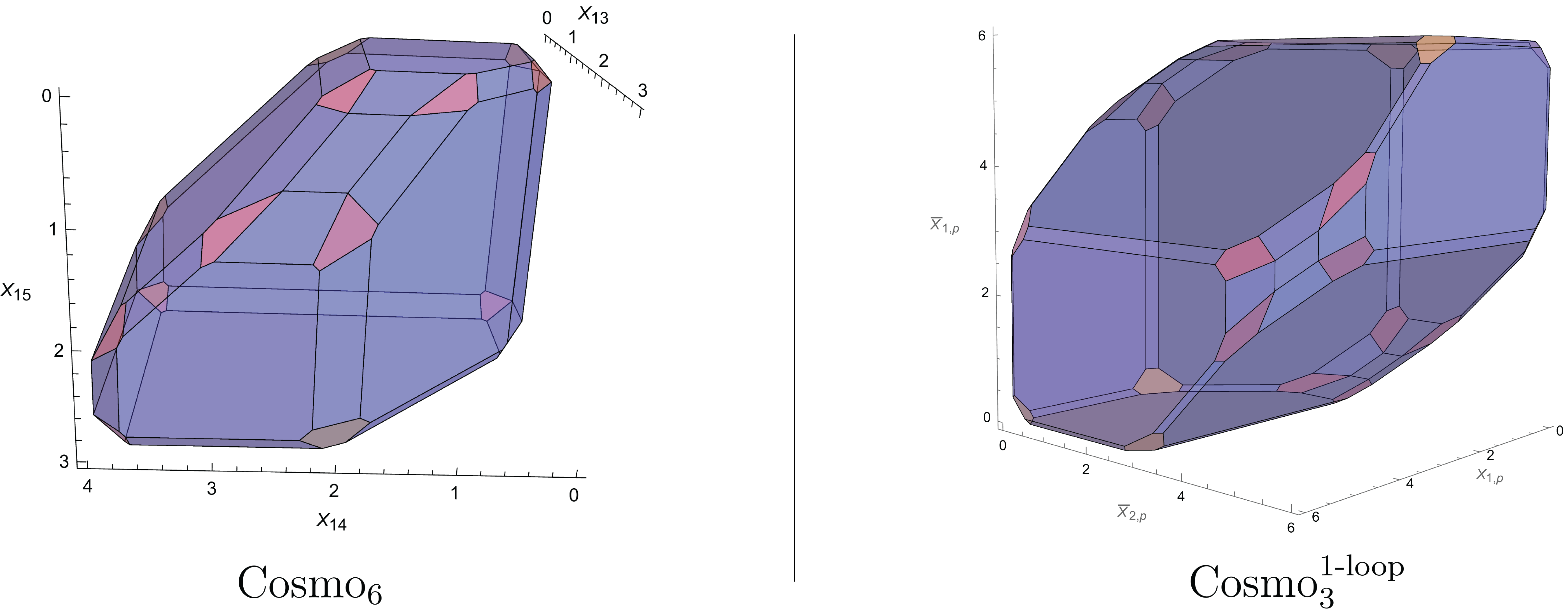}
\vspace{2mm}
    \caption{Realizations of cosmohedra. (Left) Embedding of Cosmo$_6$ with pentagonal facets highlighted in pink and hexagonal ones highlighted in yellow. (Right) Embedding of the Cosmo$_3^{\text{1-loop}}$. The purple facets correspond to partial triangulations, and the pink and yellow facets correspond to full triangulations.}
    \label{fig:embed}
\end{figure}

For the 6-point cosmohedron, we  have 44 different $\epsilon_{C}$, and we can form 105 sets $\{\epsilon_{C},\epsilon_{C'},\epsilon_{C\cup C'},\epsilon_{C\cap C'}\}.$ From these, 12 will be equalities\footnote{There is one for each triangulation whose graph associahedron is a pentagon, as in all such cases we have a non-simple vertex.}, for example:
\begin{equation}
    \epsilon_{\{(1,3),(1,4)\}}+\epsilon_{\{(1,4),(1,5)\}} = \epsilon_{\{(1,4)\}}+\epsilon_{\{(1,3),(1,4),(1,5)\}}\, .
    \label{eq:6pt_equality}
\end{equation}
Note that in this case we have $C=\{(1,3),(1,4)\}$, $C^\prime = \{(1,4),(1,5)\}$, and $C \cap C^\prime = \{(1,4)\}$. So we have that $(1,4)$ divides the hexagon into two smaller squares and $C$ fills one of the squares (the one to the left of $C \cap C^\prime$)  while $C^\prime$ fills the other (the one to the right of $C \cap C^\prime$). Therefore, we have that $C$ is to the left of $C \cap C^\prime$ and $C^\prime$ is to the right of $C \cap C^\prime$, and therefore we must have an equality.

This equality follows from saturating the four facet inequalities:
\begin{equation}
\begin{aligned}
    &X_{(1,4)}\geq \epsilon_{\{(1,4)\}}\, ,  \quad
    X_{(1,3)}+X_{(1,4)}+X_{(1,5)}\geq \epsilon_{\{(1,3),(1,4),(1,5)\}}\, ,\\
    &X_{(1,3)}+X_{(1,4)}\geq \epsilon_{\{(1,3),(1,4)\}}\, , \quad
    X_{(1,4)}+X_{(1,5)}\geq \epsilon_{\{(1,4),(1,5)\}}\, ,
\end{aligned}  
\end{equation}
thus ensuring the existence of the vertex touched by the four facets (which is precisely the one highlighted in figure \ref{fig:non-simple}). From figure \ref{fig:sixcosmo}, it is clear there are 12 such vertices in total, which in the embedding come from the 12 equalities. The remaining 93 sets will form inequalities, for example:
\begin{equation}
\begin{aligned}
    \epsilon_{\{(1,3)\}}+\epsilon_{\{(1,4)\}} & < \epsilon_{\{(1,3),(1,4)\}}\, , \\
    \epsilon_{\{(1,3),(1,4)\}}+\epsilon_{\{(1,3),(1,5)\}} & < \epsilon_{\{(1,3)\}}+\epsilon_{\{(1,3),(1,4),(1,5)\}}\, .
\end{aligned}
\label{eq:6pt_inequalities}
\end{equation}
In the first case, we have that $C\cap C^\prime = \varnothing$, and therefore we have an inequality. In the second case, we have that both $C$ and $C^\prime$ are to the right of $C\cup C^\prime$ and so we also have an inequality. 

Finding $\epsilon_{C}$ which satisfy all 105 relations will ensure that the facet inequalities \eqref{eq:ineqs} define the cosmohedron for the 6-point wavefunction. Finding such a solution is simpler if we impose the map \eqref{eq:eps_to_polys}. For example, 
\begin{equation}
\begin{aligned}
\epsilon_{\{(1,3),(1,4)\}} = \delta_{(1,2,3)}+\delta_{(1,3,4)}+\delta_{(1,4,5,6)} \, ,\quad
\epsilon_{\{(1,4),(1,5)\}} = \delta_{(1,2,3,4)}+\delta_{(1,4,5)}+\delta_{(1,5,6)} \, , \\ 
\epsilon_{\{(1,4)\}} = \delta_{(1,2,3,4)}+\delta_{(1,4,5,6)}\, , \quad
\epsilon_{\{(1,3),(1,4),(1,5)\}} = \delta_{(1,2,3)}+\delta_{(1,3,4)}+\delta_{(1,4,5)}+\delta_{(1,5,6)}\, ,
\end{aligned}
\end{equation}
which immediately satisfies \eqref{eq:6pt_equality}, as all $\delta_{P}$ in the first line match the ones in the second line. This mapping will take \eqref{eq:6pt_inequalities} to,
\begin{align*}
    \delta_{(1,3,4,5,6)}+\delta_{(1,2,3,4)} & < \delta_{(1,3,4)}+\delta_{(1,2,3,4,5,6)}\, ,\nonumber \\
     \delta_{(1,4,5,6)} +\delta_{(1,4,5,6)} &< \delta_{(1,3,4,5,6)}+\delta_{(1,4,5)}\, ,
\end{align*}
which are precisely of the form \eqref{eq:ineqs_polys}. This mapping imposed on all 105 relations will satisfy all 12 equalities and will make several of the 93 inequalities linearly dependent on each other. Thus, we will have only 57 inequalities of the form \eqref{eq:ineqs_polys}, which will be satisfied if we parametrize the $\delta_P$ with the convex function \eqref{eq:param_polys}, $$\delta_P = \delta (6 - \#P)^2\, .$$ 

Therefore, imposing the mapping \eqref{eq:eps_to_polys} in the facet inequalities \eqref{eq:ineqs}, with the parametrization \eqref{eq:param_polys}, defines the 6-point cosmohedron. A picture of the embedded object is presented on the right of figure \ref{fig:embed}.

\subsection{Higher-point examples}
Beyond $6$-points, the cosmohedron will be $4$-dimensional, or higher. Its complexity increases rapidly, as it is shown by the counting of vertices and facets in eqs. \eqref{eq:Vert_count} and \eqref{eq:Facet_count}. Nevertheless, the construction of these polytopes follows exactly the same procedure, and below we list the different $\mathcal{F}$-vectors (\textit{i.e.} the numbers of the different codimension faces) of the cosmohedra up to 9-points:

\begin{center}
\begin{tabular}{  c | c | c | c | c | c | c   } 
             & codim-1 & codim-2 & codim-3 & codim-4 & codim-5 & codim-6\\ 
  \hline
  $4$-points & 2 & --- & --- & --- & --- & ---\\ 
  \hline
  $5$-points & 10 & 10 & --- & --- & --- & ---\\ 
  \hline
  $6$-points & 44 & 114 & 72 & --- & --- & ---\\ 
  \hline
  $7$-points & 196 & 952 & 1400 & 644 & --- & ---\\ 
  \hline
  $8$-points & 902 & 7116 & 18040 & 18528 & 6704 & --- \\ 
  \hline
  $9$-points & 4278 & 50550 & 194616 & 332664 & 262728 & 78408  \\  
\end{tabular}
\end{center}
where codim stands for the codimension of the faces. 

As a comparison, we can list the $\mathcal{F}$-vector for the  associahedron of the respective amplitudes:
\begin{center}
\begin{tabular}{  c | c | c | c | c | c | c   } 
             & codim-1 & codim-2 & codim-3 & codim-4 & codim-5 & codim-6\\ 
  \hline
  $4$-points & 2 & --- & --- & --- & --- & ---\\ 
  \hline
  $5$-points & 5 & 5 & --- & --- & --- & ---\\ 
  \hline
  $6$-points & 9 & 21 & 14 & --- & --- & ---\\ 
  \hline
  $7$-points & 14 & 56 & 84 & 42 & --- & ---\\ 
  \hline
  $8$-points & 20 & 120 & 300 & 330 & 132 & --- \\ 
  \hline
  $9$-points & 27 & 225 & 825 & 1485 & 1287 & 429 \\  
%%  \hline
%%% $10$-points & 35 & 385 & 1925 & 5005 & 7007 & 5005 & 1430 \\  
\end{tabular}
\end{center}

As a quick check, one can add all the entries of the $\mathcal{F}$-vector for one of the $n$-points associahedron above, and confirm that will match the number of codimension-$1$ faces (\textit{i.e.} facets) in the corresponding cosmohedron. We know this is the case because the cosmohedron is obtained by ``shaving" each face of the associahedron, and the facets of the cosmohedron are associated to partial/full triangulations (which is precisely the information encoded by the different codim faces of the associahedron).

\section{Permuto-cosmohedra}
\label{sec:permuto_cosmo}

In the previous section, we described the set of inequalities that carve out the cosmohedron, together with the set of constraints on $\epsilon_C$ required to produce the correct polytope. As we saw,  in addition to the inequalities \eqref{eq:IneqEps}, we also had equalities, which ultimately imply that the polytope we have is not \textit{simple}. We now want to explain a systematic way to blow up the polytope into another polytope which is simple -- the permuto-cosmohedron\footnote{This object already appeared earlier when we explained the ``blow-up" of the non-simple vertex for the $n=6$ case.} -- which will be the object from which we can ultimately extract the wavefunction (as we explain in the next section). 

Let's go back to the fan definition of the polytope. As explained previously, we can go from the associahedron fan to the cosmohedron fan by adding rays corresponding to all possible subsets of chords entering on a given triangulation -- corresponding therefore to all possible partial triangulations. However, not all rays are connected to each other, which is why the cosmohedron is not simple. 

We have already explored in detail the non-simple vertex at $6$-points where facets $\{(1,4)\}$, $\{(1,3),(1,4)\}$, $\{(1,4),(1,5)\}$, $\{(1,3),(1,4),(1,5)\}$ meet in section \ref{sec:CosmoNotSimple}. In particular, we explained how the two different ``blow-up" led to objects that did not consistently describe the combinatorics of russian dolls. 

However, let's now go back to the blow up in which we produce an edge between facets $\{(1,4)\}$ and $\{(1,3),(1,4),(1,5)\}$ -- this corresponds to the full \textit{barycentric subdivision} of the associahedron fan into a permutohedron fan, leading to what we called the \textit{permuto-cosmohedron}. For this new object, we can think of each vertex as labeling the ways in which we can get a full triangulation by listing chords in a particular order. In particular, for two vertices produced in the blow up, these correspond to cases in which we start with $(1,4)$ and then we have two possible ways to continue to the full triangulation:
\begin{equation}
\begin{aligned}
     \{(1,4)\}\to \{(1,4),(1,3)\} \to  \{(1,4),(1,3),(1,5)\} , \\
      \{(1,4)\}\to \{(1,4),(1,5)\} \to  \{(1,4),(1,3),(1,5)\} ,
\label{eq:6pt_vertex_blowup}
\end{aligned}
\end{equation}
each of which corresponds to one of the vertices we obtain after \textit{simplifying} the non-simple vertex of the original cosmohedron. 

The permuto-cosmohedron is then a simple polytope whose vertices label all the possible orderings of building full triangulations out of partial triangulations. For general $n$, the fan definition of the permuto-cosmohedron is simply given by the full barycentric subdivision of the respective Assoc$_n$ fan. This object has manifestly more vertices than the cosmohedron and therefore is not precisely tailored to the wavefunction. Nonetheless, as we will see momentarily, this permutuhedral blow-up will provide us a natural way of extracting the full wavefunction from the geometry.

Before proceeding to the extraction of the wavefunction let's discuss the embedding of the permuto-cosmohedron. For the cosmohedron we saw that for each facet associated with a given collection of chords, $C$, we have an inequality of the form of \eqref{eq:ineqs}, where the $\epsilon_C$'s satisfy both equalities \eqref{eq:epsEq} and inequalities \eqref{eq:IneqEps}. To produce the full permuto-cosmohedron all we need to do is turn the \textit{equalities} \eqref{eq:epsEq} into inequalities, with the same sign, $i.e.$ we have that for \textit{any} collection of chords $C$ and $C^\prime$:
\begin{equation}
    \epsilon_C + \epsilon_{C^\prime} < \epsilon_{C \cup C^\prime} + \epsilon_{C \cap C^\prime}, 
\end{equation}
this then turns all the non-simple vertices into simple ones and gives us precisely the blow-up corresponding to the permuto-cosmohedron.

\section{Wavefunction from geometry}
\begin{figure}[t]
    \centering
    \includegraphics[width=\linewidth]{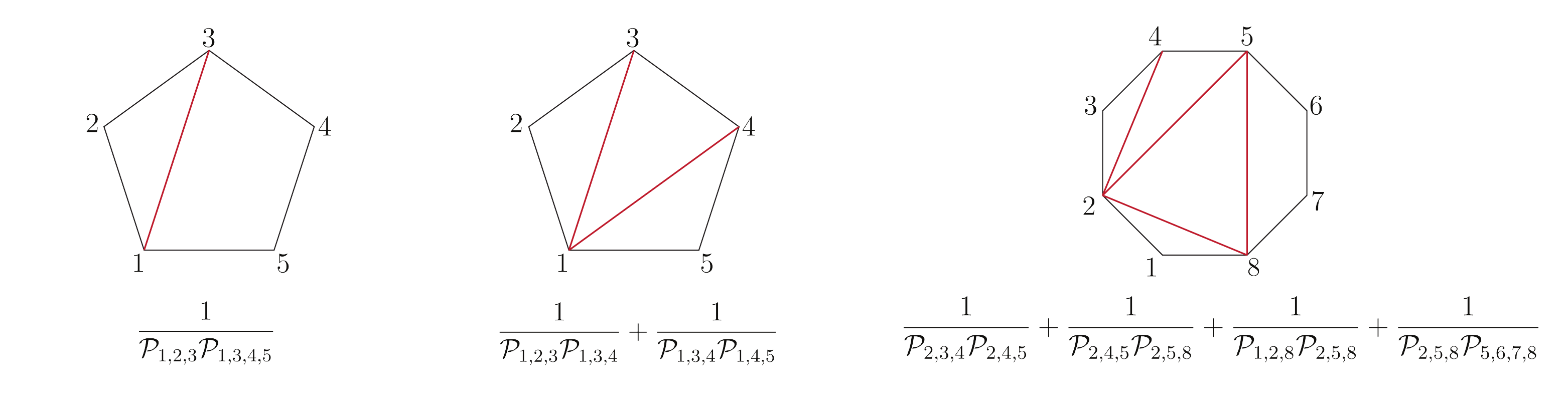}
    \caption{Examples of $1/X_C$ for different faces of $n=5$ and $n=8$ cosmohedron.}
    \label{fig:PerimeterXc}
\end{figure}

Let's now discuss how to extract the wavefunction from the geometry. We will start by defining the canonical form of the graph associahedron for a single graph, and then proceed to the generalization that gives us the full wavefunction from the cosmohedron. 

The connection between the wavefunction and geometry for single graphs is by now the familiar one. For a single diagram/$n-$pt triangulation, with $(n-3)$ chords, we have a $(n-4)$-dimensional graph associahedron. The graph associahedron is simple, hence one computation of the canonical form of the graph associahedron is given by summing over all vertices -- corresponding to complete tubings/russian dolls -- and multiplying by $1/{\cal P}$'s for all the tubes corresponding to the facets meeting at the vertex. This gives us a term with $(n-4)$ poles. Of course, {\it every} tubing associated with the diagram has the $E_{\rm total}$ tube surrounding the entire graph, as well as the small circles encircling every vertex -- corresponding to the triangles entering the triangulation dual to the graph. Hence, we have 
\begin{equation}
\Psi_G = \frac{1}{{\cal P}_{\rm tot}} \times \prod_{v \subset G} \frac{1}{{\cal P}_v} \times \Omega({\cal A}_G),
\end{equation}
where ${\cal P}_{\rm tot}$ is the perimeter of the full $n$-gon corresponding to $E_t$, and ${\cal P}_v$ the perimeter of each triangle entering the underlying triangulation.

The extraction of the wavefunction for the sum over all diagrams is much more interesting. Let's consider the simple polytope we get by blowing-up the cosmohedron as described in the previous section -- the permuto-cosmohedron. Each facet of this polytope is associated a partial triangulation given by a collection of non-overlapping chords $C$. Let $n_C$ be the number of non-triangle subpolygons entering in the partial triangulation defined by $C$, then we define 
\begin{equation}
\frac{1}{X_C} \equiv \frac{1}{n_C} \sum_{P, P^\prime \, {\rm meeting \, on \, edge}} \frac{1}{{\cal P}_P {\cal P^\prime}_{P^\prime}},
\label{eq:WF_permuto_sings}
\end{equation}
where we consider the products of the perimeters of the subpolygons entering in $C$ that share an edge, and sum over them (see figure \ref{fig:PerimeterXc}). So this means that to each facet, instead of associating a single singularity (like we do to extract the amplitude from the associahedron), we associate \textit{pairs of singularities}. It is clear that we could not associate a single singularity to each facet simply because the dimensionality of the Cosmo$_{n}$ does not match the number of singularities on $\Psi_n$. This new feature reflects that even the way we extract the wavefunction from the canonical form of cosmohedra requires a generalization from what is done in the amplitudes case.

We now look at the canonical form for the permutohedron, associating $X_C$ \eqref{eq:WF_permuto_sings}). Since the permuto-cosmohedron is simple, the canonical form is the sum over all vertices weighted by the product of all $\frac{1}{X_C}$'s for the facets that meet on the vertex. While this manifestly has only simple poles in terms of $\frac{1}{X_C}$, it will clearly have terms with simple poles as well as double and higher poles when written in terms of the $\frac{1}{{\cal P}_P}$. But the claim is that the wavefunction is given by the part of the canonical form with only simple poles: 
\begin{equation}
\Psi = \frac{1}{E_t} \times  \Omega(X_C)\vert_{{\rm single \, poles \,in {\cal P}_P}}.
\label{eq:WF_from_geo}
\end{equation}

\subsection{5-point example}
At five points the cosmohedron is simple, therefore it coincides with the permuto-cosmohedron. Then, we can directly compute the poles of each facet, $X_C$, according to \eqref{eq:WF_permuto_sings}. Let's consider the facet labelled by the cords $\{(1,3),(1,4)\}$, the singularity pairs we associate to it are (see figure \ref{fig:PerimeterXc}):
$$
\frac{1}{X_{\{(1,3),(1,4)\}}}=\frac{1}{\mathcal{P}_{123}\, \mathcal{P}_{134}} + \frac{1}{\mathcal{P}_{134}\, \mathcal{P}_{145}} \, .
$$
Similarly, we can compute the singularity pairs of the facets that meet facet $\{(1,3),(1,4)\}$ -- those are $\{(1,3)\}$, and $\{(1,4)\}$ -- for which we have:
\begin{equation*}
 \frac{1}{X_{\{(1,3)\}}}=\frac{1}{\mathcal{P}_{123}\, \mathcal{P}_{1345}}, \quad  \frac{1}{X_{\{(1,4)\}}}=\frac{1}{\mathcal{P}_{1234}\, \mathcal{P}_{145}} \, .  
\end{equation*}
We can now compute the contributions of each of these two vertices to the wavefunction:
\begin{align*}
    & \frac{1}{X_{\{(1,3),(1,4)\}}X_{\{(1,3)\}}} =\frac{1}{\mathcal{P}_{123}^2\, \mathcal{P}_{134}\, \mathcal{P}_{1345}} + \frac{1}{\mathcal{P}_{123}\, \mathcal{P}_{134}\, \mathcal{P}_{145}\, \mathcal{P}_{1345}}\, , \\
    & \frac{1}{X_{\{(1,3),(1,4)\}}X_{\{(1,4)\}}} =\frac{1}{\mathcal{P}_{123}\, \mathcal{P}_{134}\, \mathcal{P}_{145}\, \mathcal{P}_{1234}} + \frac{1}{\mathcal{P}_{134}\, \mathcal{P}_{145}^2\, \mathcal{P}_{1234}} \, .
\end{align*}
According to \eqref{eq:WF_from_geo}, in the first line above, we send the first term to zero, and in the second line we send the second term to zero. So we are left with precisely the russian dolls contributing to each vertex (see  figure \ref{fig:pentadeca}). When we sum these two terms, they add up to the wavefunction of the graph corresponding to the triangulation $\{(1,3),(1,4)\}$.
By computing the contributions from the remaining vertices of the decagon, we obtain the full wavefunction at $5$-points.
\subsection{6-point example}
Let us now see how the prescription in \eqref{eq:WF_from_geo} gives us the correct contribution in the blown up vertices at $6$-points. Let us use our running example of the vertices in \eqref{eq:6pt_vertex_blowup} as an example (see top right of figure \ref{fig:non-simple}). For the first line in \eqref{eq:6pt_vertex_blowup}, which corresponds to one vertex of the permuto-cosmohedron, the pairs of singularities are:

\begin{align*}
   \frac{1}{X_{\{(1,4)\}} X_{\{(1,3),(1,4)\}} X_{\{(1,3),(1,4),(1,5)\}}} = &\l\frac{1}{2 \mathcal{P}_{1234}\mathcal{P}_{1456}}\r \l\frac{1}{\mathcal{P}_{123}\mathcal{P}_{134}}+\frac{1}{\mathcal{P}_{134}\mathcal{P}_{1456}}\r\times \\
   & \times \l\frac{1}{\mathcal{P}_{123}\mathcal{P}_{134}}+\frac{1}{\mathcal{P}_{134}\mathcal{P}_{145}}+\frac{1}{\mathcal{P}_{145}\mathcal{P}_{156}}\r,
\end{align*}
where the $\frac{1}{2}$ in the first factor comes from the fact that the facet $\{(1,4)\}$ has two non-triangle subpolygons, two squares, thus $n_C=2$ in \eqref{eq:WF_permuto_sings}. As for the second line in \eqref{eq:6pt_vertex_blowup}, the other vertex coming from the blow up, the contribution will be:
\begin{align*}
   \frac{1}{X_{\{(1,4)\}} X_{\{(1,4),(1,5)\}} X_{\{(1,3),(1,4),(1,5)\}}} = &\l\frac{1}{2 \mathcal{P}_{1234}\mathcal{P}_{1456}}\r \l\frac{1}{\mathcal{P}_{1234}\mathcal{P}_{145}}+\frac{1}{\mathcal{P}_{145}\mathcal{P}_{156}}\r\times \\
   & \times \l\frac{1}{\mathcal{P}_{123}\mathcal{P}_{134}}+\frac{1}{\mathcal{P}_{134}\mathcal{P}_{145}}+\frac{1}{\mathcal{P}_{145}\mathcal{P}_{156}}\r.
\end{align*}

After sending all double poles (or higher) to zero, one can check that the added contribution of the two vertices above is:
$$
2\l\frac{1}{2\, \mathcal{P}_{123}\, \mathcal{P}_{134}\, \mathcal{P}_{145}\, \mathcal{P}_{156}\, \mathcal{P}_{1234}\, \mathcal{P}_{1456}}\r,
$$
which is precisely the russian doll term associated with the original non-simple vertex in the cosmohedron (see figure \ref{fig:non-simple}).
The remaining non-simple vertices follow the same blow up into two vertices, and all other vertices are simple. Following the same prescription as in the examples above, one can compute the $6$-point wavefunction from the permuto-cosmohedron.

\section{Loop cosmohedra}
\label{sec:Loop}

The associahedron picture for Tr$(\phi^3)$ amplitudes at tree-level was extended to one-loop polytopes\cite{CausalDiamonds,BinGeom}, and then to all loop orders in the curve-integral formalism\cite{CurveInt,CurveInt2}. Indeed, the picture of curves on surfaces most naturally gives us the ``Feynman fan", with every curve $X$ on the surface associated with a g-vector $g_X$. Beautifully, maximal collections of non-overlapping curves form cones that tile all of g-vector space. This fan is the setting for the ``global Schwinger parameterization" of the curve integral formalism; a related but distinct fact is that this fan can also be thought of as the normal fan of polytopes -- ``surfacehedra" -- that capture the combinatorics of surfaces and all their cuts in their facet structure. 

Thus, we should  expect cosmohedra to exist at loop-level as well, generalizing surfacehedra in the same way they generalized associahedra at tree-level. Of course these objects do exist, and in this section we will give a telegraphic account of loop-level cosmohedra, assuming some familiarity with the curves-on-surfaces picture for amplitudes of \cite{CurveInt,Gluons}. We will return to give  more leisurely, self-contained and systematic exposition of these objects in future work. 

\begin{figure}[t]
    \centering
    \includegraphics[width=\linewidth]{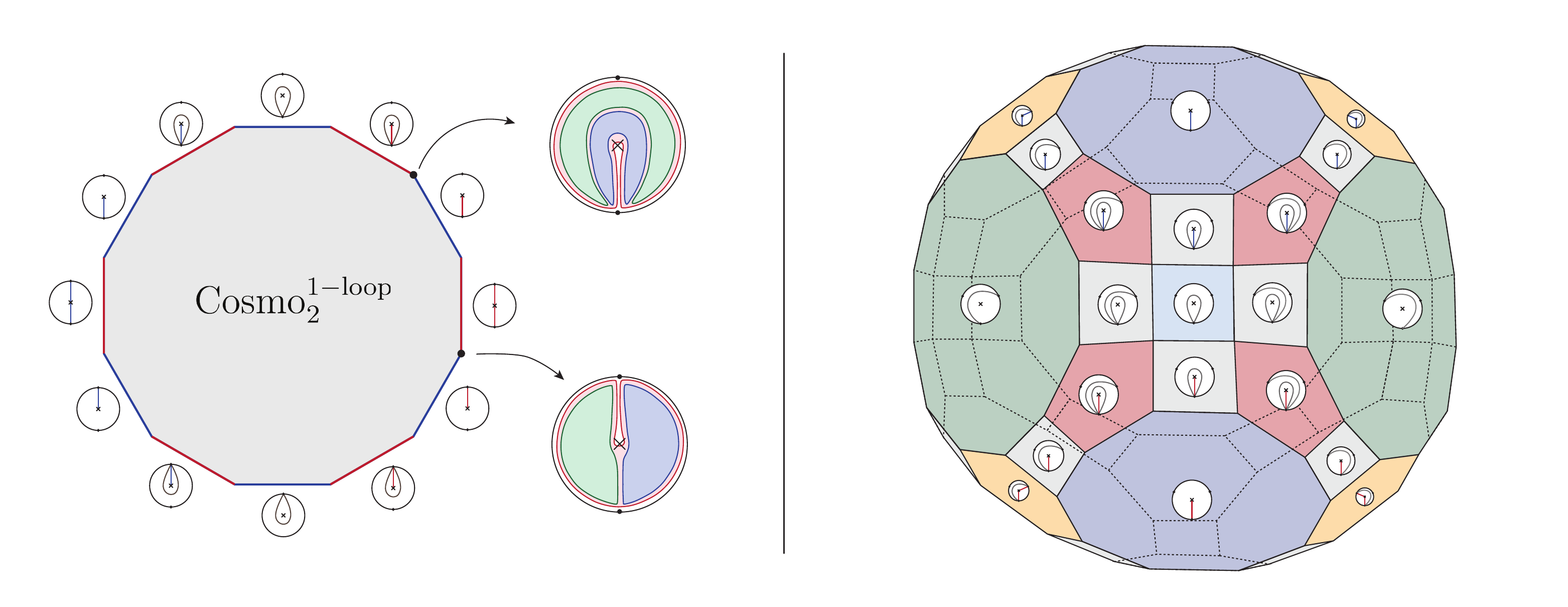}
    \caption{(Left) Cosmohedron 2-point 1-loop, edges are labelled by partial triangulations with a single curve (where we have two types of curves ending in the puncture, marked in red and blue), and vertices correspond to full triangulations. We can read off the russian doll at each vertex by taking the union of the subsurfaces entering on each edge. (Right) Cosmohedron 3-point 1-loop. Highlighted in blue and green we have facets labeled by a single curve (squares, decagons and dodecagons); in gray facets labeled by two curves (squares); and in red and yellow faces labelled by full triangulations (pentagons and hexagons) -- corresponding to the graph-associahedra for the loop graphs.}
    \label{fig:1loop}
\end{figure}

To begin with, the combinatorial definition of loop level cosmohedra is exactly the same as what we have seen at tree-level, where instead of collections of ``sub-polygons" $P$ we consider more generally collection of subsurfaces. We give simple examples of 2- and 3-dimensional cosmohedra associated with the $n=2,3$ at 1-loop, corresponding to the once-punctured disk with marked points on the boundary,  in figure \ref{fig:1loop}. Note that, as familiar for amplitudes, it is natural to include two kinds of ``loop" curves corresponding to the two kinds of ``spiraling" loop variables around the puncture. 

From the examples presented, we can also observe how the factorization \eqref{eq:factstat} holds at loop-level. Let us consider figure \ref{fig:1loop} (right), the red and yellow facets correspond to graph associahedra directly, since they are full triangulations. Then the green facets are dodecagons since they correspond to the product of a 3-point tree level cosmohedron, which is a point, a two-point loop level cosmohedron, which is a dodecagon, and a graph associahedron corresponding to the two-site chain, which is a point. The dark blue facets are decagons, since they correspond to the product of a 5-point tree level cosmohedron (a decagon), and the graph associahedron of a tadpole (a point). Then, the light blue facets are squares, since they correspond to the product of a 4-point tree-level cosmohedron, which is an interval, a one-point one-loop cosmohedron, which is also an interval, and the graph associahedron of the two-site chain (which is a point). 

Now, the most obvious picture for generalizing the cosmohedron to all loops therefore proceeds by generalizing the picture of ``cosmologizing'' the Feynman fan. These proceeds precisely in the same way as for the tree-level cosmohedron. We subdivide every cone in $g$-vector space into smaller cones, by considering all possible sums of the $g$-vectors in a given cone. This yields the fan for the loop cosmohedron, which allows us to write the facet inequalities:
$$
\sum_{{\rm chords} \, c \, {\rm in} \, C} X_c \geq \epsilon_C,
$$
where $C$ is a given partial triangulation of the punctured disk. For the loop case, the propagator variable $X_{i,j}$ differs from $X_{j,i}$, since the chord can go around the loop in two different ways\footnote{Even though when we assign momentum to these curves in the standard way, $i.e.$ by homology, they both have the same momentum. }. We will also have propagators attached to tadpoles, $X_{i,i}$. As well as the propagators in the loop, $X_{i,p}$ and $\tilde{X}_{i,p}$ (where $p$ is labelling the puncture).

The constants in the facet inequalities, $\epsilon_{C}$, obey the same equalities and inequalities as in tree level, \eqref{eq:epsEq} and \eqref{eq:IneqEps}, respectively. Also at loop level, the equalities are automatically satisfied if we map each $\epsilon_C$ to the sum of the sub-surfaces in the correspondent partial triangulation,
$$
\epsilon_C = \sum_{P \, \text{of} \, C} \delta_P\, ,
$$
And the inequalities in the $\epsilon$ are all automatically satisfied if we satisfy the inequalities:
\begin{equation}
    \delta_P + \delta_{P^\prime} < \delta_{P \cup P^\prime} + \sum_{\tilde{P}\in\{P \cap P^\prime \}}\delta_{\tilde{P}}\, ,
    \label{eq:deltaIneqsLoop}
\end{equation}
where the sum over $\delta_{\tilde{P}}$ is reflecting the fact that at loop level the intersection of two sub-surfaces can be given by two or more disjoint surfaces.

\subsection{Graph associahedra at loop-level}
\begin{figure}[t]
    \centering
    \includegraphics[width=\linewidth]{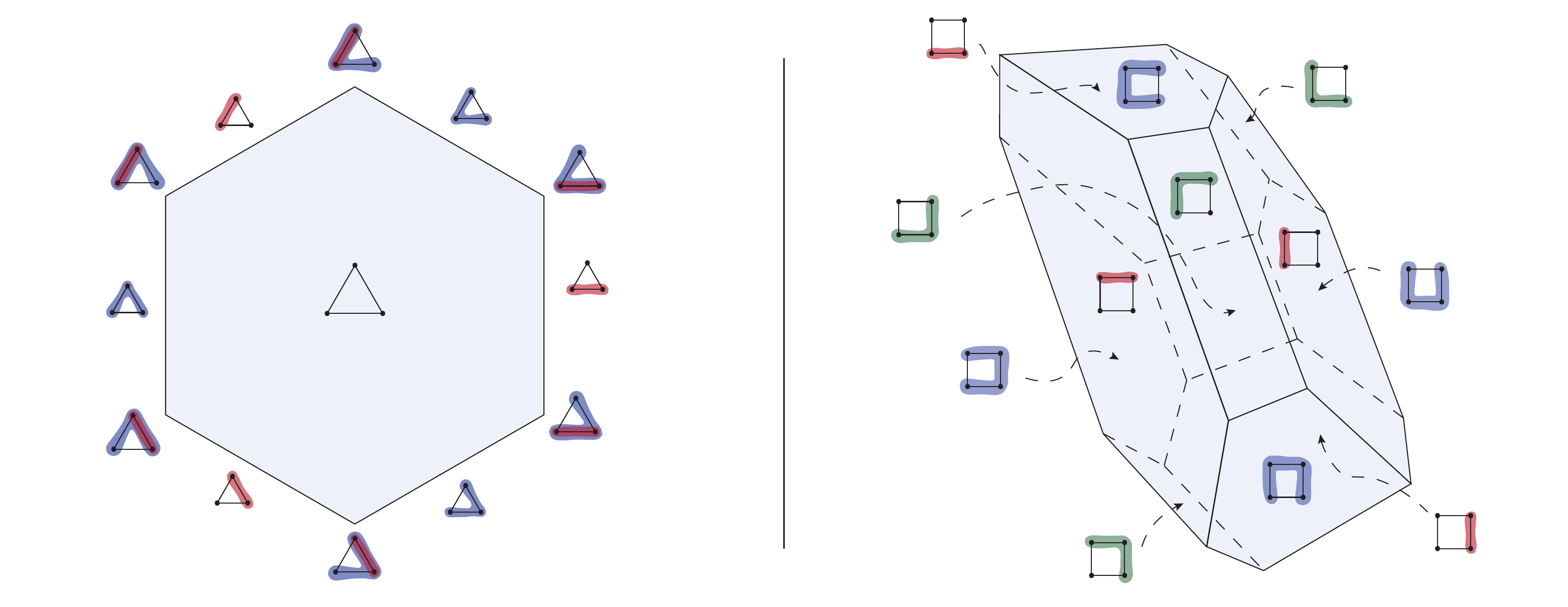}
    \caption{(Left) Graph associahedron for the triangle graph. (Right) Graph associahedron for the box graph.}
    \label{fig:GraphAssocLoop}
\end{figure}
At loop level, the graph associahedron is obtained exactly the same way as for tree-level. For each triangulation, we associate a node to each subsurface, and connect the nodes between subsurfaces that share an edge, building the dual graph, $G_T$. Then the graph associahedron, $\mathcal{A}_G$, is the polytope whose facets correspond to the different tubes of the graph (not including the tubes that enclose single vertices nor the tube that encloses the full graph), and the vertices correspond to complete tubings. The factorization property, defined by eq.\eqref{eq:fact_graphAssoc}, holds  at loop level. Let's now give some simple examples at one-loop. 

\paragraph{Three-point triangle diagram}

The graph associahedron of the triangle diagram is a hexagon (see figure \ref{fig:GraphAssocLoop}, left), precisely matching the six russian dolls one can find in the graph. The triangle graph is dual to the triangulation of the punctured disk containing curves $\{(p,1),(p,2),(p,3)\}$. The facets in figure \ref{fig:GraphAssocLoop} (left) either correspond to blue tubes or to red tubes, both are segments.

The red tube corresponds to the product of the graph associahedron of the two site chain, which is a point, with the graph associahedron of the bubble (obtained by shrinking the red tube to a node), which is a segment.

The blue tube corresponds to the product of the graph associahedron of the three-site chain, which is a segment, which the graph associahedron of the tadpole (obtained by shrinking the blue tubes to a node), which is a point.

All the terms will correspond to the product of a blue tube with a red tube, which is clear by the facet intersections in figure \ref{fig:GraphAssocLoop} (left), and respective labeling (which are the nesting of the red tube in the blue tube). One such term, after factoring out the total energy and the triangles, is:
\begin{equation*}
    \frac{1}{\mathcal{P}_{(1,2),(2,3),(3,1),(p,1)}\, \mathcal{P}_{(1,2),(2,3),(p,1),(p,3)}}\, ,
\end{equation*}
the remaining 5 terms are just variations of this one, as one can see by the labels of figure \ref{fig:GraphAssocLoop} (left).

\paragraph{Four-point box diagram}

The graph associahedron for the four-point box diagram has $20$ vertices, $32$ edges and $14$ facets, as can be seen in the right of figure \ref{fig:GraphAssocLoop}. This corresponds to the triangulation $\{(p,1),(p,2),(p,3),(p,4)\}$ of the punctured disk. The graph associahedron will have three types of facets, the tubes with two sites (red tubes in figure \ref{fig:GraphAssocLoop}) will be hexagons, the tubes with three sites (green tubes in figure \ref{fig:GraphAssocLoop}) will be squares, the tubes with four sites (blue tubes in figure \ref{fig:GraphAssocLoop}) will be pentagons.

The red tubes will correspond to the product graph associahedron of the two-site chain (subgraph inside the red tube), which is a point, with the graph associahedron of the triangle diagram (obtained after shrinking any red tube in the box), which we can see from the left of figure \ref{fig:GraphAssocLoop}, that is a hexagon.

The green tubes will correspond to the product of the graph associahedron of the three-site chain, which is an interval, with the graph associahedron of the bubble (obtained by shrinking any green tube in the box diagram in the right of figure \ref{fig:GraphAssocLoop}), which is also a segment. The product is a square.

Finally, the blue tubes will correspond to the product of the graph associahedron of the four-site chain, which is a pentagon (as can be verified in the left of figure \ref{fig:5_6GraphAssoc}), with the graph associahedron of the tadpole, which is a point.

In total, the polytope has 20 vertices, precisely matching the number of russian dolls in the graph. There will be 16 terms which correspond to a tubing which has a blue, a green and a red tube. One such term, after factoring out the total energy and the triangles, is:
\begin{equation*}
    \frac{1}{\mathcal{P}_{(1,2),(2,3),(3,4),(4,1),(p,1)}\, \mathcal{P}_{(1,2),(2,3),(3,4),(p,1),(p,4)}\, \mathcal{P}_{(1,2),(2,3),(p,1),(p,3)}}\, ,
\end{equation*}
and one can consider 16 similar tubings. In the polytope of figure \ref{fig:GraphAssocLoop}, one can identify these terms by finding the vertices that are intersections of facets labeled by a blue tube, a green tube and a red tube. The other 4 terms correspond to the product of a blue tube with two red tubes, one such example is:
\begin{equation*}
    \frac{1}{\mathcal{P}_{(1,2),(2,3),(3,4),(4,1),(p,1)}\, \mathcal{P}_{(1,2),(2,3),(p,1),(p,3)}\, \mathcal{P}_{(3,4),(4,1),(p,1),(p,1)}}\, .
\end{equation*}

Again one can count four vertices in figure \ref{fig:GraphAssocLoop} which are the intersection of two facets labeled by a red tube and one facet labeled by a blue tube.

\subsection{One-loop cosmohedra realization}
As described earlier, the embedding of the loop cosmohedra is done exactly in the same way as in the tree-level case. We now proceed to give some explicit examples.

\paragraph{Two-points}
The one-loop two-points associahedron is a hexagon, thus the corresponding cosmohedron will be a dodecagon. The Feynman fan is given by the g-vectors:
$$
g_{1,1},\, g_{2,2}, \, g_{p,1}, \, g_{p,2}, \, g_{\tilde{p},1}, \, g_{\tilde{p},2}, 
$$
and we ``cosmologize" it by adding the following linear combinations of g-vectors:
$$
g_{1,1} + g_{p,1}, \, g_{2,2} + g_{p,2}, \,g_{p,1}+g_{p,2},
$$
as well as the other three rays with $g_{p,i}\to g_{\tilde{p},i}$. Now that we have the form of our facet inequalities, we only need to parametrize the $\epsilon$ constants which will ``shave off" the underlying loop associahedron polytope. The $\epsilon$ will have to satisfy 6 inequalities in order to yield the correct polytope and since in this case the polytope is two-dimensional, and thus simple, there are no equalities to be imposed on the $\epsilon$-space,
$$
    \epsilon_{\{(1,1)\}}+\epsilon_{\{(p,1)\}}< \epsilon_{\{(1,1),(p,1)\}}, \quad \epsilon_{\{(p,1)\}} + \epsilon_{\{(p,2)\}} < \epsilon_{\{(p,1),(p,2)\}},
$$
and the remaining for are obtained by the mappings $p\to\tilde{p}$ and/or $1\to 2$. These inequalities transform into intersections and unions of sub-surfaces when using the mapping \eqref{eq:eps_to_polys}:
\begin{align*}
       \delta_{\{(1,1)\}}+\delta_{\{(1,2),(2,1),(p,1)\}} &< \delta_{\{(1,1),(p,1)\}}, \, \\\delta_{\{(1,2),(2,1),(p,1)\}} + \delta_{\{(1,2),(2,1),(p,2)\}} &< \delta_{\{(1,2),(p,1),(p,2)\}}+\delta_{\{(2,1),(p,1),(p,2)\}}, 
\end{align*}
respectively. Here, the $\delta_P$ are labelled by the cords that bound the sub-surface. The second inequality is an example of the case where the intersection of the surfaces on the left-hand side is given by multiple disjoint sub-surfaces.

\paragraph{Three-points}
At three-points the cosmohedron is three-dimensional, it has $108$ vertices, $168$ edges and $62$ facets. This means we will have 62 $\epsilon_C$, which will form $138$ inequalities, and $12$ equalities. One such equality is:
$$
\epsilon_{\{(1,1),(1,3)\}}+\epsilon_{\{(1,1),(p,1)\}}=\epsilon_{\{(1,1)\}}+\epsilon_{\{(1,1),(1,3),(p,1)\}}\, ,
$$
and the other $11$ equalities are variations of this one.  On the other hand, two examples of inequalities are:
\begin{align*}
    \epsilon_{\{(1,1)\}}+\epsilon_{\{(1,3)\}} < &\: \epsilon_{\{(1,1),(1,3)\}}\, , \\
    \epsilon_{\{(p,1),(p,2)\}}+\epsilon_{\{(p,1),(p,3)\}} < &\: \epsilon_{\{(p,1),(p,2),(p,3)\}}+\epsilon_{\{(p,1)\}}\, ,
\end{align*}
which have the corresponding form in terms of overlaps of sub-surfaces:
\begin{align*}
    \delta_{\{(1,1),(1,2),(2,3),(3,1)\}}+\delta_{\{(1,3),(3,1)\}} < &\: \delta_{\{(1,1),(1,3),(3,1)\}}\, , \\
    \delta_{\{(1,2),(2,3),(p,1),(p,3)\}}+\delta_{\{(2,3),(3,1),(p,1),(p,2)\}} < &\: \delta_{\{(2,3),(p,2),(p,3)\}}+\delta_{\{(1,2),(2,3),(3,1),(p,1)\}}\, ,
\end{align*}
where the union in the first line is the total energy sub-surface, which we set to zero.
This example also provides a good illustration of the factorization of the facets at one-loop. The facets labelled by the cords $\{(2,1)\}$, $\{(3,2)\}$ or $\{(1,3)\}$ will be dodecagons, since they constitute the factorization into the one-loop two-points cosmohedron, the tree level three-point cosmohedron and the two-site chain graph associahedron, which are a dodecagon, and two points, respectively. Thus, the cosmohedron will have three dodecagon facets. Then, the facets labelled by the cords $\{(p,1)\}$, $\{(p,2)\}$ or $\{(p,3)\}$, (as well as the facets with $p\to\tilde{p}$) will be decagons, since here the facet factorizes into one sub-surface with five boundaries and no puncture, thus it will be the cosmohedron of the five point wavefunction, which is a decagon, and the graph associahedron of the one site graph, which is a point. The cosmohedron will have $6$ decagon facets. Finally, the facets labelled by the cords $\{(1,1)\}$, $\{(2,2)\}$ or $\{(3,3)\}$ will be squares. Since they represent the factorization of the facet into a square and the one-loop one-point sub-surface, and the graph associated to it is the two-site chain. Thus, the facet is the product of two segments and a point, which is a square. Following this factorization properties, one can find the remaining facets of the cosmohedron.

\subsection{Extracting the loop wavefunction from geometry}

Extracting the wavefunction from the cosmohedron at loop level is very similar to tree level. One starts by constructing the permuto-cosmohedron, which follows from turning the equalities into inequalities, and then constructing the canonical form for the polytope and extracting the part with only simple poles. 

Firstly, we will discuss how to build the permuto-cosmohedron at loop level. We have seen in the beginning of this section that the structure of the equalities and inequalities is exactly the same. And each equality corresponds to a non-simple vertex in the cosmohedron. Then, to ``simplify" these vertices one turns the equalities into inequalities in the same way as we did at tree level:
\begin{equation*}
    \epsilon_C + \epsilon_{C^\prime} = \epsilon_{C \cup C^\prime} + \epsilon_{C \cap C^\prime} \to \epsilon_C + \epsilon_{C^\prime} < \epsilon_{C \cup C^\prime} + \epsilon_{C \cap C^\prime} \, .
\end{equation*}

Finding a parametrization of the $\epsilon$ satisfying all inequalities, will ensure we obtain the permuto-cosmohedron at loop level. 

To extract the full wavefunction, the pairs of singularities we associate to each facet have to be slightly reformulated, relative to the tree level case. If we consider a given facet and the corresponding partial triangulation, labeled by the set of cords $C$, and $n_C$ being the number of sub-surfaces with more than three bounding edges in the partial triangulation, then we still define,
\begin{equation}
\frac{1}{X_C} \equiv \frac{1}{n_C} \l \sum_{P, P^\prime \, {\rm meeting \, on \, edge}} \frac{1}{{\cal P}_P {\cal P^\prime}_{P^\prime}} \r \, .
\label{eq:WF_permuto_sings_loop}
\end{equation}

However, when in the set $C$ there is only one chord connecting the inner puncture to the disk boundary, $(p,i)$, then the partial triangulation will have a sub-surface with two edges which go around this chord -- say $\{(i,p),(p,i)\}$ -- like we see for the subsurfaces in red and blue in the (left) top russian doll depicted in figure \ref{fig:1loop}. In these cases we have to associate a triangle sub-surface to the chord, $(p,i)$, which we will define to be $\mathcal{P}_P \equiv \mathcal{T}_{(p,i)}$. And this sub-surface borders only with the sub-surface which goes around the cord $(p,i)$. The wavefunction is defined from the permuto-cosmohedron in the same way as at tree-level, except in the end, after selecting the single poles in the canonical form, we set all $\mathcal{T}_{(p,i)}\to 1$. Therefore, we can write, 
\begin{equation}
\Psi = \left. \l\frac{1}{E_t} \times  \Omega(X_C)\vert_{{\rm single \, poles \,in {\cal P}_P}}\r \right\vert_{\mathcal{T}_{(p,i)}\to 1}.
\label{eq:WF_from_geo_loop}
\end{equation}

\paragraph{One-loop two-point wavefunction}

The cosmohedron for the one-loop two-point wavefunction is simple, therefore is equivalent to the permutahedral ``blow-up". The cosmohedron is a dodecagon, and here we will discuss explicitly how to compute the contributions from three vertices, since the remaining ones are some variation of these. First, let us consider the vertex which results from the intersection of the facets $\{(p,1)\}$ and $\{(p,1),(p,2)\}$. Then according to the above discussion we can write:
\begin{equation*}
    \frac{1}{X_{\{(p,1)\}}}=\frac{1}{\mathcal{P}_{(1,2),(2,1),(p,1)}\mathcal{T}_{(p,1)}}\, , \quad
     \frac{1}{X_{\{(p,1),(p,2)\}}}=\frac{1}{\mathcal{P}_{(1,2),(p,1),(p,2)}\mathcal{P}_{(2,1),(p,1),(p,2)}}\, .
\end{equation*}
Therefore, the contribution from this vertex is:
$$
\frac{1}{E_t \mathcal{P}_{(1,2),(2,1),(p,1)}\mathcal{P}_{(1,2),(p,1),(p,2)}\mathcal{P}_{(2,1),(p,1),(p,2)}}\, ,
$$
where we have set the value of $\mathcal{T}_{(p,1)}$ to one at the end. Then, we can compute the contribution of the vertex which is the intersection of the facet $\{(p,1)\}$ and $\{(1,1),(p,1)\}$, 
\begin{equation*}
    \frac{1}{X_{\{(p,1)\}}}=\frac{1}{\mathcal{P}_{(1,2),(2,1),(p,1)}\mathcal{T}_{(p,1)}}\, , \quad
     \frac{1}{X_{\{(1,1),(p,1)\}}}=\frac{1}{\mathcal{P}_{(1,1),(1,2),(2,1)}\mathcal{P}_{(1,1),(p,1)}}+\frac{1}{\mathcal{P}_{(1,1),(p,1)}\mathcal{T}_{(p,1)}}\, ,
\end{equation*}
which in the end will lead to the contribution,
$$
\frac{1}{E_t \mathcal{P}_{(1,2),(2,1),(p,1)}\mathcal{P}_{(1,2),(2,1),(1,1)}\mathcal{P}_{(1,1),(p,1)}}\, ,
$$
keep in mind that we dropped the terms with $\mathcal{T}_{(p,1)}^2$, just like for any other sub-surface, and only in the end we set $\mathcal{T}_{(p,1)}\to 1$. And finally, we can compute the contribution from the vertex at the intersection of the facets $\{(1,1)\}$ and $\{(1,1),(p,1)\}$,
\begin{equation*}
    \frac{1}{X_{\{(1,1)\}}}=\frac{1}{\mathcal{P}_{(1,1),(1,2),(2,1)}\mathcal{P}_{(1,1)}}\, , \quad
     \frac{1}{X_{\{(1,1),(p,1)\}}}=\frac{1}{\mathcal{P}_{(1,1),(1,2),(2,1)}\mathcal{P}_{(1,1),(p,1)}}+\frac{1}{\mathcal{P}_{(1,1),(p,1)}\mathcal{T}_{(p,1)}}\, ,
\end{equation*}
and its contribution to the wavefunction is:
$$
\frac{1}{E_t \mathcal{P}_{(1,1),(1,2),(2,1)}\mathcal{P}_{(1,1)}\mathcal{P}_{(1,1),(p,1)}}\, .
$$

\paragraph{One-loop three-point wavefunction}
Now we will proceed with the three-point one-loop example. Here we will compute one contribution from a non-simple vertex and one of the terms in the triangle diagram, since these are the vertices that best illustrate the differences with the tree-level computations. Let us start with the non-simple vertex, where the facets $\{(1,1)\}$, $\{(1,1),(1,3)\}$ , $\{(1,1),(p,1)\}$, and $\{(1,1),(1,3),(p,1)\}$ meet. The permutahedral ``blow-up" splits it into two vertices, one of which, is the intersection of the facets $\{(1,1)\}$, $\{(1,1),(1,3)\}$ , and $\{(1,1),(1,3),(p,1)\}$, and another $\{(1,1)\}$,  $\{(1,1),(p,1)\}$, and $\{(1,1),(1,3),(p,1)\}$. For the first vertex, we can write,

\begin{align*}
    &\frac{1}{X_{\{(1,1)\}}}=\frac{1}{\mathcal{P}_{(1,1),(1,2),(2,3),(3,1)}\mathcal{P}_{(1,1)}}\, , \,
     \frac{1}{X_{\{(1,1),(1,3)\}}}=\frac{1}{\mathcal{P}_{(1,2),(1,3),(2,3)}\mathcal{P}_{(1,1),(1,3),(3,1)}}+\frac{1}{\mathcal{P}_{(1,1)}\mathcal{P}_{(1,1),(1,3),(3,1)}}\, , \\
      &\frac{1}{X_{\{(1,1),(1,3),(p,1)\}}}=\frac{1}{\mathcal{P}_{(1,2),(2,3),(1,3)}\mathcal{P}_{(1,1),(1,3),(3,1)}}+\frac{1}{\mathcal{P}_{(1,1),(p,1)}\mathcal{P}_{(1,1),(1,3),(3,1)}}+\frac{1}{\mathcal{P}_{(1,1),(p,1)}\mathcal{T}_{(p,1)}}\, ,
\end{align*}
for the second vertex the partial triangulation with two cords will differ, it is,
\begin{equation*}
    \frac{1}{X_{\{(1,1),(p,1)\}}}=\frac{1}{\mathcal{P}_{(1,1),(1,2),(2,3),(3,1)}\mathcal{P}_{(1,1),(p,1)}}+\frac{1}{\mathcal{P}_{(1,1),(p,1)}\mathcal{T}_{(p,1)}}\, .
\end{equation*}

Naturally, both vertices will give the same contribution, which is,
$$
\frac{1}{2 E_t \mathcal{P}_{(1,1),(1,2),(2,3),(3,1)}\mathcal{P}_{(1,1)}\mathcal{P}_{(1,1),(p,1)}\mathcal{P}_{(1,2),(2,3),(1,3)}\mathcal{P}_{(1,1),(1,3),(3,1)}}\, .
$$

Since they are two, the one-half will cancel.
Finally, we will look at the vertex at the intersection of the facets, $\{(p,1)\}$,$\{(p,1),(p,2)\}$, and $\{(p,1),(p,2),(p,3)\}$. This is one of the 6 vertices in the facet of the cosmohedron which corresponds to the triangle diagram. For this vertex, we can write,
\begin{align*}
    &\frac{1}{X_{\{(p,1)\}}}=\frac{1}{\mathcal{P}_{(1,2),(2,3),(3,1),(p,1)}\mathcal{T}_{(p,1)}}\, , \quad
     \frac{1}{X_{\{(p,1),(p,2)\}}}=\frac{1}{\mathcal{P}_{(1,2),(p,1),(p,2)}\mathcal{P}_{(2,3),(3,1),(p,1),(p,2)}}\, , \\
    &\frac{1}{X_{\{(p,1),(p,2),(p,3)\}}}=\frac{1}{\mathcal{P}_{(1,2),(p,1),(p,2)}\mathcal{P}_{(2,3),(p,2),(p,3)}}+\frac{1}{\mathcal{P}_{(2,3),(p,2),(p,3)}\mathcal{P}_{(3,1),(p,1),(p,3)}}+\frac{1}{\mathcal{P}_{(1,2),(p,1),(p,2)}\mathcal{P}_{(3,1),(p,1),(p,3)}}\, .
\end{align*}

This leads to the contribution, for this vertex,
$$
\frac{1}{E_t \mathcal{P}_{(1,2),(p,1),(p,2)}\mathcal{P}_{(2,3),(p,2),(p,3)}\mathcal{P}_{(3,1),(p,1),(p,3)}\mathcal{P}_{(2,3),(3,1),(p,1),(p,2)} \mathcal{P}_{(1,2),(2,3),(3,1),(p,1)}}\, ,
$$
which we can check to be one of the tubings of the triangle diagram.

\section{Cosmological correlahedra}

Having found a geometry underlying the wavefunction for Tr($\phi^3$) theory, it is natural to go a step further, and ask whether there is any geometry not just for the wavefunction, but directly for the physical observable, the correlator. Taking a step back to put this question into context, while the discoveries of geometries underlying amplitudes and wavefunctions have been remarkable, it is still unsatisfying that what we are supposed to physically {\it do} with these objects -- namely, mod-square them to get probabilities, expectation values and correlation functions -- is left untouched. The burning question is simple -- given that there are autonomous combinatorial/geometric structures underlying amplitudes and wavefunctions, what sort of cousin of these objects makes it natural to discover the Born rule, and ultimately the physical observables? 

In this section, we will sketch an answer to this question by briefly introducing ``cosmological correlahedra" capturing all the contributions to the correlator in Tr($\phi^3$) theory. As we will see, they naturally combine both associahedra and cosmohedra in a single higher-dimensional polytope.

Let's begin with a quick reminder on how to compute flat-space correlators in the language of polygons. In addition to subpolygons, we simply also include the chords $k_{i,j}=|\vec{k}_{i,j}|$ in the story. A term in the correlator is determined by first giving a (possibly empty) collection of non-overlapping chords, $\mathcal{C}$, with which we associate a factor of $\prod_{(i,j)\in\mathcal{C}} (1/k_{i,j})$. This collection of chords divides the polygon into subpolygons ($\mathcal{P}$ compatible with $\mathcal{C}$), and we further multiply by the wavefunction of each of the subpolygons. We then sum over all choices for the initial set of non-overlapping chords. So we can write:
\begin{equation}
    \text{Corr}_n = \Phi_n + \sum_{\mathcal{C} \neq  \emptyset} \prod_{(i,j)\in\mathcal{C}} \frac{1}{k_{i,j}} \times \prod_{\mathcal{P} \text{ compatible } \mathcal{C}} \Psi_\mathcal{P},
    \label{eq:Corr}
\end{equation}
where here we manifestly separated the case in which the collection of chords is empty, that just gives us the full wavefunction. The combinatorics of the full correlator is then clearly a hybrid between those of amplitudes (non-crossing chords) and the wavefunction (non-overlapping sub-polygons). 
\begin{figure}[t]
    \centering
    \includegraphics[width=\linewidth]{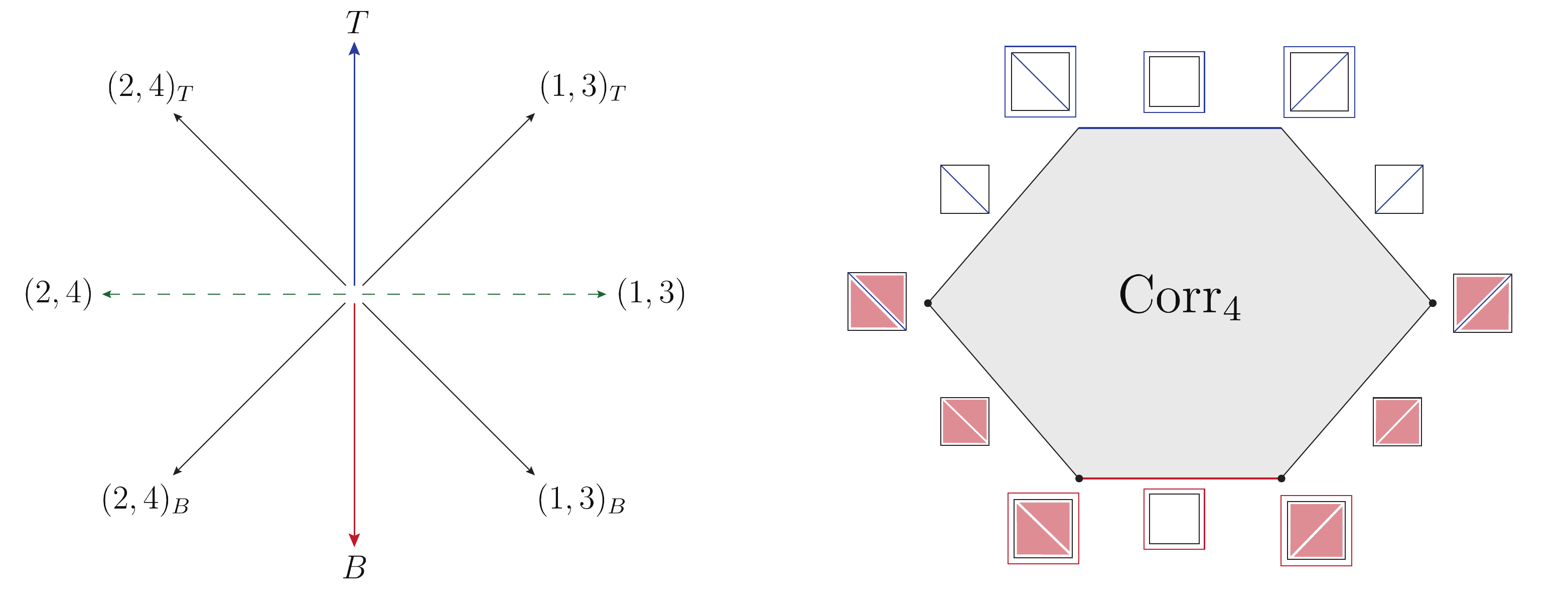}
    \caption{(Left) Fan of the cosmological correlahedron for $n=4$. In dashed, we represent the underlying associahedron fan. (Right) 4-points cosmological correlahedron.}
    \label{fig:Corr4}
\end{figure}

Now, it is natural to expect any geometry for the full correlator to live in one higher dimension than the associahedron/cosmohedron. The reason is that while all the terms in the wavefunction have an $E_t$ singularity, which is not explicitly included as a facet in the cosmohedron, this is not the case for the full correlator -- some terms have $E_t$ singularities (those coming from $\Psi_n$ in \eqref{eq:Corr}) and others don't (the remaining terms in \eqref{eq:Corr}). Thus, it stands to reason to think about an object in one higher dimension, roughly corresponding to $E_t$, with a ``bottom" facet associated with $E_t$, which looks like the cosmohedron. If this object is to include the combinatorics of non-overlapping chords, then we know that these objects alone, with no reference to sub-polygons at all, are captured by the associahedron. So it is reasonable to expect that the ``cosmological correlahedron" we are looking for should be a sort of sandwich in an extra dimension,  with the cosmohedron at the ``bottom", and an associahedron maximally far away, at the ``top" of the new direction. 

This can also be nicely motivated by trying to guess what the fan of this higher-dimensional object might look like. Let us consider the simplest possible case of $n=4$. The fan for the associahedron has the two usual rays for g-vectors $(1,3),(2,4)$, pointing in opposite directions in one dimension.  But we will introduce two new rays, ``$B$" and ``$T$" (for ``bottom" and ``top") pointing in opposite directions in a second direction. We know we want to have facets of the correlator polytope corresponding to two different kinds of single chords: one where the single chord is associated with subpolygons (like we saw earlier for the wavefunction), and another where it is associated simply with the $|\vec{k}|$ in the correlator. We will thus record images of the rays $(1,3), (2,4)$ on the bottom and top, by defining 
\begin{equation}
\begin{aligned}
    &(1,3)_B = (1,3) + B, \quad (2,4)_B = (2,4) + B, \\
    &(1,3)_T = (1,3) + T, \quad \, (2,4)_T = (2,4) + T.
\end{aligned}
\end{equation}
This gives us the six rays $T,(1,3)_T,(1,3)_B,B, (2,4)_B, (2,4)_T$, which is naturally associated with the  hexagon shown in figure \ref{fig:Corr4}. We see that this hexagon has an interval at the top and one at the bottom, naturally associated with the $n=4$ associahedron and cosmohedron respectively. Note that the top facet only has vertices of the associahedron, and does not by itself correspond to any terms in the correlator. But the remaining four vertices (highlighted in black in figure \ref{fig:Corr4}) are naturally associated with all the terms in the correlator. 

\begin{figure}[t]
    \centering
    \includegraphics[width=\linewidth]{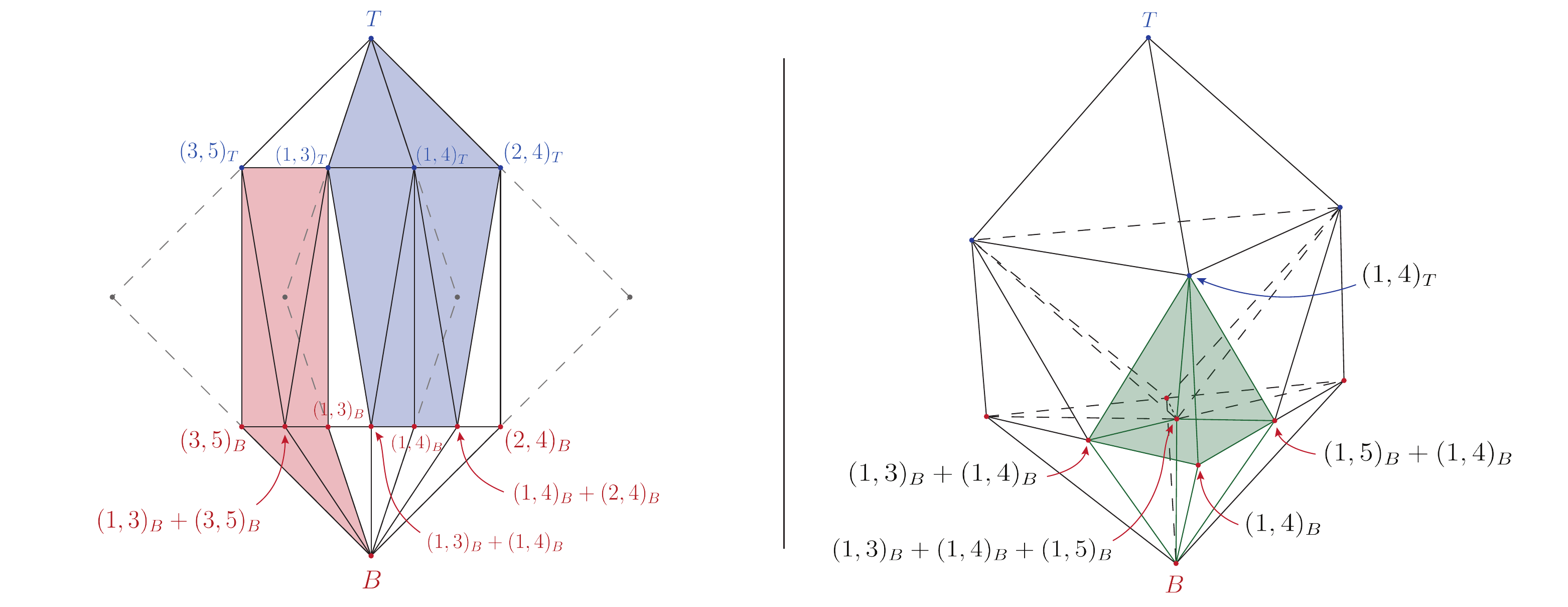}
    \vspace{2mm}
    \caption{(Left) Projection of the $n=5$ cosmological correlahedron fan. In dashed we represent the underlying associahedron fan with rays $(3,5)$, $(1,3)$, $(1,4)$ and $(2,4)$ marked in gray, with the added dimension corresponding to $E_t$. Shaded in red we highlight the pentagonal facet which is touching the base Cosmo$_5$, and in blue the hexagonal facet which is touching the top Assoc$_5$. (Right) 3-dimensional projection of the Corr$_6$ fan, coming from the underlying 3-dimensional associahedron cone containing rays $(1,3),(1,4),(1,5)$. In green, we highlight a square pyramid corresponding to a non-simple vertex of Corr$_6$.} 
    \label{fig:corrFans}
\end{figure}

Let's move on to the next example at $n=5$, where we will see almost all the relevant structure for general $n$. We again start from the rays of the associahedron, which we can label with the chords $(1,3),(1,4),(2,4),(2,5),(2,6)$, now living in two dimensions, and add $B, T$ pointing in opposite directions in an extra third direction. We then produce the rays $(i,j)_B = (i,j) + B$ and $(i,j)_T = (i,j) + T$ as before. But on the bottom, we continue to produce the rest of the rays for the cosmohedron as we have described before, by producing the sums of the bottom rays. To produce the cones, we begin by connecting all the bottom rays to $B$ and all the top rays to $T$. Next, we connect all the bottom rays amongst each other as for the cosmohedron, while all the top rays are connected to each other as they are for the associahedron. Finally, the top and bottom and connected by a very simple rule: an $(i,j)_T$ is connected to every bottom ray that contains $(i,j)$. The fan for the $n=5$ is three-dimensional but as usual we can draw a projective picture of it two-dimensionally, and this is drawn in figure \ref{fig:corrFans} (left); a combinatorial representation of the wavefunction is shown in the top of figure \ref{fig:corrcomb} (at the end of the note). Again, we see that the ``top" facet is the associahedron, and the ``bottom" facet is a cosmohedron. All the faces in between are labelled by mixtures of the ``top" chords -- which we can think of as the $|\vec{k}|$ chords in the wavefunction, and ``bottom" chords -- which give us nested subpolygons. Apart from the vertices on the top associahedron facet (marked in blue), the rest of the vertices precisely correspond to all the terms in the correlator (marked in black).

The cosmological correlahedron has a natural combinatorial definition for all $n$. Faces are labelled by $\{C,P\}$, where $C$ is a collection of  non-overlapping chords as for associahedra, and $P$ is a collection of non-overlapping subpolygons satisfying the russian doll rule as for cosmohedra, except that we now include the ``full perimeter" as subpolygons, and we have two full perimeters labelled by $T$, $B$. There are two special faces, the ``top" facet where $\{C = {\rm empty}, P=P_{{\rm full, top}}\}$ and the ``bottom" facet where $\{C = {\rm empty}, P=P_{{\rm full, bottom}}\}$. No subpolygons, nor $P_{\rm full, \, bottom}$ are allowed to occur in the list with $P_{{\rm full, \, top}}$. Then, the cosmological correlahedron generalizes the notion of compatibility for associahedra and cosmohedra in the obvious way: 
\begin{equation}
    \{C^\prime, P^\prime\} \text{ is a face of }\{C,P\} \text{ if } C \subset C^\prime \text{ and } P \subset P^\prime.
\end{equation} 

At $n=6$ the fan is four-dimensional, but we can draw a relevant piece of it three-dimensionally, as done in figure \ref{fig:corrFans} (right). The rays are produced and connected to form cones in exactly the way we described above: starting with the rays of the associahedron $(i,j)$, producing $(i,j)_B = (i,j) + B$ and $(i,j)_T = (i,j) + T$,  producing the rest of the rays of the cosmohedron from the bottom rays, and connecting all the bottom rays as for cosmohedra, the top rays are connected as for associahedra, and every top $(i,j)_T$ ray to every bottom rays that contains $(i,j)$. Again remarkably, the cones are non-overlapping, and apart from the purely top ones giving all the triangulations of the $n$-gon, the rest of the cones are associated with every term in the correlator. 

As for cosmohedra, starting with $n=6$ we encounter the phenomenon of non-simple vertices for the cosmological correlahedron. In the figure, the five rays $(1,4)_T$ together with $(1,4)_B, (1,3)_B + (1,4)_B, (1,5)_B + (1,4)_B, (1,3)_B + (1,4)_B + (1,5)_B$ form a square-pyramid, associated with a single term in the correlator. 

The picture for the fan of the cosmological correlahedron can clearly be extended to loops, and an example of a three-dimensional polytope for the 1-loop bubble is shown in the bottom of figure \ref{fig:corrcomb} (at the end of the note).

\begin{figure}[t]
    \centering
    \includegraphics[width=\linewidth]{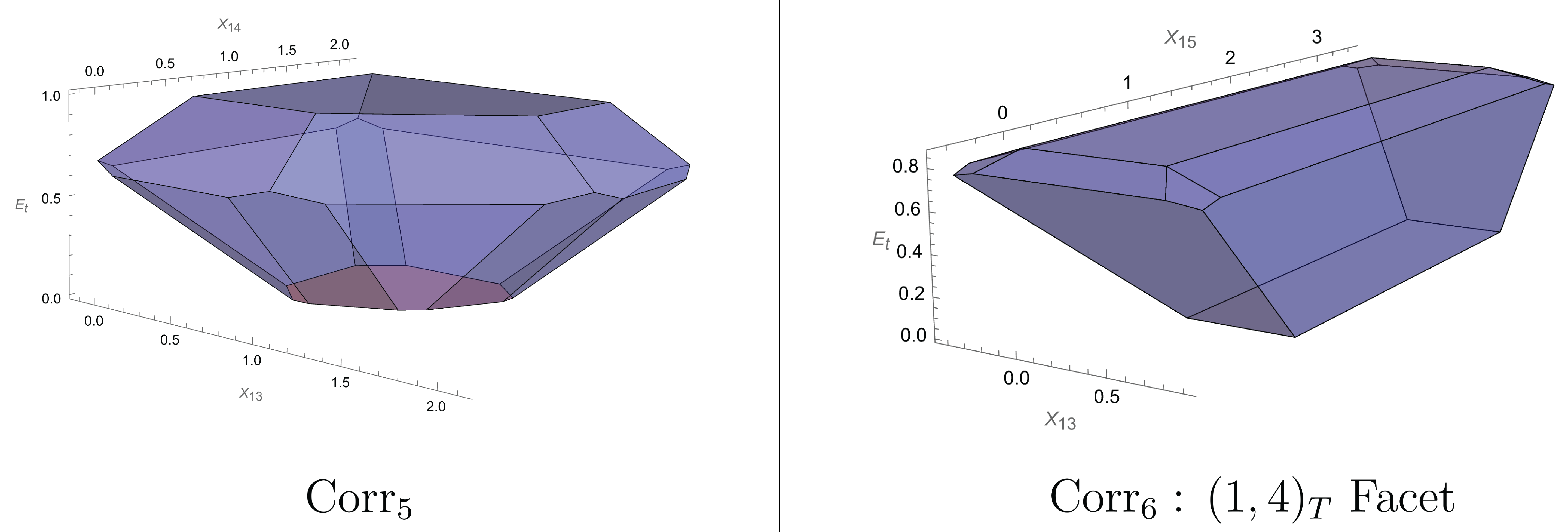}
    \vspace{2mm}
    \caption{(Left) Embedding of the Corr$_5$. (Right) Embedding of the $(1,4)_T$ facet of Corr$_6$.}
    \label{fig:embedCorr}
\end{figure}

It is also natural to cut out the cosmological correlahedron by inequalities, extending those of associahedra and cosmohedra in the obvious way, involving ``shaving parameters" $\epsilon_{T,B}$ for both the top and bottom rays. We have checked that the polytopes produced in this way have exactly the correct combinatorics for $n=6$, and that they have the correct number of vertices to account for the correlator up to $n=8$. In figure \ref{fig:embedCorr} we show the embedding of the $n=5$ cosmological correlahedron as well as the embedding of the facet $(1,4)_T$ of the $n=6$ one. For general $n$, we expect an interesting relation between $\epsilon_{T,B}$ to produce the correct combinatorics. We leave an exploration of this question, as well as the systematics of extracting the correlator from the geometry, to future work. 

\section{Outlook}

This note has concerned itself simply with introducing the cosmohedra and explaining some of the most basic physics and mathematics associated with them. 
As with amplituhedra and associahedra, it is remarkable to find mathematical structures that autonomously ``know about" and ``discover" the wavefunction. For the associahedron itself, the magic is in the basic ABHY \cite{ABHY,CausalDiamonds} ``$X+X-X-X = c, \, X \geq 0$" equations that cut it out via inequalities. These equations can be motivated and interpreted in various ways, from arising as a sort of ``wave equation" in kinematic space to capturing the data of curves on surfaces in the simplest possible way. None of these make any reference to the collection of all Feynman diagrams, and yet they give rise to an object that unifies and discovers all diagrams. As we have seen in proceeding to cosmohedra, we must include a further set of equations associated with partial triangulations, $\sum_{c \subset C} X_c \geq \epsilon_C$. The new magic is clearly in the conditions $\epsilon_C + \epsilon_{C^\prime} \leq \epsilon_{C \cup C^\prime} + \epsilon_{C \cap C^\prime}$, which must sometimes be imposed as inequalities and sometimes as equalities depending on $C,C^\prime$. Such conditions are ubiquitous in the study of various polytopes associated with graphs, where they are known as ``submodularity conditions". In our context, when the inequalities are strictly satisfied we get the ``permuto-cosmohedron" cousin of the cosmohedron. But the ``perfect" object with the correct combinatorics needs the more subtle combination of equalities and inequalities, that as we described are captured by putting $\epsilon_C = \sum_P \delta_P$, summing over all subpolygons in $C$, where the $\delta_P$ satisfy the strict submodularity condition $\delta_P + \delta_{P^\prime} < \delta_{P \cup P^\prime} + \delta_{P \cap P^\prime}$. Again, all of these expressions treat the chords in $C$ democratically; there is no hint of any sort of russian doll structure expected for the wavefunction. Nonetheless, they emerge, as a consequence of this extremely simple yet obviously deep combinatorics and geometry.

There are many open questions surrounding simply understanding these objects better. Chief amongst them is a deeper understanding of precisely how the geometry determines the wavefunction -- we have given a novel prescription for extracting the wavefunction from the canonical form of the cosmohedron -- involving replacing the poles associated with facets with products of pairs of poles, and keeping only terms with simple poles in the resulting expression. What is the deeper origin and meaning of this prescription? Is there a different idea that directly gives us the wavefunction, with multiple poles automatically removed? And is there a bigger geometry that associates individual poles -- not pairs of them -- with facets, so that the usual notion of canonical form would give the wavefunction? 

On this last point, it is worth contrasting our story with that of cosmological polytopes~\cite{CosmoPolytopes}, which for single graphs {\it do} give a geometry with a facet associated to every pole of the wavefunction. For single graphs, cosmohedra instead tell us to work with a different geometry -- graph associahedra -- and these objects are already somewhat more interesting: a common set of poles corresponding to the total energy and the individual internal triangles are factored out, and the geometry only knows about the remaining non-trivial poles that differ between the russian dolls. Consequently, even something as basic as the emergence of the amplitude as $E_t \to 0$ is understood differently: in the cosmological polytope we simply go to the total energy facet and discover (at tree-level) a simplex, which gives the amplitude. Meanwhile, as explained in detail in appendix \ref{app:CosmoPolytopes_GraphAssoc}, as $E_t \to 0$ many of the faces of the graph associahedra shrink, so the resulting objects simplifies dramatically to a product of simplices. 

Now, the cosmohedron unifies all the graph associahedra for the different diagrams into a single object. This single object does not have a single facet for every possible singularity of the wavefunction, but for (canonical) pairs of them. If it were possible to realize the old idea of gluing all cosmological polytopes together into a bigger object, then we might have a facet for every singularity. There is still no concrete idea for how to make this work, and certainly the way the cosmohedron accomplishes this -- using the geometry of the underlying associahedron as the way to generate and combine all the diagrams -- does not mesh with cosmological polytopes simply because the dimension of associahedra and cosmological polytopes are so different. At any rate, the unusual prescription we have found for extracting the wavefunction from the canonical form of the cosmohedron geometry may come to be seen as either a feature or a bug, and deserves further exploration. 

In addition to better understanding cosmohedra, there are also a huge number of bigger questions left open by our investigations, and we close by highlighting two of them that seem especially interesting and urgent. We have focused on the cosmohedron geometry associated with the ``energy integrand" for the cosmological wavefunction, but recent work on ``kinematic flow" \cite{KinFlow,KinFlowLoop} has shown that the full integrated objects satisfy differential equations, which for single graphs have a natural interpretation associated with the growth of graph tubings. There are similar interpretations for the full integrated amplitude, in terms of growing subpolygons. These tubings and subpolygons have additional decorations relative to what we have seen there, corresponding to giving two different colorings (corresponding to $\pm$ signs of energies) for internal chords. It would be fascinating to understand whether there is an extension of the cosmohedron that captures this combinatorics. 

In another direction, recall that in a precise sense, the associahedron gives us a direct path for discovering strings starting from particle amplitudes. The deep clue to the ``strings" hiding in plain sight underlying ``particles" is that, while the associahedron is primitively cut out by a simple set of inequalities, the particular structure of these inequalities also allows us to think about the associahedron as being built out of a Minkowski sum of simple pieces. The summands of these Minkowski sums can be thought of Newton polytopes for certain polynomials, and this in turn immediately generalizes particle to string amplitudes \cite{StringyCanForms}. 
Of course history did not proceed in this way -- the Koba-Nielsen formula was written down long before the connection to the associahedron was discovered. But with cosmology, we have a new opportunity. At present, there is no useful perturbative worldsheet picture for cosmological observables (or AdS boundary correlators, which our model of conformally coupled scalars is equally well suited to describe). But we have now discovered a combinatorial/geometric object unifying all diagrams for cosmology. What is the analog or extension of the Minkowski sum picture, F-polynomials, and $u$ variables in our new setting? And what sort of ``stringy" generalization of the particle wavefunctions might it describe? 

\section*{Acknowledgements}
We thank Federico Ardilla, Paolo Benincasa, Nick Early, Pavel Galashin, Austin Joyce, Thomas Lam, Hayden Lee, Alex Postnikov and Hugh Thomas for discussions and comments.
The work of N.A.H. is supported by the DOE (Grant No. DE-SC0009988), by the Simons Collaboration on Celestial Holography, by the ERC universe+ synergy grant, and further support was made possible by the Carl B. Feinberg cross-disciplinary program in innovation at the IAS.
C.F. is supported by FCT/Portugal (Grant No.~2023.01221.BD). F.V. research is funded by the European Union (ERC, UNIVERSE PLUS, 101118787). We would also like to thank the developers of \texttt{polymake} \cite{polymake}.

\appendix
\section{Lightning review of the perturbative expansion of $\Psi$}
\label{app:pertwf}
In this appendix, we review the perturbative formulation of the wavefunction. The theory we study here is a theory of colored massless scalars interacting via a cubic interaction with the following action given in \eqref{eq:action}.

The wavefunction is then defined via the path integral as follows
\begin{equation}
    \Wf = \int{} \mathcal{D}[\varphi] e^{i \mathcal{S}[\varphi]},
\end{equation}
where we integrate over all field configurations that satisfy the boundary condition at asymptotic future $\varphi(\vec{x},\eta=0)= \phi(\vec{x})$, as well as the Bunch-Davies vacuum/$i\epsilon$ prescription in the past \cite{Bunch:1978yq}, $i.e.$ $\varphi(\vec{x},\eta=-\infty(1-i\epsilon))=0$. Since we have spatial momentum conservation, it is useful to go to Fourier space, $\vec{k}$, where the solution of the free equations of motion satisfying the boundary conditions is given by:
\begin{equation}
    \varphi_{\vec{k}} = \phi_{\vec{k}} e^{i E_{k} \eta},
\end{equation}
with $E_k = |\vec{k}|$, so that we can read the 3-point interaction in momentum space to be:
\begin{equation}
    \frac{1}{3}\int \prod_{i=1,2,3}d^d k_i \, \int_{-\infty}^0 d\eta \lambda_3(\eta) \phi_{\vec{k}_1}\phi_{\vec{k}_2}\phi_{\vec{k}_3} e^{i(E_{k_1}+E_{k_2}+E_{k_3})\eta} \delta^d(\vec{k}_1+\vec{k}_2+\vec{k}_3),
\end{equation}
to perform the $\eta$ integral it is useful to Fourier represent $\lambda_3$ and analyze each mode separately:
\begin{equation}
\begin{aligned}
   & \frac{1}{3}\int d\varepsilon \int \prod_{i=1,2,3}d^d k_i \,  \int_{-\infty}^0d\eta \lambda_3(\varepsilon) \phi_{\vec{k}_1}\phi_{\vec{k}_2}\phi_{\vec{k}_3} e^{i(E_{k_1}+E_{k_2}+E_{k_3}+\varepsilon)\eta} \delta^d(\vec{k}_1+\vec{k}_2+\vec{k}_3)\\
   &= -\frac{i}{3} \int d\varepsilon \int \prod_{i=1,2,3}d^d k_i\, \phi_{\vec{k}_1}\phi_{\vec{k}_2}\phi_{\vec{k}_3} \delta^d(\vec{k}_1+\vec{k}_2+\vec{k}_3)\lambda_3(\epsilon) \underbrace{\frac{1}{E_{k_1}+E_{k_2}+E_{k_3}+\varepsilon}}_{\Psi^{(3)}},
\end{aligned}
\end{equation}
from where we can read off the bulk-to-boundary propagator $G_{B,\partial}(E_k,\eta) = e^{i E_k \eta}$ as well as the three-point wavefunction coefficient $\Psi^{(3)}$. Here we have left $\lambda_3(\eta)$ outside $\Psi^{(3)}$ to highlight the connection of this example to the case where the couplings are simply constants: if $\lambda_3(\eta) \equiv \lambda_3$, then we would have gotten $\Psi^{(3)}=1/(E_{k_1}+E_{k_2}+E_{k_3})$, so we can get the time-dependent case by shifting the sum of the energies entering the 3-point vertex by $\varepsilon$ and integrating it against some kernel, $\lambda_3(\varepsilon)$, which depends on the precise time dependence of the problem we want to study. For this reason, from now on we will focus on the simpler case $\lambda_3(\eta) \equiv \lambda_3$ and later come back to the way to transform back into the most general time-dependent case.

By expanding around the free solution, $\varphi_{\vec{k}}=\phi_{\vec{k}} e^{i E_{k} \eta} + \delta \varphi_{\vec{k}}$, we can perform the path integral in $\delta \varphi_{\vec{k}}$. The bulk-to-bulk propagator coming from $\delta \varphi_k$ reads
\begin{equation}
  G_{B,B}(E_{k};\eta_1,\eta_2) = \frac{1}{2 E_k} \left( e^{iE_k(\eta_1-\eta_2)}\theta(\eta_1-\eta_2)+ e^{iE_k(\eta_2-\eta_1)}\theta(\eta_2-\eta_1)- e^{iE_k(\eta_2+\eta_1)}\right),
  \label{eq:bulktobulk}
\end{equation}
where the first two terms are the standard Feynman propagator and the last one ensures $\delta_{\varphi} \to 0$ as $\eta_{1,2} \to 0$. 

The wavefunction can be decomposed in the wavefunction coefficients, $\Psi_n$, as follows:
\begin{equation}
    \Psi = \exp\left\{ \sum_{n\geq 2} \frac{1}{n!} \int \prod_{i=1}^n\, d^d k_i \, \Psi_n[\vec{k}_{i} ]\, \delta^d\left(\textstyle{\sum_i} \vec{k}_i\right)  \right\},
\end{equation}
each $\Psi_n$ can be represented via a diagrammatic expansion in momentum space, and these are the object of study throughout the note. 

For example at 4-points, fixing the color ordering of the external $\phi$'s to be $\Psi^{(4)}(\phi_{\vec{k_1}},\phi_{\vec{k_2}},\phi_{\vec{k_3}},\phi_{\vec{k_4}})$, we have contributions from two diagrams:
\begin{equation}
\begin{aligned}
    &\text{$s$-channel:} \quad \int_{-\infty}^0 d \eta d \eta^\prime \prod_{i=1,2}G_{B,\partial}(E_{i},\eta) \prod_{j=3,4}G_{B,\partial}(E_{j},\eta^\prime) G_{B,B}(E_{1,2};\eta,\eta^\prime),\\
    &\text{$t$-channel:} \quad \int_{-\infty}^0 d \eta d \eta^\prime \prod_{i=1,4}G_{B,\partial}(E_{i},\eta) \prod_{j=2,3}G_{B,\partial}(E_{j},\eta^\prime) G_{B,B}(E_{2,3};\eta,\eta^\prime),
\end{aligned}
\end{equation}
where $E_{1,2}= |\vec{k}_1+\vec{k}_2|=|\vec{k}_3+\vec{k}_4|$ and $E_{2,3}= |\vec{k}_2+\vec{k}_3|=|\vec{k}_1+\vec{k}_4|$. Now, from each channel, since each $G_{B,B}$ has three terms, we would naively expect to get three terms. However, remarkably these terms nicely cancel to give a single contribution from each channel:
\begin{equation}
\begin{aligned}
    &\text{$s$-channel:} \, \frac{1}{(E_1+E_2+E_{1,2})(E_3+E_4+E_{1,2})(E_1+E_2+E_3+E_4)},\\
    &\text{$t$-channel:} \, \frac{1}{(E_2+E_3+E_{2,3})(E_1+E_4+E_{2,3})(E_1+E_2+E_3+E_4)}. 
\end{aligned}
\end{equation}

A remarkable feature of the wavefunction is that it \textit{contains} the amplitude in the total energy, $E_t$, pole, $i.e.$ when we extract the residue at $E_t=\sum_{i=1}^n E_i=0$ we should obtain the scattering amplitude. We can already observe this feature for this simple 4-point example, by noting that when we have $E_1+E_2+E_3+E_4=0$, we have 4-momentum conservation and in particular $(E_1+E_2+E_{1,2})(E_3+E_4+E_{1,2})= (E_1+E_2+E_{1,2})(-E_1-E_2+E_{1,2}) = (p_1+p_2)^2=s$, where $p_i^\mu$ stands for the 4-momentum; and similarly for the $t$-channel contribution. 
\begin{figure}[t]
    \centering
    \includegraphics[width=\linewidth]{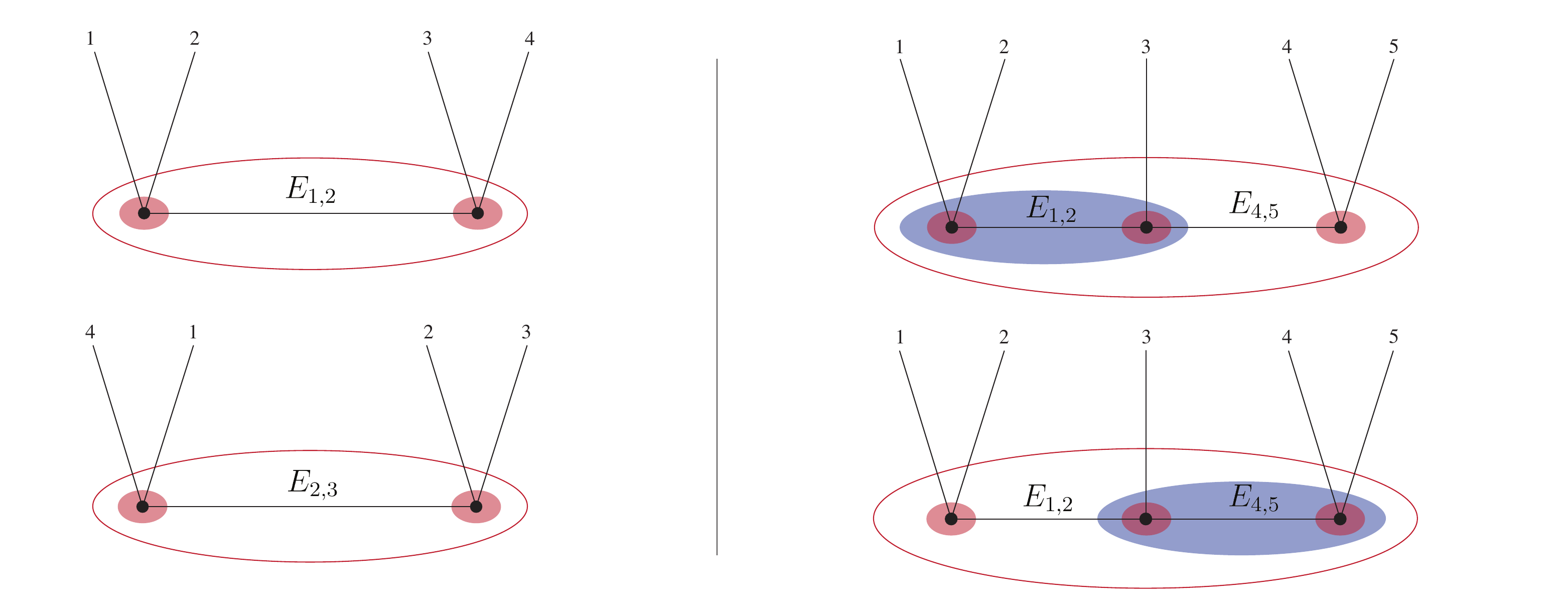}
    \caption{Tubings at 4 and 5 points}
    \label{fig:tubings}
\end{figure}

In addition, we can see that the contribution we get from each diagram corresponds is a product of three factors: the first two correspond to the sum of the energies entering each vertex, and the last one, to the sum of the energies entering the full diagram $E_t$ (energies entering the tubes depicted on the left of figure \ref{fig:tubings}). 

Similarly, if we do the same exercise at 5-points for the diagram depicted in figure \ref{fig:tubings} (right), we get two terms (instead of the naive nine terms from the two $G_{B,B}$):

\begin{equation}
\begin{aligned}
    &\frac{1}{E_t (E_1+E_2+E_{1,2})(E_{1,2}+E_3+E_{4,5})(E_{4,5}+E_4+E_5)(E_1+E_2+E_3+E_{4,5})} \\
    &+  \frac{1}{E_t (E_1+E_2+E_{1,2})(E_{1,2}+E_3+E_{4,5})(E_{4,5}+E_4+E_5)(E_{1,2}+E_3+E_4+E_5)},
\end{aligned}
\label{eq:5ptExample}
\end{equation}
where once more we get the factors corresponding to the sums of the energies entering each vertex, the total energy $E_t$, and now a new factor corresponding to the energy entering a subgraph $(E_1+E_2+E_3+E_{4,5})$ in the first one and $(E_{1,2}+E_3+E_4+E_5)$. Each contribution is then naturally associated with a \textit{tubing} of the 5-point graph we are studying -- this is a maximal collection of subgraphs inside a graph (see right of figure \ref{fig:tubings}).

For example, at 5-points the diagram considered above corresponds to the triangulation of the pentagon containing chords $(1,3)$ and $(1,4)$. Now, let's look back at the first term in \eqref{eq:5ptExample}, and notice that each term appearing can be associated to a perimeter of a subpolygon inside this momentum $5$-gon (as explained in section \ref{sec:Def_Wf}):
\begin{equation}
\begin{aligned}
    &\mathcal{P}_{1,2,3} =E_1 +E_2 +E_{1,2} = |\vec{k}_1| +|\vec{k}_2| +|\vec{k}_1+\vec{k}_2|,\\
     &\mathcal{P}_{1,3,4} = E_{1,2} +E_3 +E_{4,5} =  |\vec{k}_1+\vec{k}_2|+|\vec{k}_3|+|\vec{k}_4+\vec{k}_5|,\\ 
     &\mathcal{P}_{1,4,5} = E_4 +E_5 +E_{4,5} = |\vec{k}_4| +|\vec{k}_5| +|\vec{k}_4+\vec{k}_5|,\\
     &\mathcal{P}_{1,2,3,4} = E_1 +E_2+E_3 +E_{4,5} = |\vec{k}_1| +|\vec{k}_2|+|\vec{k}_3| +|\vec{k}_4+\vec{k}_5|,\\
    &\mathcal{P}_{1,2,3,4,5} = E_1 +E_2+E_3 +E_{4}+E_5 = |\vec{k}_1| +|\vec{k}_2|+|\vec{k}_3| +|\vec{k}_4|+|\vec{k}_5|,\\
\end{aligned}
\end{equation}
where $\mathcal{P}_{i,\dots,j}$ is the perimeter of subpolygon with vertices $\{i,\dots,j\}$. So we have that each blob entering the tubing is mapped to a subpolygon on the momentum $n$-gon and a complete tubing corresponds to a maximal collection of nested non-overlapping subpolygons, which we call a russian doll.

The full wavefunction will be given purely in russian doll terms, that is, it will be a sum of terms where each term is one collection of non-overlapping subpolygons (tubes). This is also known as the \textit{old-fashioned perturbation theory} (OFPT) representation \cite{CosmoPolytopes,DualCosmoPolyTriangs,CosmoReview} of the wavefunction. To understand this representation, let's consider the time integral representation of a given graph contributing to $\Psi$:
\begin{equation}
    \psi_{\mathcal{G}}=\int_{-\infty(1- i \epsilon)}^{0}\left[\prod_{v\in\mathcal{V}}d\eta_v\right]\left[\prod_{v\in\mathcal{V}}e^{i \l\sum_{i_v\in \mathcal{E}^{\prime}_v} E_{i_v}\r \eta_v}\right]\left[\prod_{e\in\mathcal{E}}G(E_{k_e}; \eta_e,\eta_{e^{\prime}})\right]\, ,
    \label{eq:feynman_int}
\end{equation}
where $\mathcal{V}$ is the set of vertices in the graph, $\mathcal{E}$ is the set of internal edges of the graph, $\mathcal{E}^{\prime}_v$ is the set of external states attached to the vertex $v$ (and $E_{i_v}$ are the respective moduli of the momenta), and $E_{k_e}$ is the momentum flowing in the edge $e$. We now consider the action of the operator:
$$
\Delta = -i \sum_{v\in \mathcal{V}} \partial_{\eta_v}\, ,
$$
on the integrand of \eqref{eq:feynman_int}. We can start by applying integration-by-parts. The total derivative vanishes, since the bulk-to-bulk propagators vanish at the upper boundary (see \eqref{eq:bulktobulk} when sending $\eta_1$ or $\eta_2$ to zero), and on the other hand the $i \epsilon$ prescription (Bunch-Davies condition) ensures the integrand vanishes in the lower boundary. Then we consider the action of $\Delta$ separately in the external propagators, and in the product of bulk-to-bulk propagators, $G(y_e; \eta_e,\eta_{e^{\prime}})$. It is clear that,
$$
\Delta \left[\prod_{v\in\mathcal{V}}e^{i \l\sum_{i_v\in \mathcal{E}^{\prime}_v} E_{i_v}\r \eta_v}\right] = \l\sum_{v\in\mathcal{V}}\sum_{i_v\in\mathcal{E}^{\prime}_v} E_{i_v}\r \left[\prod_{v\in\mathcal{V}}e^{i \l\sum_{i_v\in \mathcal{E}^{\prime}_v} E_{i_v}\r \eta_v}\right]\, ,
$$
where the quantity in parentheses is the total energy of $\psi_{\mathcal{G}}$. Then the action of $\Delta$ on the bulk-to-bulk propagators is essentially only the action on the boundary term, since when acting on the time-ordered terms, these vanish. This is true because $\Delta$ is the time-translation operator, and the time-ordered terms are time translation invariant. Practically, one can simply see that by acting with $\partial_{\eta_1}+\partial_{\eta_2}$ on the time ordered terms in \eqref{eq:bulktobulk}, that the derivatives of the exponentials in $\eta_1$ and $\eta_2$ will cancel each other. Therefore, we can say that:
$$
\Delta \left[\prod_{e\in\mathcal{E}}G(y_e; \eta_e,\eta_{e^{\prime}})\right]= - \l\sum_{\tilde{e}\in \mathcal{E}} e^{i E_{k_{\tilde{e}}} (\eta_{\tilde{e}}+\eta_{{\tilde{e}}^{\prime}})}\left[\prod_{e\in\mathcal{E}/\{\tilde{e}\}}G(E_{k_e}; \eta_e,\eta_{e^{\prime}})\right]\r \, .
$$
Putting everything together, the exponentials in the expression above will be just like bulk-to-boundary propagators. Then, we can write,
\begin{equation}
    E_t\, \psi_{\mathcal{G}} (E_1, ..., E_n) = \sum_{e\in\mathcal{E}_{\text{Tree}}} \psi_{\mathcal{G}_L}(\mathcal{E}_{L_e}; E_{k_e})\:\psi_{\mathcal{G}_R}(\mathcal{E}_{R_e}; E_{k_e}) + \sum_{e\in\mathcal{E}_{\text{Loop}}} \psi_{\tilde{\mathcal{G}}}(E_1, ..., E_n, E_{k_e},E_{k_e})\, .
    \label{eq:OFTP}
\end{equation}
Where in the first term we are summing over every edge that is not in a loop (the set $\mathcal{E}_{\text{Tree}}$), and $\mathcal{G}_L$ and $\mathcal{G}_R$ correspond to the subgraphs to the left and right of the edge $e$. $\mathcal{E}_{L_e}$ are the external states of the left subgraph, and similarly for the right subgraph. Additionally, both $\mathcal{G}_L$ and $\mathcal{G}_R$ also have an external state with the momentum of the edge $e$ in $\mathcal{G}$. In the second term, we are summing over the remaining edges, which are part of a loop (the set $\mathcal{E}_{\text{Loop}}$). $\tilde{\mathcal{G}}$ stands for the graph obtained by cutting the edge $e$ in the graph $\mathcal{G}$. It will have all the $n$ external states plus two more, both with momentum of the edge $e$. Equation \eqref{eq:OFTP}, when applied recursively, allows us to construct the OFPT representation of the wavefunction. In \cite{DualCosmoPolyTriangs}, the authors showed that there is one triangulation of the dual of the cosmological polytope which yields the OFPT representation.

From \eqref{eq:OFTP}, we have that in the sum on the right-hand side, each term will be a product of singularities corresponding to tubes (equivalently, subpolygons) that do not overlap, either are fully inside one another, or are disjoint. Applying this formula recursively, we can see that this property will hold in the expansion of the different terms. Therefore, it becomes clear that we can write the wavefunction as a sum of russian dolls.

\section{Lightning review of the ABHY associahedron}
\label{appendixAssoc}

The associahedron, $\text{Assoc}_n$, is a polytope that encodes the combinatorics of triangulations of $n$-gons. Concretely, $\text{Assoc}_n$ is an $(n-3)$-dimensional simple polytope whose faces are associated to partial/full triangulations of the $n-$gon, or what is the same, collections of non-overlapping chords of the $n$-gon. The codimension-1 faces are associated to partial triangulations with a single chord, codimension 2 faces with those with two chords, and so on until we reach the vertices, which are labelled by $(n-3)$ non-overlapping chords specifying a full triangulation.

If we denote a collection of non-overlapping chords by $\mathcal{C}$, then the associahedron is the polytope whose face structure reflects the combinatorics of compatible chords, which can be stated as the fundamental property that: 
\begin{center}
    $\mathcal{C}^\prime$ is a face of $\mathcal{C}$ if $\mathcal{C} \subset \mathcal{C}^\prime$. 
\end{center}

One simple way of constructing the associahedron combinatorially is via \textit{mutations}. This is if we start on a given vertex of the associahedron, corresponding to a full triangulation of the $n$-gon, we can generate the vertices that are connected to it by performing  mutations: given the collection of chords in a triangulation, each chord is then a diagonal of a square defined by the boundary edges and the remaining chords on the triangulation. A mutation flips one of the chords to the other diagonal of the square in which it is contained. Since a triangulation contains $n-3$ chords, starting at a given vertex we can mutate in $n-3$ different ways, which means that at any vertex of our polytope $(n-3)$ edges meet, which tells us the polytope is \textit{simple}. Following this procedure, we can generate the full polytope, and we further conclude that any two vertices are connected via an edge if and only if their triangulations are related by a mutation. 

For example, suppose we do this exercise for $n=4$. In that case, there are only two triangulations and the geometry is one-dimensional -- the Assoc$_4$ is a line interval with two vertices, one at each boundary of the interval, labeling the two possible triangulations. In this case it is trivial, but indeed we see a mutation relates the two triangulations. 

At $n=5$, we should find a two-dimensional geometry, which ends up being a pentagon as depicted in the left of figure \ref{fig:5_6GraphAssoc}. We see that each edge is associated with partial triangulations with a single chord, and the vertices are labeling all the possible $5$ triangulations of the pentagon.

Similarly, for $n=6$, the $\text{Assoc}_6$ is a three-dimensional polytope, with 9 codimension one facets -- one associated to each chord of the hexagon -- 21 codimension-2 facets -- associated to collections of two non-overlapping chords -- and finally 14 vertices, each labeling one triangulation of the hexagon (see figure \ref{fig:sixcosmo}, left).

One remarkable property of the associahedron is the factorization structure associated with its boundaries -- the boundaries of associahedra are given by products of lower-point associahedra. This feature stands as the geometric avatar of factorization of tree-level scalar amplitudes into products of lower-point amplitudes. For example, if we look at the 5-point associahedron (figure \ref{fig:pentadeca}, left) then we see that each boundary -- the edges -- are naturally associated with a partial triangle with a single chord which divides the pentagon into a square and a triangle. Indeed, the boundary is then given by the product of the lower-point associahedron associated to the smaller polygons appearing in the partial triangulation. In this case, the $3$-point associahedron is a point and the $4$-point is the line segment described above, so we get simply a line interval. 
Similarly, at $6$-points, we see that the polytope has $6$ pentagonal facets and $3$ square facets. The first six, correspond to partial triangulations including a single chord $(i,i+2)$ which divides the hexagon into a pentagon and a triangle, therefore we expect to get $\text{Assoc}_5 \times \text{Assoc}_3$, which is indeed what we have since these facets are pentagons. As for the square facets, these correspond to partial triangulations with a single chord of the type $(i,i+3)$ which divides the hexagon into two squares, and therefore we get that these facets are $\text{Assoc}_4 \times \text{Assoc}_4$, which is precisely a square. 

We stress that is not at all obvious a priori that the combinatorics of partial triangulations can be captured by a polytope. It's existence, and it's factorization properties on facets are most naturally understood from a particular realization in terms of a simple set of inequalities we will review in a moment. 

Now that we have understood how the combinatorial information associated to cubic tree graphs is organized in this polytope, and in particular how its boundary structure encodes the basic factorization features of amplitudes, let's see how we can connect these physical observables to this geometry. 

In a theory of colored scalars interacting via cubic interactions -- Tr$(\phi^3)$ theory -- we can write the amplitudes perturbatively over sums of cubic diagrams. Namely at leading order, once we fix an ordering for the external particles, say $e.g.$ the standard ordering $(1,2,\cdots,n-1,n)$, we get contributions from all the possible tree-level planar Feynman diagrams -- which are precisely dual to triangulations of the $n$-gon. In particular, if we associate to each edge of the $n$-gon a momentum of the particle in the scattering process, $p_1^\mu, p_2^\mu,\cdots, p_n^\mu$, then given a triangulation we have that the length$^2$ of the chords entering in the triangulation precisely give us the momentum square flowing through the propagators in the dual cubic graph. Let's denote the (length)$^2$ of a chord going from vertex $i$ to vertex $j$ by $X_{i,j}$ then we have:
\begin{equation}
    X_{i,j} = (p_i +p_{i+1} + \cdots+ p_{j-1})^2,
\end{equation}
where we have $X_{i,j} = X_{j,i}$ and $X_{i,i+1}=0$ since we are considering our particles to be massless and therefore we have $p_i^2=0$. Therefore, we can write Tr$(\phi^3)$ amplitudes at tree-level as a sum over all possible cubic Feynman diagrams -- all possible triangulations of the $n$-gon -- where for each diagram we have a factor of one over the product of the $X_{i,j}$ corresponding to the chords entering in the triangulation:
\begin{equation}
    \mathcal{A}_n(X_{i,j}) = \sum_{\text{triang. } \mathcal{T}} \prod_{X_{i,j} \in \mathcal{T}} \frac{1}{X_{i,j}}. 
\label{eq:AmpTree}
\end{equation}

This way of writing the amplitude makes manifest that it is a function exclusively of the $X_{i,j}$'s which are usually called the planar variables -- as they correspond to the invariants associated to momentum flowing through propagators of planar tree diagrams. Note, however, that the planar variables are \textit{not} all the possible Lorentz invariants dot product of momentum one can consider, for example we also have the dot products $p_i \cdot p_j$ with $i$ and $j$ non-adjacent. In particular, at $n$-points we have $n$ chosen $2$ dot products of momenta, but due to momentum conservation only $n(n-3)/2$ of these are actually independent. Quite nicely $n(n-3)/2$, is precisely the number of $X_{i,j}$ we have for an $n$-gon, and therefore we have that the planar variables form a \textit{basis of kinematic space}, where momentum conservation is automatically implemented by the fact that they lived in a \textit{closed} momentum polygon. 

This means that the non-planar variables -- corresponding to dot products of non-adjacent particles --- can be written in terms of the planar ones. Let us call the non-planar invariants by $c_{i,j} = -2 p_i \cdot p_j$ with $i,j$ not adjacent. Then we have:
\begin{equation}
    c_{i,j} = X_{i,j} + X_{i+1,j+1} -X_{i,j+1}-X_{i+1,j}.
\end{equation}

Now that we have defined the kinematic space the amplitude lives in as well as given a precise definition of the amplitude in this space \eqref{eq:AmpTree}, we can proceed to understand how to connect this object to the geometry of the associahedron.  The first step is to embed the associahedron in kinematic space — where the amplitude is defined — this is we want to define a set of inequalities in $X_{i,j}$ space that carve out this polytope. This embedding was introduced in \cite{ABHY} and we will summarize it here. As explained above, each facet of this geometry is associated with a partial triangulation with a single chord, and therefore is naturally associated with a given $X_{i,j}$. Therefore, to each facet we associate the inequality:
\begin{equation}
X_{i,j} \geq 0.
\label{eq:ABHY}
\end{equation}

So we have that all $X$’s are positive inside the polytope and vanish in the respective facets. However, as explained earlier the $\text{Assoc}_{n}$ is an $n-3$-dimensional object, and the current inequalities naively define a cone in an $n(n-3)/2$ dimensional space. So in order to bring it to the correct dimension we intersect this cone with the “ABHY” plane defined as follows: Pick a triangulation say $\{X_{1,3},X_{1,4},\cdots,X_{1,n-1}\}$ and consider the kinematic basis containing the $X$’s in the triangulation as well as the collection of non-planar variables $\mathcal{C} = \{c_{1,3},c_{1,4},\cdots,c_{1,n-1},c_{2,4},c_{2,5}, \cdots, c_{2,n-1}, \cdots c_{3,5},$ $ \cdots, c_{n-3,n-1}\}$ which contains exactly $n(n-3)/2-(n-3)$ $c$’s. Then since this forms a basis we can write all $X$’s in terms of the $X$’s in the chosen triangulation and the $c_{i,j}$’s in this collection. If we fix the non-planar variables in $\mathcal{C}$ to be \textit{positive} then \eqref{eq:ABHY} defines an $n-3$ dimensional geometry in the space spanned by $X_{1,3},X_{1,4},\cdots,X_{1,n-1}$ which is precisely the associahedron. 

Given this embedding, the amplitude is given by the canonical form of this polytope. There are various motivations for the ABHY inequalities, from a ``causal diamond" picture \cite{CausalDiamonds} in kinematic space to recording the data of curves on surfaces \cite{CurveInt}. It is striking that none of these refer to summing over all diagrams. The connection with the usual Feynman diagrams arises from a particular way of computing the canonical form of the associahedron. Indeed for any simple polytope, there is a natural triangulation, taking the inverse product of the facet inequalities meeting at each vertex and summing over all vertices, corresponding to an especially obvious triangulation of the {\it dual} polytope. Since each vertex of the associahedron corresponds to a complete triangulation, this expression for the canonical form of a simple polytope turns into the Feynman diagram expansion.

\section{Graph associahedra from cosmohedra}
\label{app:CosmoPolytopes_GraphAssoc}

The early connection between positive geometry and the cosmological wavefunction for a single graph was through cosmological polytopes~\cite{CosmoPolytopes}. In the context of cosmohedra, we have instead seen that the geometry of single graphs are instead given by graph associahedra. We have already given a combinatorial description of these graph associahedra as part of the build-up to motivating the cosmohedron itself. In this appendix, we will describe how to cut out the graph associahedra by inequalities, simply by specializing the cosmohedron to its graph-associahedron facets. Amongst other things, this will allow us to describe the simple relationship between graph associahedra and cosmological polytopes.

Note that unlike for the full cosmohedra -- where the relationship between the wavefunction and the canonical form of the geometry is the interesting ``non-linear" one described in the text -- for single graphs the story is entirely straightforward. We can associate a dual graph $G$ with a Feynman diagram/triangulation $T$ in the usual way. We associate energy variables $x_v$ with the vertices and $y_e$ with the edges; as we will see, we can think of the graph associahedron as living in $y$ space. After factoring out the total energy $E_t$, and individual triangle perimeters $t_v$, the wavefunction for $G$ is simply given by the canonical form of the graph associahedron, $\mathcal{A}_G$: 
\begin{equation}
\Psi_G = \frac{1}{E_t} \prod_v \frac{1}{t_v} \Omega_{\mathcal{A}_G} (y;t_v,E_t),
\end{equation}
whereas we will see, $E_t, t_v$ appear as parameters cutting out the graph associahedron in $y$ space. 

The graph associahedra (and certain degenerations we will describe when writing all perimeters in terms of standard $(x_v,y_e)$ energy variables)  also give a novel understanding of the emergence of the amplitude on the scattering facet at tree-level, and also explain some remarkable features of wavefunction residues that had resisted a transparent understanding to date. 

\subsection{Inequalities for the graph associahedron}
\label{sec:EmbedGraphAssoc}

The graph associahedra are particular facets of cosmohedra, corresponding to complete triangulations $T$ of the $n$-gon. For a given $T$, we can associate a dual graph $G$ in the usual way. This facet of the cosmohedron is associated with the inequality $\sum X_I \geq \sum \delta_{T_i}$, where $X_I$ are all the chords in the triangulation and $\delta_{T_i}$ are associated with each triangle of the triangulation (as given in \eqref{eq:ineqs} and \eqref{eq:eps_to_polys}). By going on this facet, we are saturating this to the equality 
\begin{equation}
\sum_I X_I = \sum_i \delta_{T_i}.
\label{eq:GraphAssocFacet}
\end{equation}
In terms of the dual graph $G$, we can think of $\delta_{T_i}$ as associated with a small circle surrounding the $i$-th vertex of $G$. 

Obviously, the other facets of the cosmohedron that meet the one associated with $T$ must correspond to partial triangulations that are coarsenings of $T$. These will become facets of the graph associahedron for $T$, so the inequalities cutting out the graph associahedron are all of the form $\sum_J X_J \geq \sum \delta_{p}$, with $J$ depending on the partial triangulation we're considering. 
We can denote these inequalities easily in the language of the tubes. The partial triangulation gives a collection of non-overlapping sub-polygons $p$, which can be denoted on $G$ by a collection of non-overlapping tubes we will also label by $p$. 

Then, $\sum_J X_J$ is the sum over all the edges of $G$ that are cut by the tubes. Clearly, the smallest tubes, which encircle a single vertex, corresponding to triangle sub-polygons, are special. We can label our partial triangulation by specifying a collection of larger (not triangle) tubes, $P$, and having done this, understanding that the vertices not encircled by tubes, are encircled with small ones. The inequalities are then 
\begin{equation}
\sum_{e \text{ not in } P} X_e \geq \sum \delta_P + \sum \delta_{t}\, ,
\end{equation}
where the sum is over the edges $e$ that are not contained in the interior of any big tube in $P$, and $t$ are the tubes encircling the single vertices not encircled by the set of big tubes $P$. 

But it is now trivial to see that the inequalities associated with more than one of these larger tube are all redundant, following from those for single tubes. Consider the simple example of the triangulation $\{(1,3),(1,4),(1,5)\}$ for $n=6$. We are working on the support of 
\begin{equation}
X_{1,3} + X_{1,4} + X_{1,5} = \delta_{1,2,3} + \delta_{1,3,4} + \delta_{1,4,5}.
    \label{eq:graphFacet}
\end{equation} 
The partial triangulations $\{(1,3),(1,4)\}$ is associated with a single tube -- corresponding to square $(1,4,5,6)$--, as is that for $(14,15)$, and the inequalities are 
\begin{equation}
X_{1,3} + X_{1,4} \geq \delta_{1,2,3} + \delta_{1,3,4} + \delta_{1,4,5,6}, \quad  X_{1,4} + X_{1,5} \geq \delta_{1,2,3,4} + \delta_{1,4,5} + \delta_{1,5,6}.
\end{equation}
But adding these inequalities and using \eqref{eq:graphFacet} we have that 
\begin{equation}
\begin{aligned}
&X_{1,3} + X_{1,4} + X_{1,4} + X_{1,5} \geq \delta_{1,2,3} + \delta_{1,3,4} + \delta_{1,4,5,6} + \delta_{1,2,3,4} + \delta_{1,4,5} + \delta_{1,5,6} \\
&\Rightarrow X_{1,4} \geq \delta_{1,2,3,4} + \delta_{1,4,5,6},
\end{aligned}
\end{equation}
which is the two-tube inequality associated with the partial triangulation $\{(1,4)\}$. This obviously extends to any number of tubes: on the support of $\sum_J X_J = \sum_i \delta_{T_i}$, the sum of the inequalities for single tubes implies the inequality for multi-tubes, and so these are redundant. 

We can naturally define the variables for the graph without referring to the underlying triangulation. Thus, we associate variables $X_e$ with the edges of $G$, and constants $\delta_P$ with single tubes $P$. The graph associahedron is then cut out by the inequalities: 
\begin{equation}
\sum_{e \text{ not in } P} X_e \geq \delta_P + \sum_{v \, {\rm not \, in \,} P} \delta_v,
\label{eq:originalGA}
\end{equation}
where $\delta_v$ is associated to the small encircling of vertex $v$. The $\delta_P$ satisfy our ubiquitous inequalities 
\begin{equation}
\delta_P + \delta_{P^\prime} < \delta_{P \cup P^\prime} + \delta_{P \cap P^\prime}
\end{equation}
where, as for the full cosmohedron, $\delta_{{\rm all}} = 0$. 

It is amusing that the graph associahedron for linear chains are simply associahedra; in this case, all the myriad properties of Tr($\phi^3$) amplitudes following from the connection with the associahedron are inherited by the wavefunction for these single graphs. Indeed, the Minkowski sum decomposition and corresponding stringy integrals exist for all graph associahedra, so cousins of full {\it stringy} Tr($\phi^3$) amplitudes exist, associated with single graphs for the wavefunction! This gives an interesting entry-point into possible stringy formulations for cosmological wavefunctions we leave to future work. 

\subsection{Relation to cosmological polytopes}

We now discuss the connection between graph associahedra and the cosmological polytope. As a natural entry into the discussion, we note again that the natural variables we are using to describe the wavefunction are {\it perimeters}, thought of as independent variables. This is an extension of the usual kinematic variables, in the general spirit of the extensions of kinematics for amplitudes given by curves on surfaces~\cite{SurfaceKin}. For a single graph, this is equivalent to working with independent variables $p$ associated to every tube. But to compare with the standard wavefunction and with cosmological polytopes, we work with the standard energy variables. As it is well-known for a single graph, these are conventionally denoted by variables $x_v$ for the sum of the energies entering each vertex and $y_e$ for the energy of each internal edge of $G$. Of course, the $x,y$ variables determine the $p$ associated with every tube, via 
\begin{equation}
{\cal P}_p=\sum_{v \, {\rm in }\, p} x_v + \sum_{e \, {\rm entering} \, p} y_e,
\end{equation} 
the sum over the energies of the vertices contained inside $p$ together with the external edges entering $p$ -- the familiar energy pole associated with the tube $p$. Note that this enforces certain equalities between the ${\cal P}_p$'s: 
\begin{equation}
{\cal P}_p + {\cal P}_{p^\prime} = {\cal P}_{p \cup p^\prime} + {\cal P}_{p \cap p^\prime}.
\end{equation}

In fact, it is easy to see that we can work backwards -- imposing these natural equalities on the perimeters ${\cal P}_p$ implies that they can be expressed in terms of $x_v,y_e$ variables associated with the graph. 

Let us now discuss the cosmological polytope for a graph with $V$ vertices and $E$ edges.  It is usually described as a projective polytope in $E + V - 1$ dimensions; of course this is equivalently thought of as a cone over this polytope in $E+V$ dimensions. This cone is cut out by the simple inequalities, for every tube $p$
\begin{equation}
{\cal P}_p \geq 0, \quad  {\rm or} \quad \sum_{v \,  {\rm in} \,  p} x_v + \sum_{e \, {\rm entering} \, p } y_e \geq 0.
\label{eq:Ineqs_GraphAssoc}
\end{equation}

Now, in the story of the graph associahedron, we are factoring out the total-energy singularity, as well as those associated with the small tubes encircling each vertex. Thus, it is natural to expect that the relationship with the graph associahedron and the cosmological polytope is revealed when we slice the cosmological polytope on the plane
\begin{equation}
\sum_v x_v = E_t, \quad x_v + \sum_{e \, {\rm connected\, to } \, v} y_e= t_v,
\label{eq:SlicedCosm}
\end{equation}
where we hold $E_t, t_v > 0$ as constants. 

Indeed, as we now see, this slice of the cosmological polytope is very closely related to the graph associahedron. For instance, for the simplest cases of the 2-dimensional graph associahedra, corresponding to the 4-site chain and star graphs (shown in figure \ref{fig:5_6GraphAssoc}, right), this slice of the cosmological polytope gives us precisely the familiar pentagon and hexagon. But more generally, it is obvious that this sliced cosmological polytope cannot be precisely the same as the graph associahedron -- the graph associahedron knows about the general perimeters, not about the specialization associated with working with $x$'s and $y$'s. 

There is a beautiful resolution of this discrepancy. The sliced cosmological polytope is obtained by a {\it degeneration} of the graph associahedron, when the $\delta_P$ occurring in the inequalities cutting out the graph associahedron saturate most of the inequalities they satisfy. Indeed, we have that
\begin{equation}
\begin{cases}
    \delta_P + \delta_{P^\prime} < \delta_{P \cup P^\prime} + \delta_{P \cap P^\prime},
\quad  {\rm if \,} P \cap P^\prime \, {\rm is \, a \, single \, vertex \, or} \, P \cup P^\prime \, {\rm is \, everything} \, , \\
\delta_P + \delta_{P^\prime} = \delta_{P \cup P^\prime} + \delta_{P \cap P^\prime},
\quad  {\rm otherwise}\, .
\end{cases}
\label{eq:ineqsDegGA}
\end{equation}
Imposing these equalities has the effect of shrinking some of the faces of the graph associahedra -- the graph associahedra are all simple polytopes, but the sliced cosmological polytopes are not. For instance, for the case of the 5 site chain where the graph associahedron is the usual three-dimensional associahedron (see figure \ref{fig:7GraphAssoc}, right), while the sliced cosmological polytope still has 9 faces, but only has 12 vertices and 19 edges instead of the usual 14 vertices and 21 edges; two of the edges of the usual associahedron are then contracted to a point in the sliced cosmological polytope. 

It is straightforward to establish the connection between the sliced cosmological polytope and this degeneration of the graph associahedron. We simply take the graph, and solve for the $x_v$ variables by setting all the perimeters associated with the small tubes encircling the vertices to $t_v$. We are left only with $y_e$ variables. These satisfy a single equality (from the $E_t$ equation in \eqref{eq:SlicedCosm}), and the rest of the inequalities coming from \eqref{eq:Ineqs_GraphAssoc}. 
Then the inequality for any large tube $P$ becomes 
\begin{equation} 
\sum_{e\, {\rm not \, in}\, P} (2 y_e) \geq \sum_{v \, {\rm not \, in}\, P} t_v - E_t.
\label{eq:yE_Ineqs}
\end{equation}
If we identify $2y_e \equiv X_e$, these are just the inequalities for cutting out the graph associahedron \eqref{eq:originalGA}, but with a special choice for the RHS of the inequality. Comparing with the RHS of the $P$ inequality for the graph associahedron $\delta_P + \sum_{v \, {\rm not \, in}\, P} \delta_{v}$ we have that 
\begin{equation}
\delta_P = \sum_{v \,{\rm not \, in \,} P} (t_v - \delta_v) - E_t.
\label{eq:choiceofDP}
\end{equation}
Further matching $\sum_e 2 y_e = \sum_v t_v - E_t$ (from \eqref{eq:SlicedCosm}) with $\sum_e X_e = \sum_v \delta_v$ (from the facet defined by the graph associahedron in the cosmohedron \eqref{eq:GraphAssocFacet}) lets us identify 
\begin{equation}
\sum_v (t_v - \delta_v) = E_t.
\end{equation}
It is then very easy to see that the choice for $\delta_P$ in \eqref{eq:choiceofDP} satisfies the inequalities and equalities given for the degenerated graph associahedron we defined above \eqref{eq:ineqsDegGA}. 

\subsection{$E_t$ and all-vertex singularities}

It is a beautiful fact that the residue on the $E_t \to 0$ pole of the wavefunction gives the scattering amplitude. It is important to emphasize that this fact crucially depends on using the usual kinematics for the wavefunction, and does not arise when treating the perimeters as independent variables. Indeed, in the russian doll picture, {\it every} term has an $E_t$ pole! So there is no special simplification to the wavefunction when $E_t \to 0$; for instance, at large $n$ there are still the same number of factorially many terms as there are for the full wavefunction. And yet, when we express all the perimeters in terms of $(x_v,y_e)$'s, there is a vast simplification -- the amplitude appearing as the residue on the pole is a single term! 

This is clearly an important phenomenon, and it is interesting to understand it thoroughly. Of course the time-integral representation for the wavefunction makes this fact rather obvious, but this representation has many other defects, amongst other things being riddled with spurious $1/y_e$ poles that only cancel in the full sum.  The russian doll picture only has physical poles term-by-term, but as we have just said, does not make the appearance of the amplitude on the $E_t \to 0$ pole manifest. 

The cosmological polytope gives us a very satisfying understanding of what happens as $E_t \to 0$. Starting from the defining representation as a convex hull of points, it is easy to see that (at tree-level) there are very few vertices lying on the scattering facet where $E_t \to 0$, and that this facet is in fact a simplex; since the canonical form of a simplex is trivially given by the product of its facet inequalities this makes the emergence of the amplitude obvious. 

Instead, $E_t$ does not appear as a facet of the graph associahedron, nor of its degenerated cousin enforcing working with $(x_v,y_e)$ variables; $E_t$ has been ``factored out" and only affects the polytope through its role in defining the inequalities. It is therefore interesting to understand how the scattering amplitude arises in this language. 

There is yet another striking feature of the wavefunction for single graphs, which has been observed since the earliest literature on the subject, but which has resisted a simple understanding. Namely, if we take the residue of the wavefunction in all the vertex variables, the result is extremely simple, given by the product $\prod_e \frac{1}{2 y_e}$. The residue for individual terms in the russian doll sum are in general much more complicated, but the full sum simplifies dramatically. 

The degenerated graph associahedra beautifully explain both the appearance of the amplitude as $E_t \to 0$ as well as the simplicity of the total vertex residue, in a uniform way. It is easy to see that in both cases, the relevant limit ends up greatly simplifying the degenerated graph associahedron. When the vertex residues are taken, the degenerated graph associahedron turns into a simplex. Instead as $E_t \to 0$, the graph associahedron degenerates to a product of simplices, such that the full canonical form is again a single term. In both cases, when the canonical form is multiplied by the appropriate prefactors factored out in $\Psi_G$, we get precisely the correct result.

\paragraph{Total Vertex Residue} Let's first see what happens on the total-vertex residue. This is taking the residue on all $t_v \to 0$ in $\Psi_G$, that leaves us with the canonical form for the degenerated graph associahedron when all the $t_v \to 0$. Using that $\sum_e (2 y_e) = \sum t_v - E_t$ (from \eqref{eq:SlicedCosm}), we can equivalently write the inequalities for any tube $P$ \eqref{eq:yE_Ineqs} as 
\begin{equation}
\sum_{e \, {\rm in} \, P} (2 y_e) \leq \sum_{v \, {\rm in} \, P} t_v \quad \xrightarrow[]{t_v \to 0} \sum_{e \, {\rm in} \, P} (2 y_e) \leq 0,
\end{equation}
which greatly simplify when setting all $t_v \to 0$. Now we always have tubes $P$ enclosing a single edge $e$, and for these the inequality is simply $(2 y_e) \leq 0$. But these then imply all the other inequalities! Thus, as the $t_v \to 0$, the degenerated graph associahedron, $\tilde{\mathcal{A}}_G$, turns into the simplex cut out by $(2 y_e) \leq 0$, with $\sum (2 y_e) = -E_t$. The canonical form of this simplex is precisely:
\begin{equation}
    \Omega_{\tilde{\mathcal{A}}_G} = E_t \times \prod_e \frac{1}{2 y_e}  \quad \Rightarrow \quad \Psi_G = \frac{1}{E_t} \times  \Omega_{\tilde{\mathcal{A}}_G} = \prod_e \frac{1}{2 y_e},
\end{equation} 
so when we multiply by the prefactor in $\Psi_G$ we get precisely the simple residue we are looking to explain. 

\paragraph{$E_t$ residue} The emergence of the amplitude on the $E_t \to 0$ facet is more interesting. The first observation is simple: when we set $E_t \to 0$, only the inequalities for a small subset of tubes are relevant, as they imply all the remaining. Given any tree-graph, for a given edge $e$, there are two special tubes, the ``left" and ``right" tubes, $L_e,R_e$, which cross the edge and encircle all the vertices to one or other part of the full tree graph. Note that some of these $L,R$, which correspond to circling just one vertex at the very end of the graph, are part of the circlings that have been factored out, and are not facets of the polytope -- corresponding to triangles in the subpolygon picture. These are obviously important for the emergence of the amplitude -- on the support of $E_t =0$, the Lorentz-invariant propagator is simply the product $L_e R_e$. It is very easy to see that in general, any tube $P$ is given as a positive sum over $L/R$ tubes, minus some multiple of $E_t$. Thus, when $E_t = 0$, any other tube is a positive sum of $(L,R)$'s, and hence all the other inequalities (\eqref{eq:Ineqs_GraphAssoc}, for $P$ not a $L,R$ type tube) are redundant. 

But there are still many $(L,R)$'s, and it is not trivially obvious that the degenerated graph associahedron has gotten much ``simpler" as $E_t \to 0$. But it has! Indeed, we will now see that the canonical form of the degenerated graph associahedron as $E_t \to 0$, $\mathcal{A}^{E_t}_G$, is 
\begin{equation}
\Omega_{\mathcal{A}^{E_t}_G} = \frac{\prod_{\text{internal} \, v} t_v}{\prod_{\text{internal}} L \prod_{\text{internal}} R},
\end{equation}
where internal $v$ stands for every vertex in the graph that is connected to more than one internal edge, and internal $L/R$ stand for the $L,R$ tubings that enclose more than a single vertex. The numerator of this form kills the factors of $1/t_{\text{int}}$ factored out in front on $\Psi$, for all the $t_v$ associated to internal vertices, while the denominator combines with the $1/t_{\text{ext}}$ factors, so that for each internal edge we get a factor $L_e R_e$, which precisely turns in to the Lorentz-invariant dot product of the 4-momentum flowing through that edge, to giving us the amplitude.

It remains to establish this simple expression for the canonical form. The canonical form for any polytope is given by the product of all the poles corresponding to the polytope facets, with a numerator factor $N$, which enforces the fact that the form only has unit residues on the vertices of the polytope. For a simplex, this numerator factor is just a constant, depending on the constants occurring in the inequalities cutting out the facets of the polytope. For a general complicated polytope, the numerator is a complicated function. But for products of simplices, the numerator is similarly just a constant. We will now see that the residue of the form on all possible vertices of the degenerated associahedron is so simple that we can determine the residue to be precisely the one shown above, showing both that the polytope has turned into a product of simplices and giving us the amplitude as desired. 

\begin{figure}[t]
    \centering
    \includegraphics[width=\linewidth]{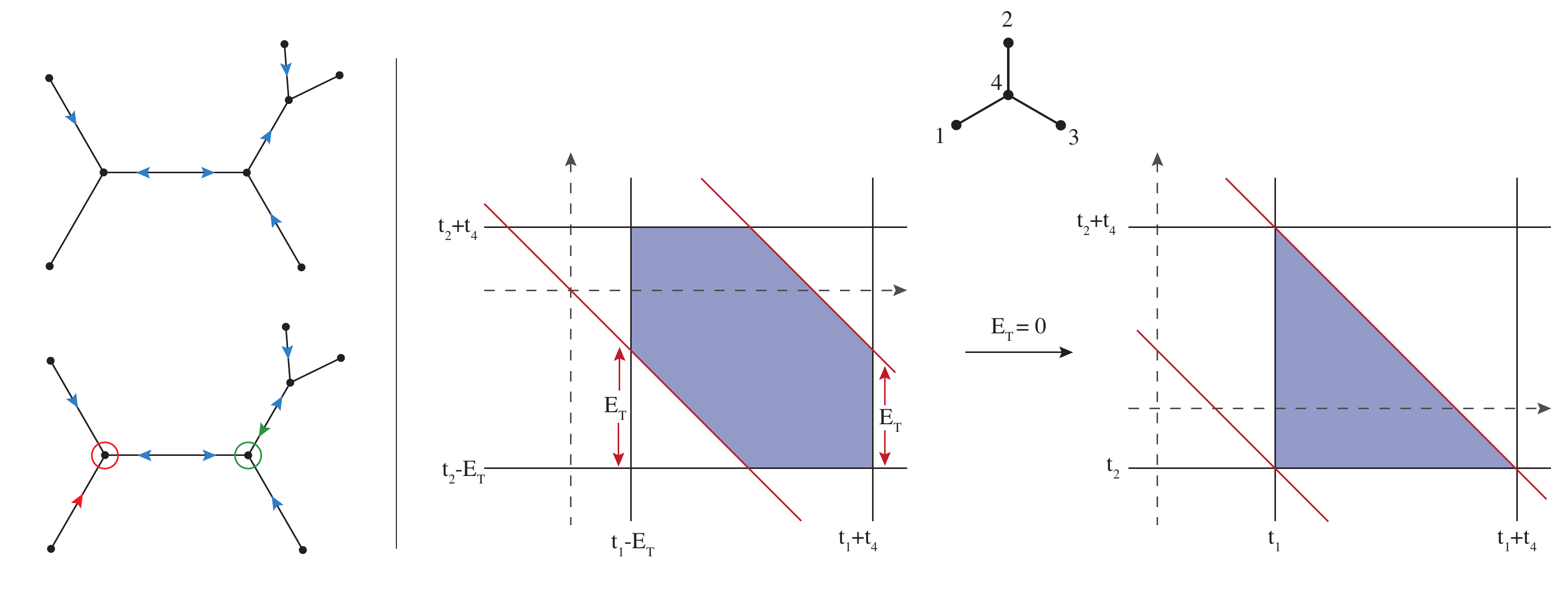}
    \caption{Left top: an example of an ``arrowing" of a tree-graph with $(E-1)$ arrows. The graph has seven edges, and we have placed six arrows on the edges, with no arrow pointing towards a boundary vertex. Left bottom: if we place any seventh arrow on the graph, there is a unique vertex with all incoming arrows. Examples are highlighted for adding the red arrow with the circled red with all incoming arrows, and the same with green arrow and circled vertex. Right: the (degenerated) graph associahedron for the Mercedes-Benz graph is a hexagon. But when $E_t \to 0$, it further degenerates into a triangle; this is the mechanism for the emergence of the amplitude on the $E_t \to 0$ pole.}
    \label{fig:EtGraphAssoc}
\end{figure}

We will begin by representing any $L_e,R_e$ tube by placing an arrow on the edge $e$, pointing left or right, respectively (see figure \ref{fig:EtGraphAssoc}, top left). Note for the case of an edge touching an outside vertex of the graph, we don't include the arrow pointing towards the vertex, since these are the small tubes that have been factored out. 

We would like to compute the residues when we set $(E-1)$ of the $L/R$ tubes to zero. It is useful to represent this choice of $(E-1)$ tubes by putting $(E-1) $ arrows on the graph.  It is then easy to see that any ``arrowing" of the graph that avoids having vertices with all incoming arrows gives a non-zero residue. For the cases where some vertices have all incoming arrows, it is impossible to find a solution for the $y$'s, and so the residue vanishes. For the ``legal" configurations, the $(E-1)$ equalities fully localize the $y$'s, and it is easy to read off the value of $y_e$ for any edge $e$ touching vertices $v,v^\prime$. If $e$ does not have any arrows, then $y_e = t_v + t_v^\prime$. If there is an arrow pointing from $v$ to $v^\prime$, then $2 y_e = t_v$. Finally, if there are two arrows pointing in opposite directions on $e$, then $y_e = 0$. This allows us to compute the value of any {\it other} $L/R$ tube (that is not among the ones we took residues on to begin with) in a simple way. Take this $L/R$ tube, and represent it by drawing one more $E$'th arrow on the graph, along the edge associated to the $L/R$ tube. Then the vertex the arrow points to, call it $v_*$,  will now have all incoming arrows (see figure \ref{fig:EtGraphAssoc}, left bottom, where we have added a red/green extra arrow). The value of this $L/R$ tube (associated to this $E$ th arrow), on the support of the residue solution defined by the first $(E-1)$ tubes,  is simply given by $t_{v^*}$! 

In this way we see that remarkably, 
\begin{equation}
\left(\prod_{\substack{\text{all $L$ not in the}\\\text{  $(E-1)$ residue}}} L \right) \left(\prod_{\substack{\text{all $R$ not in the}\\\text{  $(E-1)$ residue}}}  R  \right)= \prod_{{\rm all \, internal}} t_{{\rm internal}},
\end{equation}
and this is completely independent of which $(E-1)$ set we pick. Thus, simply choosing this product for the numerator correctly normalizes the canonical form, leading to the correct final result to get the amplitude as the residue when $E_t \to 0$. 

In figure \ref{fig:EtGraphAssoc} (right), we show what happens when we send $E_t\to 0$ for the case of the graph associahedron of the star graph at $6$-points in the $(2y_1),(2y_2)$ plane -- the hexagon becomes a triangle. The canonical form for this triangle is 
\begin{equation} 
\Omega_{\text{triangle}}^{E_t}= \frac{t_4}{R_1 R_2 R_3} \quad \Rightarrow \quad \Psi_{\text{triangle}}^{E_t} = \frac{1}{L_1 L_2 L_3 t_4} \times \Omega_{\text{triangle}}^{E_t} = \prod_{i=1}^3 \frac{1}{L_i R_i}
\end{equation} 
where $R_1 = x_2+x_3+y_{1,4}$, $R_2 = x_1+x_3+y_{2,4}$, $R_3 = x_1+x_2+y_{3,4}$ and $L_1 = x_1+y_{1,4}$, $L_2 = x_2+y_{2,4}$, $R_3 = x_3+y_{3,4}$ (where, since $E_t=0$, we have $x_1+x_2+x_3=0$).
The factor of $t_4$ in the numerator is there for unit leading singularities. This $t_4$ in the numerator cancels the $1/t_4$ factored in front for the wavefunction, leaving us with the amplitude.

\bibliographystyle{unsrt}
\bibliography{references}

\begin{thebibliography}{10}

\bibitem{Amplitudhedron}
Nima Arkani-Hamed and Jaroslav Trnka.
\newblock {The Amplituhedron}.
\newblock {\em JHEP}, 10:030, 2014.

\bibitem{GrassGeoScatAmp}
Nima Arkani-Hamed, Jacob~L. Bourjaily, Freddy Cachazo, Alexander~B. Goncharov,
  Alexander Postnikov, and Jaroslav Trnka.
\newblock {\em {Grassmannian Geometry of Scattering Amplitudes}}.
\newblock Cambridge University Press, 4 2016.

\bibitem{SMatDuality}
Nima Arkani-Hamed, Freddy Cachazo, Clifford Cheung, and Jared Kaplan.
\newblock {A Duality For The S Matrix}.
\newblock {\em JHEP}, 03:020, 2010.

\bibitem{WittenTwistorString}
Edward Witten.
\newblock {Perturbative gauge theory as a string theory in twistor space}.
\newblock {\em Commun. Math. Phys.}, 252:189--258, 2004.

\bibitem{ABHY}
Nima Arkani-Hamed, Yuntao Bai, Song He, and Gongwang Yan.
\newblock {Scattering Forms and the Positive Geometry of Kinematics, Color and
  the Worldsheet}.
\newblock {\em JHEP}, 05:096, 2018.

\bibitem{CurveInt}
N.~Arkani-Hamed, H.~Frost, G.~Salvatori, P-G. Plamondon, and H.~Thomas.
\newblock {All Loop Scattering As A Counting Problem}.
\newblock 9 2023.

\bibitem{CurveInt2}
N.~Arkani-Hamed, H.~Frost, G.~Salvatori, P-G. Plamondon, and H.~Thomas.
\newblock {All Loop Scattering For All Multiplicity}.
\newblock 11 2023.

\bibitem{TropLag}
Nima Arkani-Hamed, Carolina Figueiredo, Hadleigh Frost, and Giulio Salvatori.
\newblock {Tropical Amplitudes For Colored Lagrangians}.
\newblock 2 2024.

\bibitem{Zeros}
Nima Arkani-Hamed, Qu~Cao, Jin Dong, Carolina Figueiredo, and Song He.
\newblock {Hidden zeros for particle/string amplitudes and the unity of colored
  scalars, pions and gluons}.
\newblock 12 2023.

\bibitem{Gluons}
Nima Arkani-Hamed, Qu~Cao, Jin Dong, Carolina Figueiredo, and Song He.
\newblock {Scalar-Scaffolded Gluons and the Combinatorial Origins of Yang-Mills
  Theory}.
\newblock 12 2023.

\bibitem{NLSM}
Nima Arkani-Hamed, Qu~Cao, Jin Dong, Carolina Figueiredo, and Song He.
\newblock {NLSM $\subset$ Tr$(\phi^3)$}.
\newblock 1 2024.

\bibitem{CHY}
Freddy Cachazo, Song He, and Ellis~Ye Yuan.
\newblock {Scattering of Massless Particles in Arbitrary Dimensions}.
\newblock {\em Phys. Rev. Lett.}, 113(17):171601, 2014.

\bibitem{CosmoCollider}
Nima Arkani-Hamed and Juan Maldacena.
\newblock {Cosmological Collider Physics}.
\newblock 3 2015.

\bibitem{Maldacena:2011nz}
Juan~M. Maldacena and Guilherme~L. Pimentel.
\newblock {On graviton non-Gaussianities during inflation}.
\newblock {\em JHEP}, 09:045, 2011.

\bibitem{Raju:2012zr}
Suvrat Raju.
\newblock {New Recursion Relations and a Flat Space Limit for AdS/CFT
  Correlators}.
\newblock {\em Phys. Rev. D}, 85:126009, 2012.

\bibitem{Benincasa:2018ssx}
Paolo Benincasa.
\newblock {From the flat-space S-matrix to the Wavefunction of the Universe}.
\newblock 11 2018.

\bibitem{Arkani-Hamed:2018bjr}
Nima Arkani-Hamed and Paolo Benincasa.
\newblock {On the Emergence of Lorentz Invariance and Unitarity from the
  Scattering Facet of Cosmological Polytopes}.
\newblock 11 2018.

\bibitem{CosmoPolytopes}
Nima Arkani-Hamed, Paolo Benincasa, and Alexander Postnikov.
\newblock {Cosmological Polytopes and the Wavefunction of the Universe}.
\newblock 9 2017.

\bibitem{Anninos:2014lwa}
Dionysios Anninos, Tarek Anous, Daniel~Z. Freedman, and George Konstantinidis.
\newblock {Late-time Structure of the Bunch-Davies De Sitter Wavefunction}.
\newblock {\em JCAP}, 11:048, 2015.

\bibitem{CosmoBoot}
Nima Arkani-Hamed, Daniel Baumann, Hayden Lee, and Guilherme~L. Pimentel.
\newblock {The Cosmological Bootstrap: Inflationary Correlators from Symmetries
  and Singularities}.
\newblock {\em JHEP}, 04:105, 2020.

\bibitem{DiffEq_CosmoCorr}
Nima Arkani-Hamed, Daniel Baumann, Aaron Hillman, Austin Joyce, Hayden Lee, and
  Guilherme~L. Pimentel.
\newblock {Differential Equations for Cosmological Correlators}.
\newblock 12 2023.

\bibitem{KinFlow}
Nima Arkani-Hamed, Daniel Baumann, Aaron Hillman, Austin Joyce, Hayden Lee, and
  Guilherme~L. Pimentel.
\newblock {Kinematic Flow and the Emergence of Time}.
\newblock 12 2023.

\bibitem{Hillman:2019wgh}
Aaron Hillman.
\newblock {Symbol Recursion for the dS Wave Function}.
\newblock 12 2019.

\bibitem{GeometryCosmoCorr}
Paolo Benincasa and Gabriele Dian.
\newblock {The Geometry of Cosmological Correlators}.
\newblock 1 2024.

\bibitem{Benincasa:2024lxe}
Paolo Benincasa and Francisco Vaz\~ao.
\newblock {The Asymptotic Structure of Cosmological Integrals}.
\newblock 2 2024.

\bibitem{Goodhew:2020hob}
Harry Goodhew, Sadra Jazayeri, and Enrico Pajer.
\newblock {The Cosmological Optical Theorem}.
\newblock {\em JCAP}, 04:021, 2021.

\bibitem{CosmoReview}
Paolo Benincasa.
\newblock {Amplitudes meet Cosmology: A (Scalar) Primer}.
\newblock 3 2022.

\bibitem{DualCosmoPolyTriangs}
Paolo Benincasa and William J.~Torres Bobadilla.
\newblock {Physical representations for scattering amplitudes and the
  wavefunction of the universe}.
\newblock {\em SciPost Phys.}, 12(6):192, 2022.

\bibitem{KinFlowLoop}
Daniel Baumann, Harry Goodhew, and Hayden Lee.
\newblock {Kinematic Flow for Cosmological Loop Integrands}.
\newblock 10 2024.

\bibitem{IRSub}
Paolo Benincasa and Francisco Vaz\~ao.
\newblock {Cosmological Infrared Subtractions \& Infrared-Safe Computables}.
\newblock 5 2024.

\bibitem{MelvillePajer}
Scott Melville and Enrico Pajer.
\newblock {Cosmological Cutting Rules}.
\newblock {\em JHEP}, 05:249, 2021.

\bibitem{PajerUnitarityLocality}
Sadra Jazayeri, Enrico Pajer, and David Stefanyszyn.
\newblock {From locality and unitarity to cosmological correlators}.
\newblock {\em JHEP}, 10:065, 2021.

\bibitem{GoodhewCutCosmoCorr}
Harry Goodhew, Sadra Jazayeri, Mang Hei~Gordon Lee, and Enrico Pajer.
\newblock {Cutting cosmological correlators}.
\newblock {\em JCAP}, 08:003, 2021.

\bibitem{LoopInts}
Paolo Benincasa, Giacomo Brunello, Manoj~K. Mandal, Pierpaolo Mastrolia, and
  Francisco Vaz\~ao.
\newblock {On one-loop corrections to the Bunch-Davies wavefunction of the
  universe}.
\newblock 8 2024.

\bibitem{PokrakaDEs}
Shounak De and Andrzej Pokraka.
\newblock {A physical basis for cosmological correlators from cuts}.
\newblock 11 2024.

\bibitem{CosmoSMatrix}
Scott Melville and Guilherme~L. Pimentel.
\newblock {A de Sitter S-matrix from amputated cosmological correlators}.
\newblock {\em JHEP}, 08:211, 2024.

\bibitem{SimpCosmoCorr}
Chandramouli Chowdhury, Arthur Lipstein, Jiajie Mei, Ivo Sachs, and Pierre
  Vanhove.
\newblock {The Subtle Simplicity of Cosmological Correlators}.
\newblock 12 2023.

\bibitem{Sachs2}
Till Heckelbacher, Ivo Sachs, Evgeny Skvortsov, and Pierre Vanhove.
\newblock {Analytical evaluation of cosmological correlation functions}.
\newblock {\em JHEP}, 08:139, 2022.

\bibitem{CosmoTreeTheo}
Santiago Agui~Salcedo and Scott Melville.
\newblock {The cosmological tree theorem}.
\newblock {\em JHEP}, 12:076, 2023.

\bibitem{Meltzer}
David Meltzer.
\newblock {The inflationary wavefunction from analyticity and factorization}.
\newblock {\em JCAP}, 12(12):018, 2021.

\bibitem{Bittermann}
Noah Bittermann and Austin Joyce.
\newblock {Soft limits of the wavefunction in exceptional scalar theories}.
\newblock {\em JHEP}, 03:092, 2023.

\bibitem{CespedesScott}
Sebasti\'an C\'espedes, Anne-Christine Davis, and Scott Melville.
\newblock {On the time evolution of cosmological correlators}.
\newblock {\em JHEP}, 02:012, 2021.

\bibitem{circles}
Nima Arkani-Hamed and Carolina Figueiredo.
\newblock {Circles and Triangles, the NLSM and Tr($\Phi^3$)}.
\newblock 3 2024.

\bibitem{UnivSplit}
Qu~Cao, Jin Dong, Song He, Canxin Shi, and Fanky Zhu.
\newblock {On universal splittings of tree-level particle and string scattering
  amplitudes}.
\newblock {\em JHEP}, 09:049, 2024.

\bibitem{CosmoLightStates}
Paolo Benincasa.
\newblock {Cosmological Polytopes and the Wavefuncton of the Universe for Light
  States}.
\newblock 9 2019.

\bibitem{SurfaceKin}
Nima Arkani-Hamed, Qu~Cao, Jin Dong, Carolina Figueiredo, and Song He.
\newblock {Surface Kinematics and ''The'' Yang-Mills Integrand}.
\newblock 8 2024.

\bibitem{CosmoFull}
N.~Arkani-Hamed, C.~Figueiredo, and F.~Vazão.
\newblock {to appear}.
\newblock 2025.

\bibitem{PostnikovGA1}
Alexander Postnikov.
\newblock Permutohedra, associahedra, and beyond, 2005.

\bibitem{PostnikovGA2}
Alexander Postnikov, Victor Reiner, and Lauren Williams.
\newblock Faces of generalized permutohedra, 2007.

\bibitem{CausalDiamonds}
Nima Arkani-Hamed, Song He, Giulio Salvatori, and Hugh Thomas.
\newblock {Causal diamonds, cluster polytopes and scattering amplitudes}.
\newblock {\em JHEP}, 11:049, 2022.

\bibitem{BinGeom}
Nima Arkani-Hamed, Song He, Thomas Lam, and Hugh Thomas.
\newblock {Binary geometries, generalized particles and strings, and cluster
  algebras}.
\newblock {\em Phys. Rev. D}, 107(6):066015, 2023.

\bibitem{StringyCanForms}
Nima Arkani-Hamed, Song He, and Thomas Lam.
\newblock {Stringy canonical forms}.
\newblock {\em JHEP}, 02:069, 2021.

\bibitem{polymake}
Ewgenij Gawrilow and Michael Joswig.
\newblock {\em {polymake: a Framework for Analyzing Convex Polytopes}}.
\newblock 2000.

\bibitem{Bunch:1978yq}
T.~S. Bunch and P.~C.~W. Davies.
\newblock {Quantum Field Theory in de Sitter Space: Renormalization by Point
  Splitting}.
\newblock {\em Proc. Roy. Soc. Lond. A}, 360:117--134, 1978.

\end{thebibliography}

\newpage

\begin{figure}[t]
    \centering
    \includegraphics[width=\linewidth]{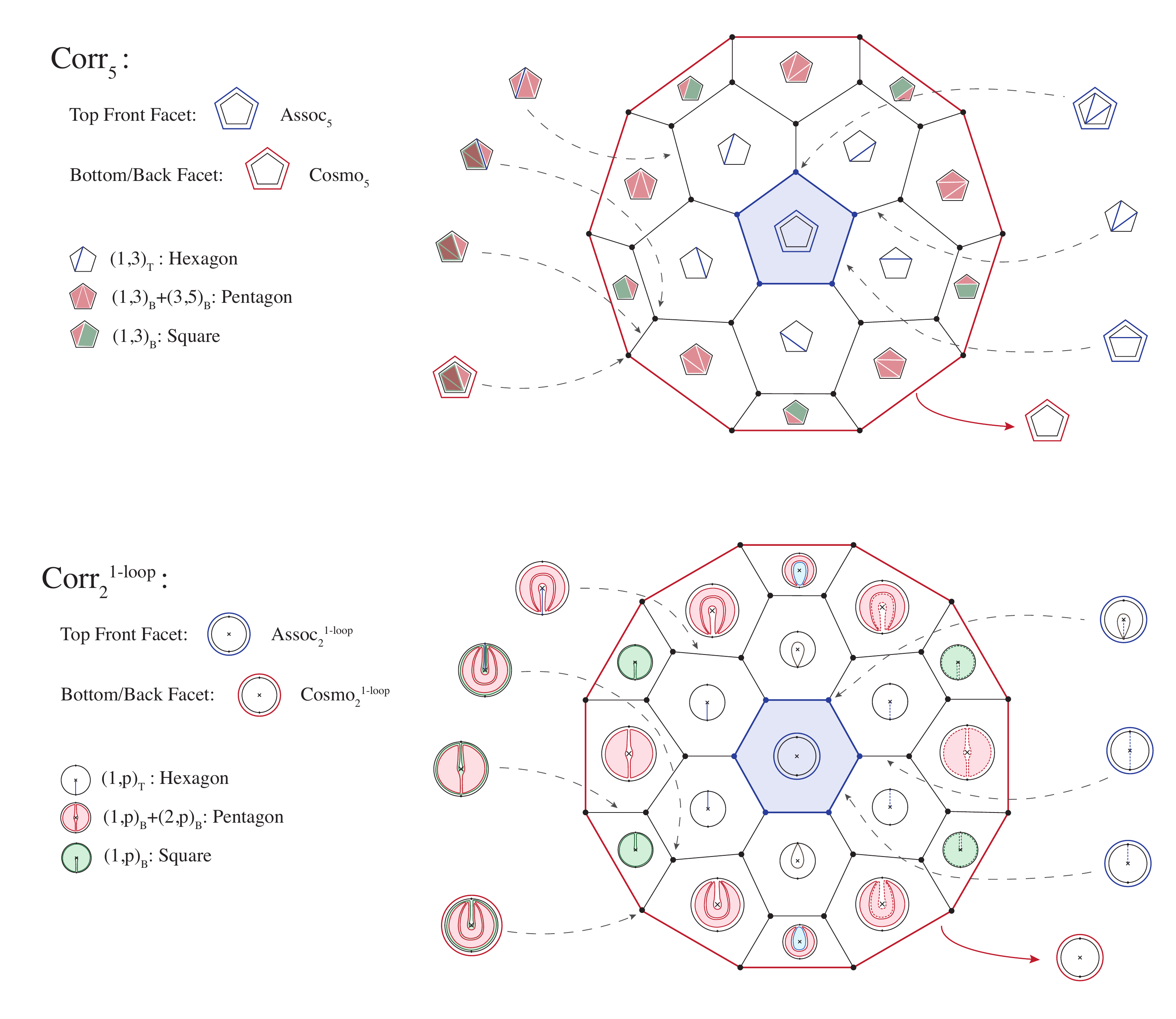}
    \vspace{2mm}
    \caption{The top figure shows the combinatorial structure of the three-dimensional $n=5$ cosmological correlahedron, looked at from above. We see the top pentagon facet as the $n=5$ associahedron, and the bottom decagon as the $n=5$ cosmohedron. A number of other facets, edges and vertices are labelled by collections $(C,P)$ of non-overlapping chords and subpolygons. There are 30 vertices. The  top 5 vertices (marked in blue) all correspond to triangulations of the associahedron, the rest of the vertices are all the terms in the correlator. The bottom figure shows exactly the same for the $n=2$, 1-loop correlator. As for the amplitude polytopes, there are two kinds of ``loop" variable, touching the puncture. The top facet is the hexagon familiar from the amplitude. The bottom is the dodecagon for the cosmohedron. The vertices not on the top facet all correspond to the terms in the correlator.}
    \label{fig:corrcomb}
\end{figure}

\end{document}